\documentclass[prd,aps,12pt,preprintnumbers,amssymb,amsmath,amsfonts,nofootinbib]{revtex4-2}
\usepackage{latexsym}
\usepackage{ulem}
\usepackage[latin1]{inputenc}
\usepackage{cancel}
\newcommand{\ba}{\begin{eqnarray}}
\newcommand{\ea}{\end{eqnarray}}

\newcommand{\be}{\begin{equation}}
\newcommand{\ee}{\end{equation}}
\newcommand{\pa}{\partial}

\newcommand{\nn}{\nonumber}

\newcommand{\PB}{\textrm{PB}}

\usepackage{slashed}
\usepackage{pifont}
\usepackage{dcolumn}
\usepackage{graphics}                                                                         
\usepackage[dvips]{graphicx}
\usepackage{multirow}
\usepackage{makecell}   

\usepackage[x11names,dvipsnames]{xcolor}
\usepackage{hyperref}
\usepackage{tikz-feynman}
\tikzfeynmanset{compat=1.1.0}
\usepackage{tikz}
\usetikzlibrary{trees}
\usetikzlibrary{decorations.pathmorphing}
\usetikzlibrary{decorations.markings}
\usepackage{subcaption}

\begin{document}
                                                                                
\date{\today}
\title{In-medium kinetic theory of $D$ mesons \\ and heavy-flavor transport coefficients}
\author{Juan M. Torres-Rincon$^1$, Gl\`oria Monta\~na$^2$, \`Angels Ramos$^2$, and Laura Tolos$^{3,4,5,6}$}
\affiliation{$^1$Institut f\"ur Theoretische Physik, Goethe Universit\"at Frankfurt, Max von Laue Strasse 1, 60438 Frankfurt, Germany}
\affiliation{$^2$Departament de F\'isica Qu\`antica i Astrof\'isica and Institut de Ci\`encies del Cosmos (ICCUB), Facultat de F\'isica,  Universitat de Barcelona, Mart\'i i Franqu\`es 1, 08028 Barcelona, Spain}
\affiliation{$^3$Institute of Space Sciences (ICE, CSIC), Campus UAB, Carrer de Can Magrans, 08193, Barcelona, Spain}
\affiliation{$^4$Institut d'Estudis Espacials de Catalunya (IEEC), 08034 Barcelona, Spain}
\affiliation{$^5$Faculty of Science and Technology, University of Stavanger, 4036 Stavanger, Norway}
\affiliation{$^6$Frankfurt Institute for Advanced Studies, Ruth-Moufang-Str. 1, 60438 Frankfurt am Main, Germany}
\date{\today}

\begin{abstract}

  We extend the kinetic theory of $D$ mesons to accommodate thermal and off-shell effects due to the medium modification of the heavy-meson spectral functions. From the Kadanoff-Baym approach we derive the off-shell Fokker-Planck equation which encodes the heavy-flavor transport coefficients. We analyze the thermal width (damping rate) of $D$ mesons due to their scattering off light mesons, focusing on new in-medium effects: off-shell corrections, inelastic channels, and the contribution of the Landau cut. We obtain that the latter effect (absent for vacuum scattering amplitudes) brings sizable corrections at moderate temperatures. We discuss how the heavy-flavor transport coefficients, like the drag and diffusion coefficients, are modified in matter. We find that the $D$-meson spatial diffusion coefficient matches smoothly to the latest results of lattice-QCD calculations and Bayesian analyses at higher temperatures.

\end{abstract}

\maketitle


\section{Introduction}

Heavy hadrons are considered to be an efficient and unique probe for testing the different quantum chromodynamics (QCD) phases created in heavy-ion collisions (HiCs), in both quark-gluon plasma (QGP) and hadronic phases (see Refs.~\cite{Aarts:2016hap,Prino:2016cni,Dong:2019unq,Dong:2019byy,Zhao:2020jqu} for recent reviews). Due to the large mass of the heavy (charm and bottom) quarks as compared to the mass of the light-flavor quarks, heavy quarks have large relaxation times and, thus, cannot totally relax to equilibrium during the fireball expansion in HiCs. For this reason, heavy mesons constitute ideal probes to characterize the QGP properties and determining their in-medium properties in a hadronic medium at extreme conditions is a subject that attracts a lot of interest nowadays. 

The characterization of the different QCD phases can be performed by analyzing experimental observables in HiCs, such as the nuclear modification ratio as well as the elliptic flow~\cite{Song:2015sfa,Aarts:2016hap,Prino:2016cni,Dong:2019unq,Dong:2019byy,Zhao:2020jqu}. These physical observables are strongly correlated to the behavior of the transport properties of heavy hadrons, and these depend crucially on the interaction of the heavy particles with the surrounding medium.

In particular, the diffusion of open-charm ($D$ mesons) in hadronic matter was initially obtained within an effective theory that incorporated both chiral and heavy-quark symmetries~\cite{Laine:2011is}, and also using parametrized interactions with light mesons and baryons~\cite{He:2011yi}. Moreover, the scattering amplitudes of $D$ mesons with light mesons and baryons were obtained by means of effective Lagrangians at leading order~\cite{Ghosh:2011bw}. However, the need of unitarization was later pointed out so as to avoid unphysically large transport coefficients~\cite{Abreu:2011ic}.

Following these initial works, we exploited chiral and heavy-quark symmetries to obtain the unitarized effective interaction of heavy mesons, such as $D$~\cite{Tolos:2013kva,Torres-Rincon:2013nfa} and also $\bar B$~\cite{Abreu:2012et,Torres-Rincon:2014ffa}, with light mesons and baryons. With these interactions, we obtained the heavy-meson transport coefficients as a function of temperature and baryochemical potential of the hadronic bath by means of the Fokker-Planck equation approach ~\cite{Tolos:2013kva,Torres-Rincon:2013nfa,Ozvenchuk:2014rpa}. Moreover, we examined the transport coefficients of the low-lying heavy baryons ($\Lambda_c$ and $\Lambda_b$) using a similar unitarized framework to account for the interaction of these states with light mesons~\cite{Tolos:2016slr,Das:2016llg}. 

Similar approaches using different models or effective descriptions, both below and above the phase transition, have been developed (see Refs.~\cite{
vanHees:2004gq,Moore:2004tg,Mannarelli:2005pz,vanHees:2005wb,
CasalderreySolana:2006rq,vanHees:2007me,Beraudo:2009pe,
He:2011yi,Ghosh:2011bw,He:2011qa,Das:2012ck,Berrehrah:2013mua,
Ozvenchuk:2014rpa,Das:2015ana,Lang:2016jpe,Liu:2018syc} for some references). These works exploited the Fokker-Planck (or Langevin) equation description for the heavy particles. Some studies pointed out limitations in the QGP phase and considered the Boltzmann equation instead~\cite{Das:2013kea,Tolos:2016slr}. While Fokker-Planck or Boltzmann kinetic equations seem natural starting points to address transport coefficients, the scattering amplitudes used to describe collisions (computed from a microscopic model) were completely independent of the kinetic theory. Therefore, from a purely theoretical perspective, there is a lack of internal consistency in these calculations, as it would be desirable to construct both the interaction rates and the transport equation from the same microscopic theory. 

On the other hand, in most of the previous analyses the transition amplitudes of heavy mesons with light hadrons and, hence, the transport coefficients were calculated without implementing medium corrections for the interactions. Indeed, off-shell effects cannot be accounted for in the standard Boltzmann or Fokker-Planck equations. For that, an extension using the more general Kadanoff-Baym equations is required. Second, in-medium interactions induce new kinematic domains which affect the $D$-meson properties. These new effects---which would also affect $D$-meson transport coefficients---can be naturally incorporated on the derivation of an off-shell kinetic equation. Therefore, to apply our findings of Refs.~\cite{Montana:2020lfi,Montana:2020vjg} we are forced to address the derivation of an off-shell kinetic equation and, with this, to consistently describe interaction rates and the transport equation from the same effective theory. 

In this paper we focus on $D$ mesons and analyze the effect on the transport properties of the scattering of these heavy mesons with a mesonic bath at finite temperature. To this end we make use of the in-medium unitarized amplitudes in a mesonic environment at finite temperature recently developed in Refs.~\cite{Montana:2020lfi,Montana:2020vjg}, that have been tested against lattice QCD calculations below the temperature of the QCD phase transition~\cite{Montana:2020var}. The final goal is to calculate the $D$-meson transport coefficients below the transition temperature, paying a special attention to the inclusion of off-shell effects coming from the full spectral features of the $D$ meson in a hot mesonic bath. We will compare our results with the latest results of lattice-QCD~\cite{Banerjee:2011ra,Kaczmarek:2014jga,Francis:2015daa,Brambilla:2020siz,Altenkort:2020fgs} and Bayesian analyses~\cite{Ke:2018tsh} performed for temperatures close and above $T_c$.

The paper is organized as follows. In Sec.~\ref{sec:Boltzmann} we introduce the kinetic equation for $D$ mesons incorporating the $T$-matrix approximation in the collision terms. In Sec.~\ref{sec:equil} we particularize to an equilibrium situation and describe the $T$-matrix approach and the spectral properties of $D$ mesons at finite temperature. In Sec.~\ref{sec:kinematic} we show the different kinematic contributions to the $D$-meson thermal width coming from the off-shell treatment of the $D$ meson. In Sec.~\ref{sec:transport} we obtain the different transport coefficients within the off-shell approach after re-deriving the Fokker-Planck equation. Our results are given in Sec.~\ref{sec:results}, while we conclude in Sec.~\ref{sec:conclusions}. In Appendix~\ref{app:onshell} we detail the derivation of the on-shell Fokker-Planck equation, to connect to the formalism used in previous works.

\section{Non-equilibrium description: \\ Off-shell kinetic equation with $T$-matrix approximation~\label{sec:Boltzmann}}

In this section we describe the kinetic equation for $D$ mesons interacting with light particles. The corresponding effective field theory (both in vacuum and at finite $T$) has been developed in Refs.~\cite{Montana:2020lfi,Montana:2020vjg}. Therefore we would like to be able to use quantum field theory techniques in order to achieve our goal. This can be addressed via the Kadanoff-Baym equations~\cite{kadanoff1962quantum}, which have been derived for real scalar fields several times in the literature~\cite{kadanoff1962quantum,Danielewicz:1982kk,calzetta1988nonequilibrium,Botermans:1990qi,Davis:1991zc,Blaizot:1999xk,Rammer,Juchem:2003bi,Cassing:2008nn,bonitz2016quantum} using the real-time formalism~\cite{Chou:1984es,Bellac:2011kqa}. For this reason, we skip standard steps in the derivation of the kinetic theory, and only stress the particular details concerning our heavy-light system.

We want to note that in this work the derivation of a kinetic theory---technically more involved than the calculation of equilibrium quantities---is employed to determine the particular form of the transport equation, and then, to provide a rigorous and practical definition of the transport coefficients. However, the actual calculation of these coefficients (as well as the thermal decay widths) only requires equilibrium properties, which will be mostly taken from our previous Refs.~\cite{Montana:2020lfi, Montana:2020vjg}. The eventual solution of the kinetic equation in real time is not addressed in this work and left for future studies.

Out of equilibrium, the fundamental quantities are the two Wightman functions for the $D$ meson. Their definition is
\begin{align} 
  iG_D^> (x,y) \equiv \langle D(x) D(y) \rangle \ , \label{eq:Ggreat}  \\
  iG_D^< (x,y) \equiv \langle D(y) D(x) \rangle \ , \label{eq:Gless}
\end{align}
and they correspond to the time-ordered Green's function (ordered along the real-time contour) of the $D$-meson propagator, depending on the relative ordering of the time arguments $x^0,y^0$~\cite{kadanoff1962quantum,Blaizot:1999xk,Rammer,Cassing:2008nn}. 

The momentum dependence is introduced through a Wigner transform (Fourier transform of the relative coordinates). Then we focus on $G_D^<(X,k)$, also called Wigner function ($X$ being the center of mass coordinates). This function satisfies the following kinetic equation~\cite{kadanoff1962quantum,Danielewicz:1982kk,calzetta1988nonequilibrium,Botermans:1990qi,Davis:1991zc,Blaizot:1999xk,Rammer,Juchem:2003bi,Cassing:2008nn,bonitz2016quantum},
  
\begin{align} 
& \left(  k^\mu - \frac{1}{2} \frac{\pa \textrm{Re } \Pi^R (X,k)}{\pa k_\mu} \right) \frac{\pa iG_D^< (X,k)}{\pa X^\mu} + \frac{1}{2}  \frac{\pa \textrm{Re } \Pi^R (X,k)}{\pa X^\mu} \frac{\pa iG_D^< (X,k)}{\pa k_\mu}  \nn \\
& - \frac{i}{2}  \{ \Pi^< (X,k), \textrm{Re } G_D^R (X,k) \}_{\PB}
=\frac{1}{2}  i\Pi^< (X,k)  iG_D^> (X,k) -\frac{1}{2} i\Pi^> (X,k) iG_D^< (X,k) \ , 
\label{eq:almostoffGless}
\end{align}
which is given as function of 4-position $X^\mu=(t,{\bf X})$ and 4-momentum $k^\mu=(k^0,{\bf k})$. Notice that $k^0$ and $\bf{k}$ are independent variables (although related via the spectral distribution $S_D(X,k)$, to be defined later). Such a general case where the $D$ meson is not on its mass shell---as its energy is not determined by its momentum---will be generically denoted as {\it off-shell}. Already in equilibrium, we have shown in Refs.~\cite{Montana:2020lfi, Montana:2020vjg} that the $D$ meson at $T\neq 0$ is characterized by a continuous spectral function, which represents the distribution of possible energies for a given value of the momentum (see later).

In addition, the $D$-meson properties are modified by different self-energies: $\Pi^R(X,k)$ is the retarded one, and the so-called ``greater'' and ``lesser'' $\Pi^\gtrless(X,k)$ are related to the collision processes of the $D$ mesons~(see more details in Refs.~\cite{Danielewicz:1982kk,calzetta1988nonequilibrium,Botermans:1990qi,Davis:1991zc,Blaizot:1999xk,Rammer,Juchem:2003bi,Cassing:2008nn,bonitz2016quantum}). In fact, the right-hand side of Eq.~(\ref{eq:almostoffGless}) represents the collision term of the transport equation. Also the retarded Green's function $G^R_D(x,y)$ appears in the so-called Poisson bracket of Eq.~(\ref{eq:almostoffGless}).

As in this work we will eventually apply equilibrium properties in an homogeneous background, we neglect the mean-field term in Eq.~(\ref{eq:almostoffGless}), proportional to $\pa_{X^\mu} \textrm{Re } \Pi^R$~\cite{Blaizot:2001nr}. In addition, we will not consider the Poisson bracket in the left-hand side. This term was shown to be unimportant in the quasiparticle limit and can be safely neglected~\cite{Danielewicz:1982kk,Blaizot:2001nr,Cassing:2008nn}. However, one should not forget that in the off-shell case it contributes to the out-of-equilibrium dynamics of $G^<_D (X,k)$~\cite{Botermans:1990qi,Blaizot:2001nr} (see also Ref.~\cite{Cassing:2008nn} and references therein).

With these approximations we arrive to the final form of the ``off-shell'' transport equation,
\be \left(  k^\mu - \frac{1}{2} \frac{\pa \textrm{Re } \Pi^R (X,k)}{\pa k_\mu} \right) \frac{\pa iG_D^< (X,k)}{\pa X^\mu} =\frac{1}{2}  i\Pi^< (X,k)  iG_D^> (X,k) -\frac{1}{2} i\Pi^> (X,k) iG_D^< (X,k) \ . \label{eq:offGless} \ee

For completeness, the transport equation for $G_D^>(X,k)$ is similar to Eq.~(\ref{eq:offGless}), and shares the same collision term,
\be \left(  k^\mu - \frac{1}{2} \frac{\pa \textrm{Re } \Pi^R (X,k)}{\pa k_\mu} \right) \frac{\pa iG_D^> (X,k)}{\pa X^\mu} =\frac{1}{2}  i\Pi^< (X,k)  iG_D^> (X,k) -\frac{1}{2} i\Pi^> (X,k) iG_D^< (X,k) \ . \label{eq:offGgreat} \ee

Related to the dispersion relation and the spectral function of the $D$ mesons one can consider the equations for the retarded/advanced Green's function $G_D^{R/A}$~\cite{kadanoff1962quantum,Danielewicz:1982kk,calzetta1988nonequilibrium,Botermans:1990qi,Davis:1991zc,Blaizot:1999xk,Rammer,Juchem:2003bi,Cassing:2008nn,bonitz2016quantum}. For a scalar field, these are complex conjugated, $G_D^R (X,k) = [G_D^A (X,k)]^* $, and related to the Wightman functions as
\be G_D^R (X,k) -G_D^A (X,k) =G_D^> (X,k) -G_D^< (X,k)  \ . 
\label{eq:greenrelation} \ee

After some standard manipulations of the equations of motion one can arrive to the familiar equation,
\be [ k^2  - m_{D,0}^2 -  \Pi^R(X,k) ] \ G_D^R(X,k)=1 \ , \label{eq:eqforGR} \ee
where the (out-of-equilibrium) retarded self-energy dresses the bare mass of the $D$ meson. The pole of the retarded Green's function, i.e., the zero of the left-hand side of Eq.~(\ref{eq:eqforGR}), will generically provide the dispersion relation of the $D$ meson modified by interactions. In the next section we will detail its form in the homogeneous and equilibrium case~\cite{Montana:2020lfi}.

To close the equation for $G_D^<(X,k)$ the remaining step is to detail the $D$-meson self-energies in terms of the Green's functions. In our case, in consistency with the effective approach described in Refs.~\cite{Montana:2020vjg,Montana:2020lfi}, we will incorporate exact unitarity constraints to the scattering matrix, implementing an in-medium $T$-matrix resummation of the scattering amplitudes. In the nonequilibrium context this is called the $T$-matrix approximation~\cite{kadanoff1962quantum,Danielewicz:1982kk,Botermans:1990qi}, where the $D$-meson self-energies $\Pi^\gtrless(X,k)$ can be written in terms of the (retarded) $T$-matrix element as~\cite{kadanoff1962quantum},

\begin{align}
  i\Pi^< (X,k)  = & \sum_{ \{ a,b,c \} } \int \frac{d^4 k_1}{(2\pi)^4} \int \frac{d^4 k_2}{(2\pi)^4} \int \frac{d^4 k_3}{(2\pi)^4} (2\pi)^4 \delta^{(4)} (k_1+k_2-k_3-k)  \nn \\
  &\times |T (k_1^0+k_2^0+i\epsilon, {\bf k}_1 + {\bf k}_2)|^2 \ iG_{D_a}^< (X,k_1) iG_{\Phi_b}^< (X,k_2) iG_{\Phi_c}^>(X,k_3) \ , \label{eq:Pil} \\
  i\Pi^> (X,k)  = &  \sum_{ \{ a,b,c \} } \int \frac{d^4 k_1}{(2\pi)^4} \int \frac{d^4 k_2}{(2\pi)^4} \int \frac{d^4 k_3}{(2\pi)^4} (2\pi)^4 \delta^{(4)} (k_1+k_2-k_3-k)  \nn \\
  &\times |T (k_1^0+k_2^0+i\epsilon, {\bf k}_1 + {\bf k}_2)|^2 iG_{D_a}^> (X,k_1) iG_{\Phi_b}^> (X,k_2) iG_{\Phi_c}^<(X,k_3) \ . \label{eq:Pig}
\end{align}

\begin{figure}[ht]
  \centering
  \includegraphics[width=80mm]{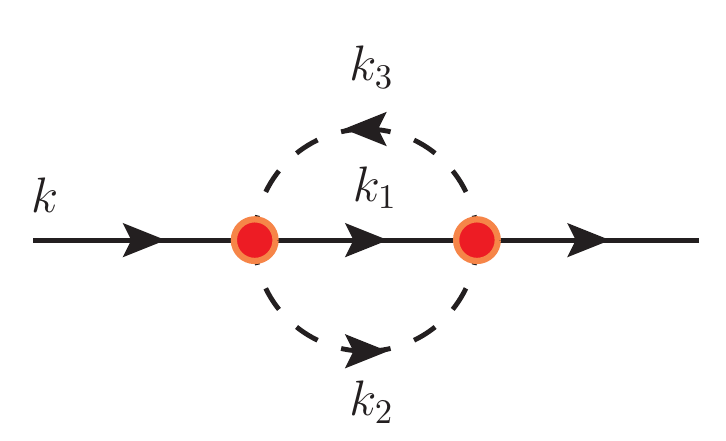}
  \caption{The 2-loop structure of $\Pi^\lessgtr(X,k)$ for the $D$ meson. Solid lines: $D$-meson propagator $G_D^\lessgtr(X,k_1)$; Dashed lines: light meson propagator $G_\Phi^\lessgtr(X,k_2),G_\Phi^\gtrless(X,k_3)$; Red circles: $T$-matrix operators.}
  \label{fig:Pilessgtr}
\end{figure}

Diagrammatically, the self-energies $\Pi^\lessgtr(X,k)$ are given by the two-loop diagram represented in Fig.~\ref{fig:Pilessgtr}. The solid lines represent heavy mesons, while dashed lines are $\Phi$ propagators, and the vertices stand for the full $T$ matrices. All of them are nonequilibrium quantities and only known after solving the transport equation.

\begin{figure}[ht]
  \centering
  \includegraphics[width=160mm]{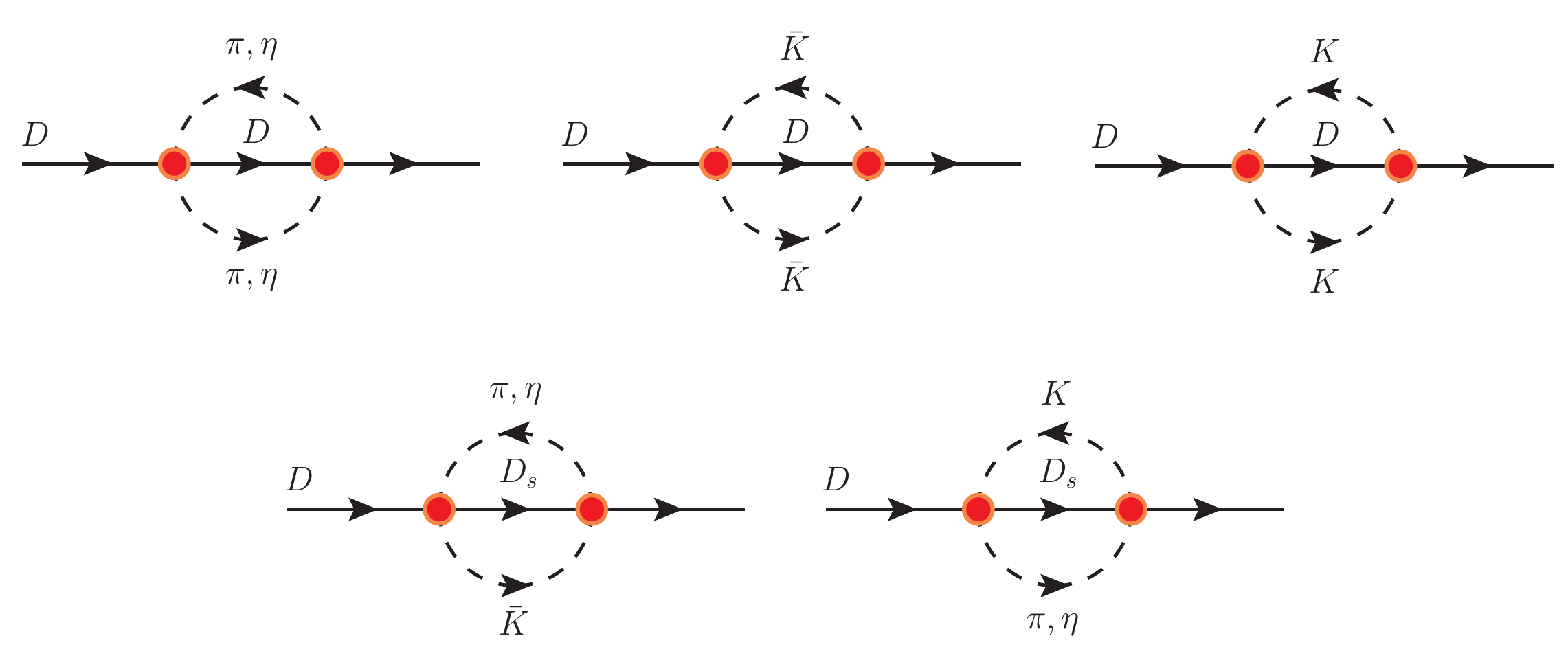}
  \caption{Diagrams contributing to $\Pi_D^\lessgtr(X,k)$ for the $D$ meson in Eqs.~(\ref{eq:Pil}) and (\ref{eq:Pig}). A total of 10 channels are needed due to the coupled-channels problem for the $D$-meson interaction. Red circles are $T$-matrix elements in the appropriate channel.
  \label{fig:Pilessgtrchannels}}
\end{figure}

The sum over $a,b$ and $c$ in Eqs.~(\ref{eq:Pil}) and (\ref{eq:Pig}) encodes the different species that can interact and that are fixed by the effective vertices. This sum is restricted to particular combinations (respecting conservation of quantum numbers) which are described in detail in Ref.~\cite{Montana:2020vjg}, where a full coupled-channels analysis was performed. In particular, $D_a$ can describe either $D$ or $D_s$ mesons, and $\Phi_b,\Phi_c$ can represent $\pi,K,\bar{K},\eta$. Figure~\ref{fig:Pilessgtrchannels} shows all 10 allowed diagrams, although not all of them are equally important. We will comment on this when addressing the effect of inelastic processes in our calculations.

We now discuss the form of the Wightman functions $G_{D,\Phi}^\lessgtr (X,k)$. The light degrees of freedom satisfy their own set of Kadanoff-Baym equations, which are coupled to those of the heavy mesons. In the context of heavy-ion collisions, the standard approach for heavy-flavor dynamics is to exploit the fact that the light degrees of freedom have reached equilibrium much before than the heavy sector, as the latter has a much longer relaxation time (roughly proportional to the mass of the particle). In our goal of accessing transport coefficients of heavy mesons, we will also assume this, so there is no need to consider the kinetic equation for light mesons. 

In addition, we apply the thermal local equilibrium solution for $G_\Phi^\lessgtr(X,k)$, which can be expressed as~\cite{kadanoff1962quantum}
\begin{align}
 iG_{\Phi}^< (X,k) & = 2\pi S_{\Phi} (X,k) f^{(0)}_{\Phi}(X,k^0)  \ , \label{eq:soleq1} \\
 iG_{\Phi}^> (X,k) & = 2\pi S_{\Phi} (X,k) \ [1+f^{(0)}_{\Phi}(X,k^0)] \ , \label{eq:soleq2}
 \end{align}
where $S_\Phi(X,k)$ is the equilibrium light-meson spectral function and $f_\Phi^{(0)}(X,k^0)$ is the equilibrium occupation number, i.e., the Bose-Einstein distribution function. Equation (\ref{eq:soleq2}) incorporates the Bose enhancement factor $1+f_\Phi^{(0)}(X,k^0)$.

Concerning the $D$ mesons, we will assume that they are not far from equilibrium (which is enough to address the calculation of the transport coefficients) and use a similar form as in Eqs.~(\ref{eq:soleq1}) and (\ref{eq:soleq2}) for the Green's function, the so-called Kadanoff-Baym ansatz,
\begin{align}
 iG_{D}^< (X,k) & = 2\pi S_{D} (X,k) f_{D}(X,k^0)  \ , \label{eq:ansatz1} \\
 iG_{D}^> (X,k) & = 2\pi S_{D} (X,k) \ [1+f_{D}(X,k^0)] \ , \label{eq:ansatz2}
 \end{align}
where the $D$-meson spectral function $S_D(X,k)$ and the distribution function $f_D(X,k^0)$ are out of equilibrium.

Inserting these ans\"atze into the kinetic equation (see Eq.~(\ref{eq:offGless})), together with the $D$-meson self-energies defined in Eqs.~(\ref{eq:Pil}) and (\ref{eq:Pig}), one obtains 
\begin{align}
  &  \left(  k^\mu - \frac{1}{2} \frac{\pa \textrm{Re } \Pi^R}{\pa k_\mu} \right) \frac{\pa}{\pa X^\mu} [S_D(X,k) f_D(X, k^0)] =  \frac{1}{2}  \int \prod_{i=1}^3 \frac{d^4 k_i}{(2\pi)^3} (2\pi)^4 \delta^{(4)} (k_1+k_2-k_3-k) \nn \\ 
  &\times |T (k_1^0+k_2^0+i\epsilon, {\bf k}_1 + {\bf k}_2)|^2 \ S_D (X,k_1) S_\Phi (X,k_2) S_\Phi (X,k_3) S_D (X,k)  \label{eq:transportG} \\
& \times \left[ f_D (X, k^0_1) f^{(0)}_\Phi (X, k^0_2) \tilde{f}^{(0)}_\Phi (X, k^0_3)  \tilde{f}_D (X, k^0) -   \tilde{f}_D (X, k^0_1) \tilde{f}^{(0)}_\Phi (X, k^0_2)  f^{(0)}_\Phi (X, k^0_3) f_D (X, k^0) \right]  \ , \nn
  \end{align}
where we defined $\tilde{f}_i(X,k^0) \equiv 1+f_i(X,k^0)$. Notice that we have not written explicitly the sum over $\{a,b,c\}$ in the right-hand side, but it should be understood to account for all possible physical processes.

Focusing on the positive-energy $D$ meson one can integrate over $dk^0$ along the positive branch in both sides,
\begin{align}
  & \int_0^{+\infty} dk^0 \left(  k^\mu - \frac{1}{2} \frac{\pa \textrm{Re } \Pi^R}{\pa k_\mu} \right) S_D(X,k) \frac{\pa f_D(X,k^0)}{\pa X^\mu} \label{eq:offshellkinetic} \\ 
  & = \frac{1}{2} \int_0^{+\infty} dk^0 \int \prod_{i=1}^3 \frac{d^4 k_i}{(2\pi)^3} (2\pi)^4 \delta^{(4)} (k_1+k_2-k_3-k) |T (k_1^0+k_2^0+i\epsilon, {\bf k}_1 + {\bf k}_2)|^2  \nn \\
  &\times S_D (X,k_1) S_\Phi (X,k_2) S_\Phi (X,k_3) S_D (X,k) \nn \\
& \times \left[ f_D (X, k^0_1) f^{(0)}_\Phi (X, k^0_2) \tilde{f}^{(0)}_\Phi (X, k^0_3)  \tilde{f}_D (X, k^0) -   \tilde{f}_D (X, k^0_1) \tilde{f}^{(0)}_\Phi (X, k^0_2)  f^{(0)}_\Phi (X, k^0_3) f_D (X, k^0) \right]  \ , \nn
  \end{align}
where the transport equation for the spectral function has been used after Eq.~(\ref{eq:transportG})\footnote{Note that from Eqs.~(\ref{eq:ansatz1}) and (\ref{eq:ansatz2}) one has $S_D(X,k)=i[G_D^>(X,k)-G_D^<(X,k)]/(2\pi)$. Then, subtracting the transport equations, Eq.~(\ref{eq:offGgreat}) minus Eq.~(\ref{eq:offGless}), one obtains that the transport equation for the spectral density reads
$$ \left(  k^\mu - \frac{1}{2} \frac{\pa \textrm{Re } \Pi^R (X,k)}{\pa k_\mu} \right) \frac{\pa S_D (X,k)}{\pa X^\mu} = 0 \ . $$}. 

This equation is very similar to the standard Boltzmann equation but the effects of the medium (temperature and density) are here incorporated into the $T$-matrix and the spectral functions of the interacting particles. The reduction of this equation to the classical Boltzmann equation makes an extra assumption for the spectral functions, the {\it quasiparticle approximation}. In that approximation the spectral function is only characterized by the quasiparticle energy $E_k$ and the thermal decay width $\gamma_k$, and $S(k)$ admits a Lorenztian shape peaked at $E_k$ and with a spectral width $\gamma_k \ll E_k$. In the limit $\gamma_k/E_k \rightarrow 0$ one can consider the {\it narrow limit},
\be S(k) \rightarrow \frac{z_k}{2E_k} [\delta(k^0-E_k)-\delta(k^0+E_k)] \ . \label{eq:Diracdelta} \ee
In this limit one is effectively treating the quasiparticle as a stable state (no thermal width), but with a medium-modified energy $E_k$ plus a correction due to the $z_k$ factor.

\begin{figure}[ht]
  \centering
  \includegraphics[scale=1]{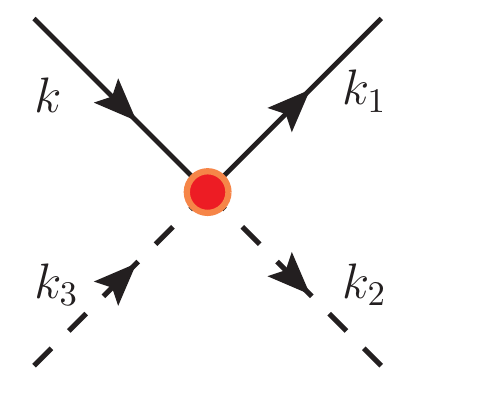}
  \caption{Labelling of incoming and outgoing momenta in a generic binary scattering. Solid lines represent heavy mesons, while dashed lines represent light mesons. The vertex corresponds to a retarded $T$-matrix element.
  \label{fig:naming}}
\end{figure}

The narrow limit allows to trivially perform the integration over $k^0_i$ variables in the kinetic equation in Eq.~(\ref{eq:offshellkinetic}) to obtain the on-shell (or Boltzmann) transport equation. When assuming this approximation for all the particles involved, different combinations of the Dirac delta functions can be taken, which describe different scattering processes~\cite{Blaizot:1999xk,Blaizot:2001nr}. However in this limit, the energy-momentum conservation only allows $2\leftrightarrow 2$ processes~\cite{Blaizot:2001nr,Juchem:2003bi}. Among those, the one with a $D \bar{D}$ pair in the initial state can be neglected due to the low $D$-meson population (and the inverse reaction is suppressed due to the high energy threshold for the two incoming pions). Therefore one only considers one type of collisions ($D\Phi \rightarrow D\Phi$). We label the momenta of a generic scattering as $k+3 \rightarrow 1+2$, see Fig.~\ref{fig:naming} for illustration.

Denoting for simplicity $f_i\equiv f_{D,\Phi} (X,E_i)$ (where the species is fixed by the value of $i$, i.e., $i=\{1,k\}$ for the heavy meson, $i=\{2,3\}$ for the light meson), we obtain
\begin{align}
 & \left[  \frac{\pa}{\pa t} - \frac{{\bf k}}{E_k} \cdot \nabla_X \right] f_k  =  \frac{z_{k}}{2E_k} \int \left( \prod_{i=1}^3 \frac{d^3 k_i \ z_i}{(2\pi)^3 2E_i} \right)  (2\pi)^4 \delta^{(3)} ({\bf k}+ {\bf k}_3-{\bf k}_1-{\bf k}_2) \nn \\ 
& \times \left\{ \delta(E_k+E_3-E_1-E_2) \ |T (E_k+E_3, {\bf k} + {\bf k}_3)|^2 
 \ \left( f_1 f_2 \tilde{f}_3 \tilde{f}_k-   \tilde{f}_1 \tilde{f}_2  f_3 f_k \right) \right.  \nn \\
& \left. + \delta(E_k-E_3-E_1+E_2) \ |T (E_k-E_3, {\bf k} + {\bf k}_3)|^2 
 \ \left( f_1 \tilde{f}_2 f_3  \tilde{f}_k -   \tilde{f}_1 f_2 \tilde{f}_3 f_k \right)  \right\} \ ,  \label{eq:onshellkinetic}
\end{align}
where the two different collision terms depend on the sign of the energy of the light meson $E_3$. The first one evaluates the scattering amplitude above the mass threshold of particles $k$ and $3$ and we will refer to it as {\it unitary contribution}. The second one implies the scattering amplitude below the energy threshold and it is nonzero due to the so-called Landau cut~\cite{Weldon:1983jn,Ghosh:2011bw} of the two-particle propagator. This term will be referred to as {\it Landau contribution}, and it will be of key importance in this work.

If only elastic collisions are considered, then one can simplify the equation by exchanging variables ${\bf k}_2$ and ${\bf k}_3$ in the last term to obtain,
\begin{align}
 & \left[  \frac{\pa}{\pa t} - \frac{{\bf k}}{E_k} \cdot \nabla_X \right] f_k  =  \frac{z_{k}}{2E_k} \int \left( \prod_{i=1}^3 \frac{d^3 k_i \ z_i}{(2\pi)^3 2E_i} \right)  (2\pi)^4 \delta^{(4)} (k+ k_3-k_1- k_2) \nn \\ 
& \times \left[ |T (E_k+E_3+i\epsilon, {\bf k} + {\bf k}_3)|^2 +
|T (E_k-E_2+i\epsilon, {\bf k} - {\bf k}_2)|^2  \right] 
 \ \left( f_1 f_2 \tilde{f}_3 \tilde{f}_k-   \tilde{f}_1 \tilde{f}_2  f_3 f_k \right) \ .   \label{eq:onshellkineticelastic}
\end{align}
This equation looks almost as the Boltzmann equation\footnote{More exactly, the Boltzmann-Uehling-Uhlenbeck equation, as quantum effects are incorporated.} considered in previous works, where the effect of the Landau contribution was neglected, vacuum scattering amplitudes were employed, and the factors $z_i$ were set to 1. A version of the transport equation where these factors were kept, is presented in Ref.~\cite{Davis:1991zc}. In our particular case, the approximation $z_i \simeq 1$ is an excellent one, given the good quasiparticle description of the $D$ mesons. Notice that the quasiparticle energies in Eq.~(\ref{eq:onshellkineticelastic}) and the $T$ matrix do contain medium modifications.

\section{$D$-meson properties at thermal equilibrium~\label{sec:equil}}

In the previous section we have sketched the derivation of a kinetic equation for the $D$ mesons, which sets a starting point in the study of the non-equilibrium evolution of these heavy particles. As explained, the analysis of the real-time dynamics will be analyzed in a future study, as our current goal is to extract transport coefficients, for which equilibrium properties are enough. 

In this section we analyze the different elements appearing in the off-shell kinetic equation of Eq.~(\ref{eq:offshellkinetic}), namely the $T$-matrix elements, the retarded $D$-meson self-energy and the spectral function, for the particular case of a system in equilibrium.

\subsection{$T$-matrix equation at finite temperature}

\begin{figure}[htbp!]
\centering 
\begin{subfigure}[b]{0.54\textwidth} \centering
\captionsetup{skip=0pt}
 \begin{tikzpicture}[baseline=(i.base)]
    \begin{feynman}[small]
      \vertex (a) {\(D_i\)};
      \vertex [below right = of a] (i) {};
      \vertex [above right = of i] (b) {\(D_j\)};
      \vertex [below right = of i] (d) {\(\Phi_j\)};
      \vertex [below left=of i] (c) {\(\Phi_i\)};
      \diagram* {
        (a) -- [fermion] (i), 
        (i) -- [fermion] (b),
        (c) -- [charged scalar] (i),
        (i) -- [charged scalar] (d),
       };
     \draw[dot,minimum size=4mm,thick,orange,fill=red] (i) circle(1.5mm);
    \end{feynman}
  \end{tikzpicture}
  $=$
  \begin{tikzpicture}[baseline=(i.base)]
    \begin{feynman}[small]
      \vertex (a) {\(D_i\)};
      \vertex [below right = of a] (i) {};
      \vertex [above right = of i] (b) {\(D_j\)};
      \vertex [below right = of i] (d) {\(\Phi_j\)};
      \vertex [below left=of i] (c) {\(\Phi_i\)};
      \diagram*{
        (a) -- [fermion] (i), 
        (i) -- [fermion] (b),
        (c) -- [charged scalar] (i),
        (i) -- [charged scalar] (d),
       } ;    
     \draw[dot,blue,fill=blue] (i) circle(.8mm);
    \end{feynman}
  \end{tikzpicture}
  $+$
  \begin{tikzpicture}[baseline=(i.base)]
    \begin{feynman}[small]
      \vertex (a) {\(D_i\)};
      \vertex [below right = of a] (i) {};
      \vertex [right = of i] (j) {};
      \vertex [above right = of j] (b) {\(D_j\)};
      \vertex [below right = of j] (d) {\(\Phi_j\)};
      \vertex [below left=of i] (c) {\(\Phi_i\)};
      \vertex [above right = of i] (b1) {\(D_k\)};
      \vertex [below right=of i] (d1) {\(\Phi_k\)};
      \diagram*{
        (a) -- [fermion] (i), 
        (j) -- [fermion] (b),
        (i) -- [orange, fermion, very thick, half left, looseness=1.2] (j),
        (i) -- [charged scalar, half right, looseness=1.2] (j),
        (c) -- [charged scalar] (i),
        (j) -- [charged scalar] (d),
       } ;    
     \draw[dot,blue,fill=blue] (i) circle(0.8mm);
     \draw[dot,minimum size=4mm,thick,orange,fill=red] (j) circle(1.5mm);
    \end{feynman}
  \end{tikzpicture}
\caption{}
\label{fig:BS-a}
\end{subfigure}
\hspace{3mm}\begin{subfigure}[b]{0.38\textwidth}\centering
\captionsetup{skip=-5pt}
 \begin{tikzpicture}[baseline=(a.base)]
    \begin{feynman}[small]
      \vertex (a) {};
      \vertex [right = of a] (b) {};
      \diagram* {
        (a) -- [orange, fermion, very thick,edge label=\(\textcolor{black}{D}\)] (b), 
       };
    \end{feynman}
  \end{tikzpicture}
  $=$
 \begin{tikzpicture}[baseline=(a.base)]
    \begin{feynman}[small]
      \vertex (a) {};
      \vertex [right = of a] (b) {};
      \diagram* {
        (a) -- [fermion, edge label=\(D\)] (b), 
       };
    \end{feynman}
  \end{tikzpicture}
  $+$
  \begin{tikzpicture}[baseline=(a)]
    \begin{feynman}[small, inline=(a)]
      \vertex (a) {};
      \vertex [right = of a] (i) {};
      \vertex [right = of i] (b) {};
      \vertex [below = 0.4cm of i] (d) {};
      \vertex [below = 0.9cm of i] (e) {\(\pi\)};
      \diagram* {
        (a) -- [fermion,edge label=\(D\)] (i), 
        (i) -- [orange, fermion, very thick,edge label=\(\textcolor{black}{D}\)] (b),
       } ;   
     \draw[dashed] (d) circle(0.3cm);
     \draw[dot,minimum size=4mm,thick,orange,fill=red] (i) circle(1.5mm);
    \end{feynman}
  \end{tikzpicture}
\caption{}
\label{fig:BS-b}
\end{subfigure} \\
\begin{subfigure}[b]{\textwidth}\centering
\captionsetup{skip=-5pt}
  \begin{tikzpicture}[baseline=(a)]
    \begin{feynman}[small, inline=(a)]
      \vertex (a) {};
      \vertex [right = of a] (i) {};
      \vertex [right = of i] (b) {};
      \vertex [below = 0.4cm of i] (d) {};
      \vertex [below = 0.9cm of i] (e) {\(\pi\)};
      \diagram* {
        (a) -- [fermion,edge label=\(D\)] (i), 
        (i) -- [fermion,edge label=\(D\)] (b),
       } ;   
     \draw[dashed] (d) circle(0.3cm);
     \draw[dot,minimum size=4mm,thick,orange,fill=red] (i) circle(1.5mm);
    \end{feynman}
  \end{tikzpicture}
  $=$
  \begin{tikzpicture}[baseline=(a)]
    \begin{feynman}[small, inline=(a)]
      \vertex (a) {};
      \vertex [right = of a] (i) {};
      \vertex [right = of i] (b) {};
      \vertex [below = 0.4cm of i] (d) {};
      \vertex [below = 0.9cm of i] (e) {\(\pi\)};
      \diagram* {
        (a) -- [fermion,edge label=\(D\)] (i), 
        (i) -- [fermion,edge label=\(D\)] (b),
       } ;   
     \draw[dashed] (d) circle(0.3cm);
     \draw[dot,blue,fill=blue] (i) circle(0.8mm);
    \end{feynman}
  \end{tikzpicture}
  $+$
    \begin{tikzpicture}[baseline=(a)]
    \begin{feynman}[small, inline=(a)]
      \vertex (a) {};
      \vertex [right = of a] (i) {};
      \vertex [right = of i] (j) {};
      \vertex [right = of j] (b) {};
      \vertex [above right = of i] (c) {};
      \vertex [below right = of i] (d) {};
      \diagram* {
        (a) -- [fermion,edge label=\(D\)] (i), 
        (i) -- [orange, fermion, very thick] (j),
        (j) -- [fermion,edge label=\(D\)] (b),
        (i) -- [charged scalar, half left] (j),
        (j) -- [charged scalar, half left,edge label=\(\pi\)] (i),
       } ;   
     \draw[dot,blue,fill=blue] (i) circle(0.8mm);
     \draw[dot,blue,fill=blue] (j) circle(0.8mm);
    \end{feynman}  
  \end{tikzpicture}
  $+$
    \begin{tikzpicture}[baseline=(a)]
    \begin{feynman}[small, inline=(a)]
      \vertex (a) {};
      \vertex [right = of a] (i) {};
      \vertex [right = of i] (j) {};
      \vertex [right = of j] (k) {};
      \vertex [right = of k] (b) {};
      \vertex [above right = of i] (c) {};
      \vertex [below right = of i] (d) {};
      \vertex [above right = of j] (e) {};
      \vertex [below right = of j] (f) {};
      \diagram* {
        (a) -- [fermion,edge label=\(D\)] (i), 
        (i) -- [orange, fermion, very thick] (j),
        (j) -- [orange, fermion, very thick] (k),
        (k) -- [fermion,edge label=\(D\)] (b),
        (i) -- [charged scalar, half left] (j),
        (j) -- [charged scalar, half left] (k),
        (k) -- [charged scalar, half left,edge label=\(\pi\),looseness=0.9] (i),
       } ;   
     \draw[dot,blue,fill=blue] (i) circle(0.8mm);
     \draw[dot,blue,fill=blue] (j) circle(0.8mm);
     \draw[dot,blue,fill=blue] (k) circle(0.8mm);
    \end{feynman}  
  \end{tikzpicture}
  $+ \ ...$ 
\caption{}
\label{fig:BS-c}
\end{subfigure}\hspace{-1cm}
\caption{(a) Bethe-Salpeter equation for the $T$ matrix (red circle) of the scattering $D_i\Phi_i\rightarrow D_j\Phi_j$ at finite temperature, with a dressed internal heavy-meson propagator (thick orange line).  (b) Heavy-meson propagator dressed with its thermal self-energy. (c) Pionic contribution to the $D$-meson self-energy, showing the infinite sum of diagrams resulting from the expansion of the $T$ matrix. (Diagrams adapted from Ref.~\cite{Montana:2020lfi})}
\label{fig:BS}
\end{figure}
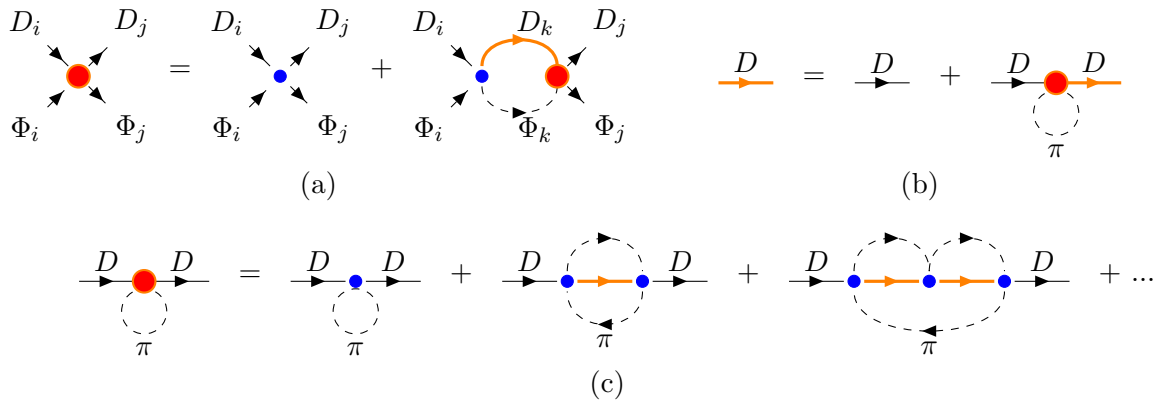

The $T$ matrix appearing in Eq.~(\ref{eq:Pil}) and Eq.~(\ref{eq:Pig}) is a retarded 4-point amplitude which follows from a Bethe-Salpeter equation in coupled channels,
\begin{align}
T_{ij} = V_{ij}+V_{ik}G_{D\Phi,k}T_{kj} \ .
\end{align}
In the equilibrium case, it was solved at finite temperature in Refs.~\cite{Montana:2020vjg,Montana:2020lfi} taking potentials $V_{ij}$ that were obtained from an effective theory for the interaction of charmed mesons with the light pseudoscalar degrees of freedom $\Phi$, respecting both chiral and heavy-quark spin-flavor symmetry. The two-meson loop function $G_{D\Phi,k}$ was regularized with a cutoff, with which we reproduced lattice-QCD data of scattering lengths in vacuum~\cite{Guo:2018tjx}, and at finite temperature it was calculated employing the imaginary-time formalism. The Bether-Salpeter equation at finite temperature is depicted schematically in Fig.~\ref{fig:BS-a}, where the heavy-meson propagator in $G_{D\Phi,k}$ is dressed with the self-energy obtained from closing the pion line of the $T$-matrix element. The Dyson equation for the dressed heavy-meson propagator is graphically represented in Fig~\ref{fig:BS-b}, while the contribution to the $D$-meson self-energy coming from the in-medium pions is shown in Fig~\ref{fig:BS-c} (analogous figures apply for the contribution of other $\Phi$ mesons and the unlabeled intermediate particles can be any $D\Phi$ combination with the proper quantum numbers). Note that the diagramatic equations depicted in Fig.~\ref{fig:BS-a} and Fig~\ref{fig:BS-b} are coupled to each other. In Refs.~\cite{Montana:2020vjg,Montana:2020lfi} the $T$ matrix and the heavy-meson self-energy are calculated self-consistently in thermal equilibrium.

From Fig.~\ref{fig:BS-c} one can see that the unitarity cut of the $T$ matrix provides a source for the charmed-meson vacuum decay width through the imaginary part of the corresponding self-energy. As a matter of example, when a $D$ meson interacts with a pion it can suffer an elastic scattering or, if the energy of the collision is large enough, then the pair can also convert into a $D\eta$ or $D_s {\bar K}$ pair. At finite temperature the bath is populated by $\Phi$ mesons, although their relative importance is weighted by the appropriate Bose-Einstein distribution functions. Consequently, at $T\neq 0$ the contribution of the unitary cut to the decay width is convoluted by statistical weight factors, and additional physical processes appear in the kinematic region of the so-called Landau cut, which are forbidden in vacuum due to kinematic restrictions, such as the absorption of in-medium real light mesons by the $D$ meson.
 
On the other hand, the structure of the $T$ matrix is smeared at finite temperature and the thresholds of the unitary ($\sqrt{s}\geq(m_D+m_\Phi)$) and Landau ($\sqrt{s}\leq|m_D-m_\Phi|$) cuts are smoothened with increasing temperatures as a result of the widening of the $D$-meson spectral function (see Refs.~\cite{Montana:2020vjg,Montana:2020lfi} for details).

\subsection{Quasiparticle approximation at finite temperature}

Let us now consider the $D$-meson self-energy and spectral function at finite temperature. These equilibrium quantities can be computed using the imaginary-time formalism, as it was done in Ref.~\cite{Montana:2020vjg}, and a subsequent analytic continuation for real energies. 

The $D$-meson retarded propagator reads [cf. Eq.~(\ref{eq:eqforGR}) in equilibrium],
\be G^R_D (k^0,{\bf k})= \frac{1 }{(k^0)^2-{\bf k}^2-m_D^2 -\textrm{Re } \Pi^R(k^0,{\bf k};T) - i \textrm{Im } \Pi^R(k^0,{\bf k};T)} \ , \label{eq:GretT} \ee
where $m_{D}$ is the (vacuum) $D$-meson mass, renormalized by the vacuum contribution of the retarded $D$-meson self-energy $\Pi^R$. Please notice that after mass renormalization, the real and imaginary parts of the self-energy in Eq.~(\ref{eq:GretT}) only contain thermal corrections.

The spectral function can be obtained from the imaginary part of the retarded Green's function as~\cite{kadanoff1962quantum,Blaizot:1999xk}
\be S_D (k^0,{\bf k}) \equiv \frac{iG_D^> (k^0,{\bf k}) - iG_D^< (k^0,{\bf k})}{2\pi}  = -\frac{1}{\pi} \textrm{Im } G^R_D (k^0,{\bf k}) \ , \ee
where we have used Eq.~(\ref{eq:greenrelation}). This definition of $S_D(k^0,{\bf k})$ is compatible with the conventions used in Refs.~\cite{Montana:2020lfi,Montana:2020vjg} for the equilibrium case.

In terms of the $D$-meson retarded self-energy, the spectral function reads
\be S_D(k^0,{\bf k}) =- \frac{1}{\pi} \frac{ \textrm{Im } \Pi^R (k^0,{\bf k};T) }{[(k^0)^2-{\bf k}^2-m_{D}^2 -\textrm{Re } \Pi^R(k^0,{\bf k};T)]^2 + [ \textrm{Im } \Pi^R(k^0,{\bf k};T)]^2} \label{eq:spectralfunc} \ . \ee

Unless otherwise stated, this generic form is the one that we implement in our calculations, and the function is extracted from Ref.~\cite{Montana:2020vjg} (see Figs. 6 and 7 in that reference). In the quasiparticle approximation the pole of the retarded Green's function in Eq.~(\ref{eq:GretT}) is not far from the vacuum one and the spectral function can be written as
\be S_D (k^0,{\bf k}) \simeq \frac{z_k}{2\pi E_k} \frac{\gamma_k}{(k^0-E_k)^2 +\gamma_k^2} \ , \label{eq:quasiparticle} \ee
where $E_k$ is the quasiparticle energy, solution of
\be E^2_k -{\bf k}^2-m_{D}^2 -\textrm{Re } \Pi^R(E_k,{\bf k};T) =0 \ , \label{eq:quasienergy} \ee
and the damping rate $\gamma_k$ is defined as
\be \gamma_k = -\frac{z_k}{2E_k} \textrm{Im } \Pi^R(E_k,{\bf k};T) \ , \label{eq:gammak} \ee
with the $z_k$ factor,
\be z_k^{-1} = 1 -\frac{1}{2E_k} \left. \left( \frac{\pa \textrm{Re } \Pi^R(k^0,{\bf k};T)}{\pa k^0} \right) \right|_{k^0=E_k} \ . \ee

In particular, this approximation entails that the $D$-meson damping rate should be much smaller than the quasiparticle energy $E_k$. We conclude this section, by discussing how reliable is the quasiparticle approximation using improved results with respect to those in Ref.~\cite{Montana:2020vjg}\footnote{The differences essentially consist in an increase of the values of the cutoffs in the integrals of the imaginary time formalism.}. 

\begin{figure}[ht]
\centering
\includegraphics[width=74mm]{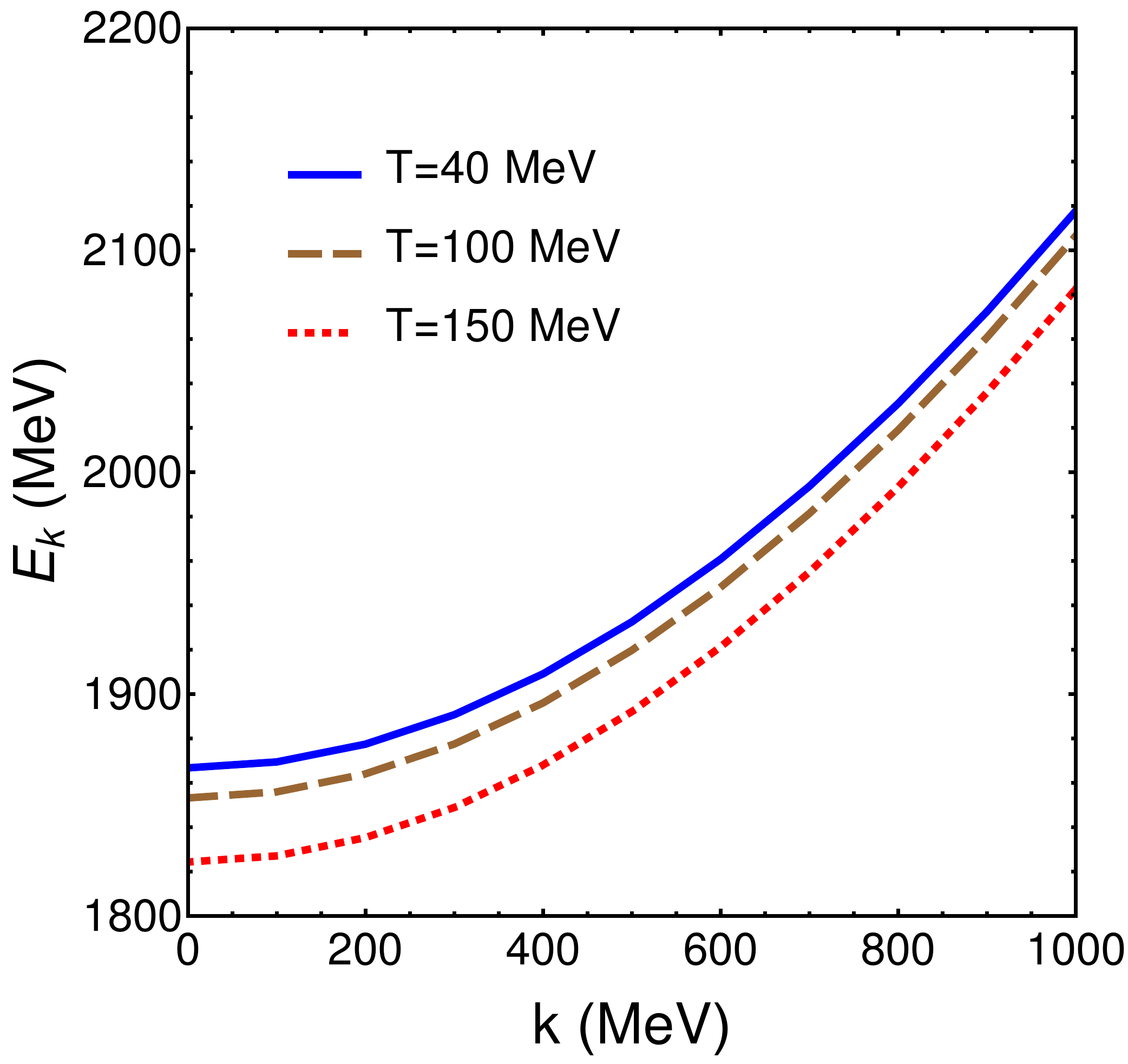}
\includegraphics[width=74mm]{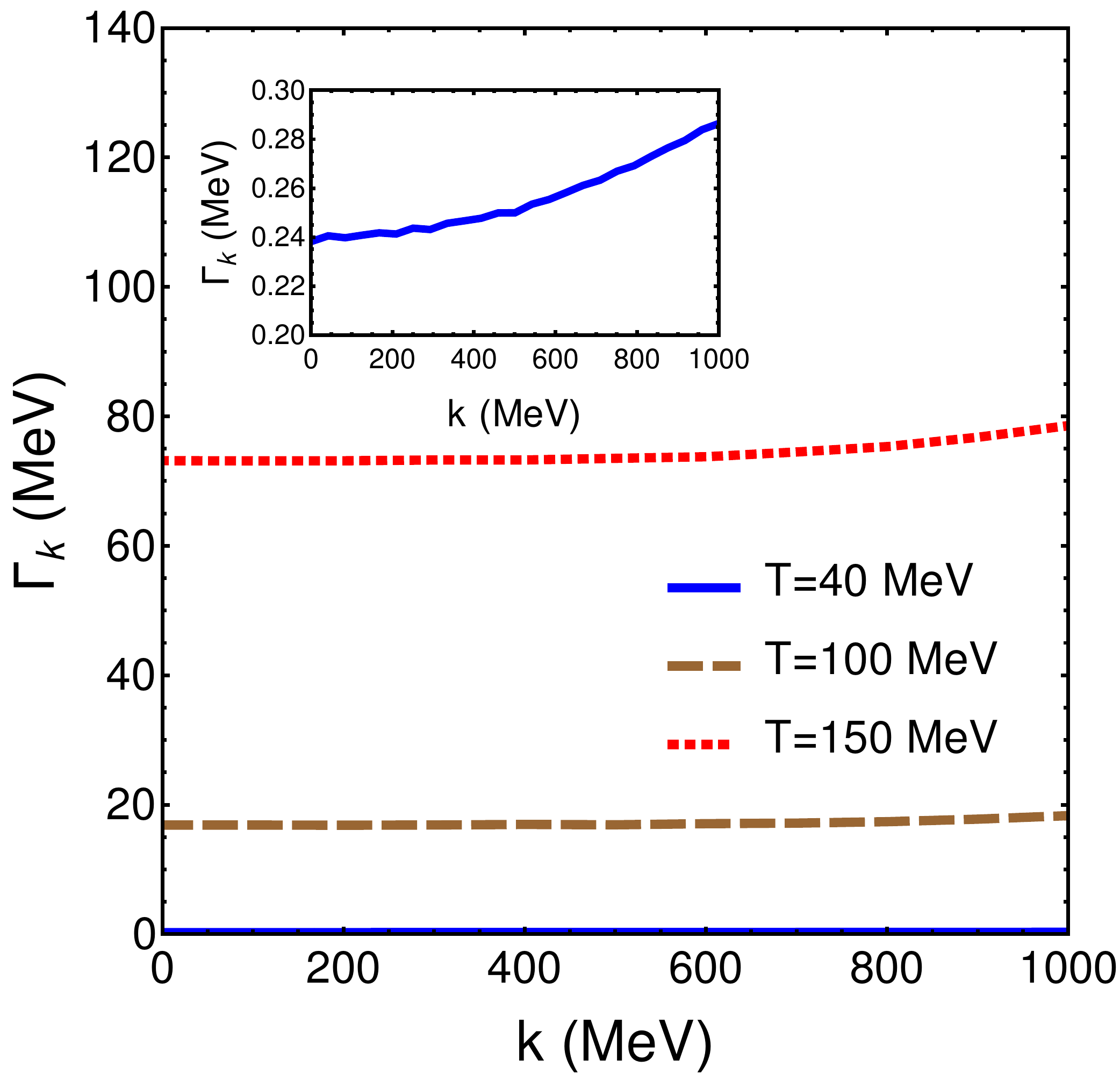}
\caption{Left panel: $D$-meson quasiparticle energy in a thermal pion gas. Right panel: Thermal width (or twice the damping rate) of $D$ mesons in a thermal pion gas.}
\label{fig:Ekzk}
\end{figure}

Firstly, we extract the quasiparticle energies $E_k$ for three different temperatures $T=40,100,150$ MeV, spanning the range considered in Ref.~\cite{Montana:2020vjg}. The quasiparticle energies are computed by solving Eq.~(\ref{eq:quasienergy}), and they are given in the left panel of Fig.~\ref{fig:Ekzk} as a function of the $D$-meson momentum. The value of $E_k$ at $k=0$ is interpreted as the $D$-meson thermal mass $m_D(T)=E_{k=0}$, and it decreases with temperature. Indeed, we have checked that $m_D(T)$ coincide almost perfectly with the values obtained by looking at the peak of the spectral function calculated from Eq.~(\ref{eq:spectralfunc}). This agreement is already an indication of the good quasiparticle approximation. Notice that, in agreement with our previous calculation~\cite{Montana:2020vjg}, the results of Fig.~\ref{fig:Ekzk} only contain the pion contribution into the $D$-meson self-energy.

We have numerically checked that the $z_k$ factors are very close to 1. Rather independent of the quasiparticle momentum $k$, we observe up to 2 \% (1 \%) deviations from unity for $T=150$ MeV ($T=100$ MeV). For $T=40$ MeV, the factor $z_k$ is fully compatible with 1. Therefore the approximation $z_k \simeq 1$ is an excellent one, and will be used in what follows.

As shown in Eq.~(\ref{eq:gammak}), the damping rate $\gamma_k$ is computed from the imaginary part of the $D$-meson self-energy. Instead of $\gamma_k$, we show in the right panel of Fig.~\ref{fig:Ekzk} the $D$-meson thermal width $\Gamma_k$, simply defined as twice the damping rate, $\Gamma_k \equiv 2\gamma_k$. This quantity is negligible at low temperatures, but the effects of the medium makes it sizable at $T=150$ MeV, with thermal widths of the order of 70 MeV. Nevertheless, these values are still small  ($\sim$30 times smaller) compared to the corresponding $E_k$. This validates the quasiparticle approximation in our system, at least, in the equilibrium case for the temperatures and momenta considered in this work.

In the next sections we will explore different approximations to address the $D$-meson thermal width and transport coefficients. From the results presented in this section---where the quasiparticle approximation is a very good one---we can anticipate that pure off-shell effects will not contribute much to these quantities. Nevertheless, in some of the calculations we will keep the full shape of the spectral function in order to quantify the importance of its broadening due to thermal effects.

\section{Analysis of the $D$-meson thermal width in equilibrium~\label{sec:kinematic}}

In this section we analyze in great detail several effects on the $D$-meson thermal width $\Gamma_k$, defined as twice the damping rate of Eq.~(\ref{eq:gammak}). We work in the equilibrium case, and for a $D$ meson which propagates on-shell, i.e., its energy $E_k$ is fixed by the momentum $k$ through Eq.~(\ref{eq:quasienergy}). However, notice that the internal propagators of the retarded self-energy do not need to be on their mass shell. These off-shell effects will be addressed here, together with the importance of the Landau cut and the contribution of the inelastic channels.

\subsection{Thermal width from the imaginary part of the scattering amplitude}

In the right panel of Fig.~\ref{fig:Ekzk} we have already shown the $D$-meson thermal width, given by
\be \Gamma_k = - \frac{z_k}{E_k} \textrm{Im } \Pi^R (E_k,k) \ , \label{eq:Gammak} \ee
obtained from the retarded self-energy calculation of Ref.~\cite{Montana:2020vjg}.  In that work only the effect of pions was considered in the self-consistent calculation of the $D$-meson self-energy, being the contribution of the other light mesons very suppressed. Here we will also analyze these contributions as we have access to all the elements of the $T$ matrix.

Let us review the calculation of $\Gamma_k$ in  more detail and study the contributions from the different kinematic ranges. Following the method used in Refs.~\cite{Montana:2020lfi,Montana:2020vjg}, the calculation of Eq.~(\ref{eq:Gammak}) in terms of $T$-matrix elements is performed in the imaginary time formalism via the expression of the self-energy
\be \Pi (i\omega_n, {\bf k})=- T \int \frac{d^3p}{(2\pi)^3} \sum_{i\omega_m} G_\pi(i\omega_m, {\bf p}) \ T_{D\pi} (i\omega_m+i\omega_n, {\bf p}+{\bf k}) \ , \ee
which is diagrammatically represented in the lower diagram of Fig.~\ref{fig:BS}. Here, $i\omega_n$ denotes the external Matsubara frequency  while $i\omega_m$ is the internal one, which is summed over.

In accordance to the approximation made in Ref.~\cite{Montana:2020vjg}, in this work we also use the free pion propagator,
\begin{align}
  \Pi (i\omega_n, {\bf k}) & = -T \int \frac{d^3p}{(2\pi)^3} \sum_{i\omega_m} \frac{1}{(i\omega_m)^2-E_p^2} \ T_{D\pi} (i\omega_m+i\omega_n,{\bf p}+{\bf k}) \ .
\end{align}

Introducing the spectral representation of the $T$-matrix amplitude,
\be   T_{D\pi} (i\omega_m+i\omega_n,{\bf p}+{\bf k}) = - \frac{1}{\pi} \int_{-\infty}^\infty d\Omega \frac{\textrm{ Im } T_{D\pi} (\Omega,{\bf p}+{\bf k})}{i\omega_m+i\omega_n-\Omega} \ , \ee
summing over Matsubara frequencies via complex integration, by deforming the domain to encircle the three simple poles, and finally performing the analytic continuation ($i\omega_n \rightarrow k^0 + i\epsilon$), we obtain the imaginary part of the retarded self-energy,
\begin{align}
\textrm{ Im }  \Pi^R(E_k,{\bf k}) & =  \int \frac{d^3p}{(2\pi)^3}   \left[  \frac{f^{(0)}(E_p)}{2E_p}   \textrm{Im } T_{D\pi} (E_k+E_p,{\bf p}+{\bf k})  \right. \nn \\
& \left. + \frac{1+f^{(0)}(E_p)}{2E_p} \textrm{ Im } T_{D\pi} (E_k-E_p,{\bf p}+{\bf k})  \right. \nn \\ 
& - \left.   \frac{f^{(0)}(E_k+E_p)}{2E_p} \textrm{ Im } T_{D\pi}(E_k+E_p,{\bf p}+{\bf k}) \right. \nn \\
& + \left. \frac{f^{(0)}(E_k-E_p)}{2E_p} \textrm{ Im } T_{D\pi}(E_k- E_p,{\bf p}+{\bf k}) \right]  , \label{eq:ImPiR}
\end{align}
where we have already fixed the external $D$-meson energy to the quasiparticle energy $E_k$,  which is a function of $k$, cf. Eq.~(\ref{eq:quasienergy}). We have neglected the $z_k$ factors, because they are very close to one, as previously discussed (but they can be easily incorporated, if desired).

Using the result in Eq.~(\ref{eq:ImPiR}), we can write the $D$-meson thermal width in Eq.~(\ref{eq:Gammak}) as the contribution of four pieces,
\be \Gamma_k = \Gamma_k^{(1)}+\Gamma_k^{(2)}+\Gamma_k^{(3)}+\Gamma_k^{(4)} \ , \label{eq:Gammak4terms} \ee
where
\begin{align} 
  \Gamma_k^{(1)} & =  -\frac{1}{E_k} \int \frac{d^3p}{(2\pi)^3}  \frac{f^{(0)}(E_p)}{2E_p} \ \textrm{Im } T_{D\pi}(E_k+E_p,{\bf p}+{\bf k}) \ , \label{eq:Gammak1} \\ 
\Gamma_k^{(2)} & =  -\frac{1}{E_k} \int \frac{d^3p}{(2\pi)^3}   
\frac{1+f^{(0)}(E_p)}{2E_p} \ \textrm{ Im } T_{D\pi} (E_k-E_p,{\bf p}+{\bf k}) \ ,  \label{eq:Gammak2} \\ 
\Gamma_k^{(3)} & =  \frac{1}{E_k} \int \frac{d^3p}{(2\pi)^3} \frac{f^{(0)}(E_k+E_p)}{2E_p} \ \textrm{ Im } T_{D\pi} (E_k+E_p,{\bf p}+{\bf k}) \ , \label{eq:Gammak3} \\ 
\Gamma_k^{(4)} & =  -\frac{1}{E_k} \int \frac{d^3p}{(2\pi)^3}
\frac{f^{(0)}(E_k-E_p)}{2E_p} \ \textrm{ Im } T_{D\pi} (E_k- E_p,{\bf p}+{\bf k})  \ . \label{eq:Gammak4}
\end{align}

It is important to realize that both $\Gamma_k^{(1)}$ and $\Gamma_k^{(3)}$ receive a contribution from the scattering amplitude above the two-particle mass threshold; while $\Gamma_k^{(2)}$ and $\Gamma_k^{(4)}$ depend on the values of the $T$-matrix below threshold. The latter contribution is related to the Landau cut, and only appears at finite temperature when the total momentum of the collision if different from zero or when the masses of the interacting particles are different~\cite{Weldon:1983jn,Das:1997gg,Torres-Rincon:2017zbr}. 

In particular, when vacuum amplitudes are used, the Landau cut disappears, and only $\Gamma_k^{(1)}$  and $\Gamma_k^{(3)}$ contribute. Incidentally, this is the situation in the pion-pion vacuum scattering of Ref.~\cite{Schenk:1993ru}, where only $\Gamma_k^{(1)}$ appears. In our case, $\Gamma_k^{(3)}$ is in fact extremely small, because it is roughly proportional to the product of the pion and $D$-meson densities, and the latter are very scarce~\footnote{This can be shown from the relation $f^{(0)}(E_k+E_p)=f^{(0)} (E_k) f^{(0)} (E_p) / (1+f^{(0)} (E_k)+f^{(0)} (E_p))$}. However, when considering a medium-dependent interaction, $\Gamma_k^{(2)}$ and $\Gamma_k^{(4)}$ will also have a potential important contribution that we now quantify.

\begin{figure}[ht]
  \centering
 \includegraphics[width=75mm]{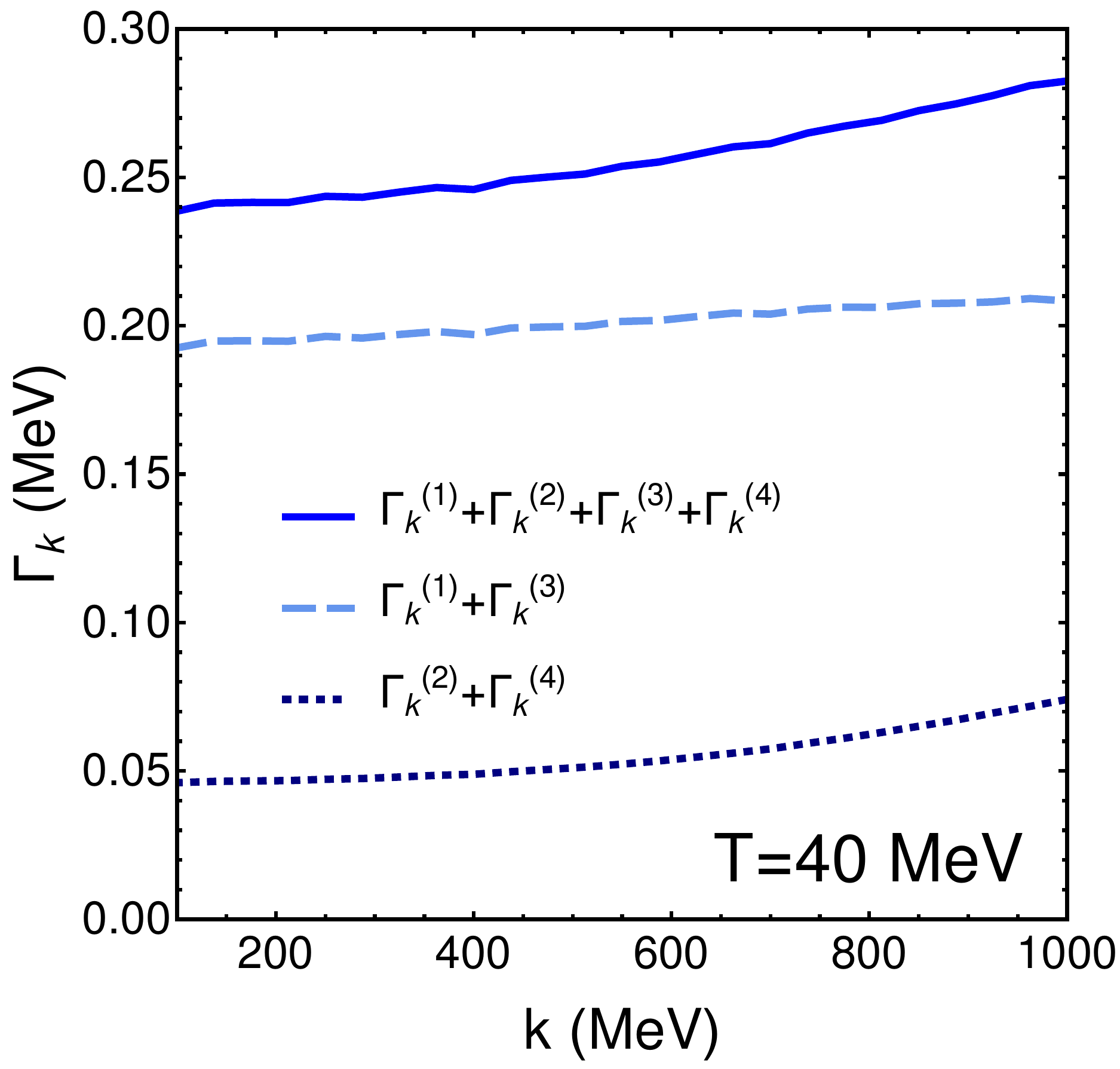}
  \includegraphics[width=75mm]{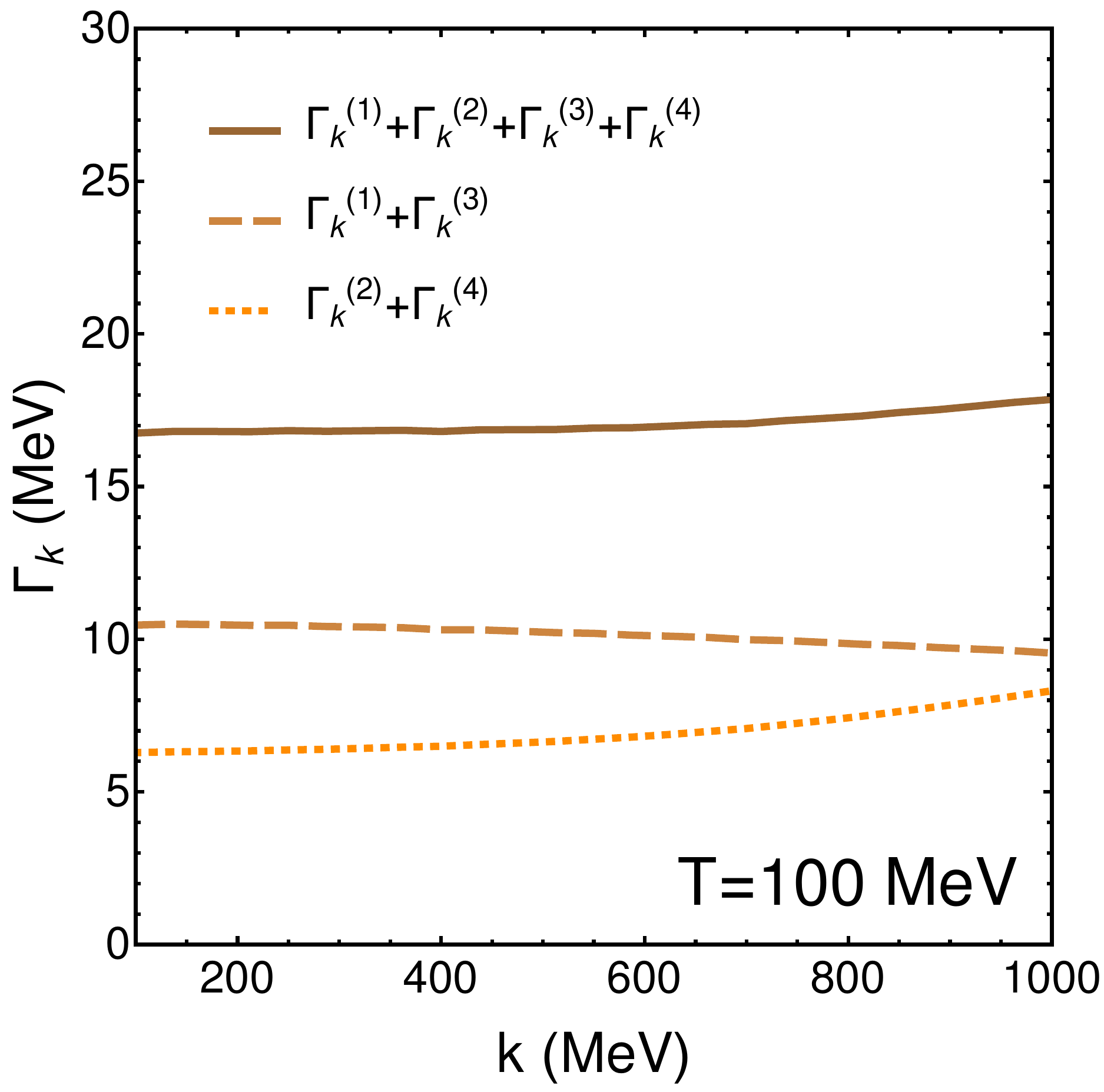}
  \includegraphics[width=75mm]{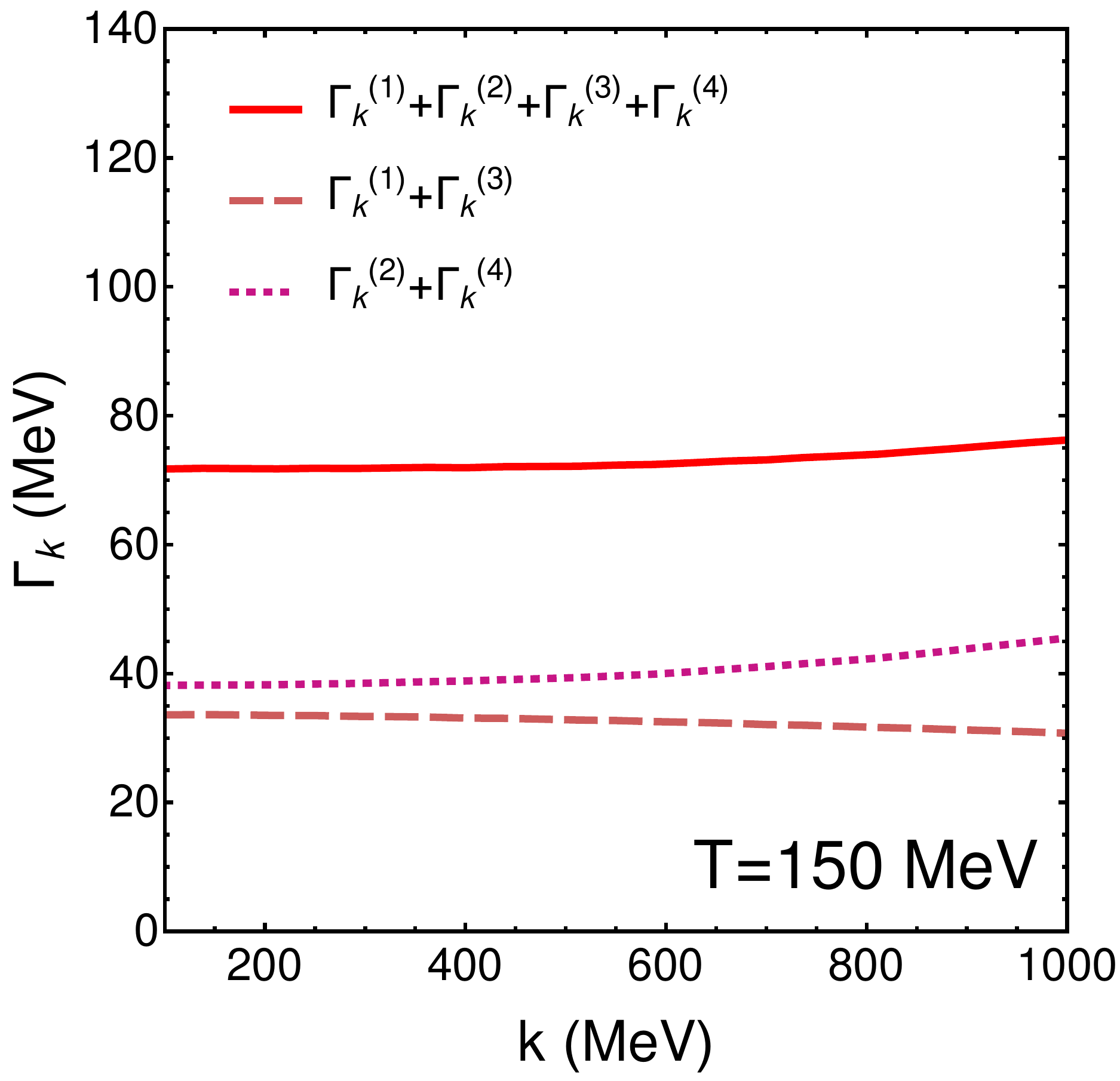}
    \caption{The $D$-meson thermal width as computed from Eq.~(\ref{eq:Gammak4terms}). In dashed lines we show  the contribution of $\Gamma^{(1)}+\Gamma^{(3)}$ (Eqs.~(\ref{eq:Gammak1})+(\ref{eq:Gammak3}) above the threshold), whereas in dotted lines, we display  the contribution of $\Gamma^{(2)}+\Gamma^{(4)}$ [Eqs.~(\ref{eq:Gammak2})+(\ref{eq:Gammak4}) below the threshold (Landau cut)].}
  \label{fig:Gammak}
\end{figure}

To gauge the weight of the different terms in Eq.~(\ref{eq:Gammak4terms}) we plot in Fig.~\ref{fig:Gammak} the different contributions at three different temperatures. The input for $T_{D\pi}$ is taken from updated results of the scattering amplitudes obtained  in Ref.~\cite{Montana:2020vjg}.

At small temperatures ($T=40$ MeV) the terms $\Gamma_k^{(1)}+\Gamma_k^{(3)}$ dominate. Nevertheless, the two pieces coming from the Landau cut $\Gamma_k^{(2)}+\Gamma_k^{(4)}$ give a nonzero contribution resulting in a 20\% of the total thermal width. At $T=100$ MeV, the contribution of the unitarity cut ($\Gamma_k^{(1)}+\Gamma_k^{(3)}$) and the Landau cut ($\Gamma_k^{(2)}+\Gamma_k^{(4)}$) are similar. For the higher temperatures $T=150$ MeV the contribution from the Landau cut already surpasses that of the unitary cut. This means that for temperatures close to $T_c$ there is a dominant contribution to the thermal width coming from the Landau cut, which would be overlooked if vacuum amplitudes were used. We have also checked that $\Gamma_k^{(3)}$ is negligible for all momenta and temperatures.

\subsection{Thermal width from the scattering amplitude squared}

The calculation of the thermal width due to thermal pions using Eqs.~(\ref{eq:Gammak4terms}) to (\ref{eq:Gammak4}) allowed us to distinguish the relative weight of the unitary and Landau contributions. However, the effect of the individual collision terms ($D\pi \rightarrow D\pi, D\pi \rightarrow D\eta, D\pi \rightarrow D_s \bar{K}$) cannot be disentangled. On the other hand, the computation of heavy-flavor transport coefficients is performed from the kinetic transport equation, where the collision rates, that are proportional to $|T_{D\pi}|^2$, are used. For these reasons, we will derive an alternative expression for $\Gamma_k$, in terms of the scattering amplitude squared, which will also serve to double check our previous determination of $\Gamma_k$.

We start from the same definition of the thermal width of Eq.~(\ref{eq:Gammak})
and use the relation
\be \textrm{Im } \Pi^R (E_k,k) = \frac{1}{2 i} \left( \Pi^> (E_k,k) - \Pi^<(E_k,k)\right) \ . \ee
In equilibrium, we can exploit the so-called Kubo-Martin-Schwinger relation between the ``lesser'' and ``greater'' self-energies~\cite{kadanoff1962quantum,Blaizot:1999xk},
\be \Pi^< (E_k,k) = e^{-\beta E_k} \Pi^> (E_k,k) \  , \ee
obtaining,
\be \Gamma_k =  \frac{i}{2 E_k } [\Pi^> (E_k,k) - \Pi^< (E_k,k) ] =  \frac{i}{2   E_k} \frac{1}{\tilde{f}_k^{(0)}} \Pi^>(E_k,k) \ , \ee
where we have denoted $\tilde{f}^{(0)}_k \equiv \tilde{f}^{(0)} (E_k)$, and already simplified $z_k\simeq 1$.

We can now insert the expression in Eq.~(\ref{eq:Pig}) for $\Pi^>(E_k,k)$, which provides an interpretation of the thermal width in terms of particle collisions. As in the derivation of the off-shell transport theory we replace the light-meson propagators by those for free particles, but keep the full spectral function of the internal $D$ meson. We can write the thermal width as
\be \Gamma_k=\Gamma_k^{(U)} + \Gamma_k^{(L)} \ , \label{eq:widthT2} \ee
with
\begin{align}
\Gamma_k^{(U)} & = \frac{1}{2E_k} \frac{1}{\tilde{f}^{(0)}_k} \sum_{\lambda=\pm} \lambda \int dk_1^0 \int \prod_{i=1}^3 \frac{d^3 k_i }{(2\pi)^3} \frac{1}{2E_2} \frac{1}{2E_3} |T (E_k+E_3,{\bf k}+{\bf k}_3)|^2 \ S_D(k_1^0,{\bf k}_1) \nn \\
&\times (2\pi)^4 \delta^{(3)} ( {\bf k}+{\bf k}_3-{\bf k}_1-{\bf k}_2) \delta(E_k+E_3-k_1^0- \lambda E_2) \tilde{f}^{(0)} (k^0_1) f^{(0)} (E_3) \tilde{f}^{(0)} ( \lambda E_2)  \ ,   \label{eq:widthT2U} \\
\Gamma_k^{(L)} & = \frac{1}{2E_k} \frac{1}{\tilde{f}^{(0)}_k} \sum_{\lambda=\pm} \lambda \int dk_1^0 \int \prod_{i=1}^3 \frac{d^3 k_i }{(2\pi)^3} \frac{1}{2E_2} \frac{1}{2E_3} |T (E_k-E_3,{\bf k}+{\bf k}_3)|^2 \ S_D(k_1^0,{\bf k}_1) \nn \\
&\times (2\pi)^4 \delta^{(3)} ( {\bf k}+{\bf k}_3-{\bf k}_1-{\bf k}_2) \delta(E_k-E_3-k_1^0- \lambda E_2) \tilde{f}^{(0)} (k^0_1) \tilde{f}^{(0)} (E_3)  \tilde{f}^{(0)} (\lambda E_2)   \ , \label{eq:widthT2L}
 \end{align}
where, like in the expression in Eq.~(\ref{eq:Pig}), there is an implicit restricted summation over particle species, according to the allowed scattering channels. In particular, if one focuses on the pion contribution to the $D$-meson thermal width---hence particle 3 being a $\pi$---then the remaining sum over species 1 and 2 contains three possibilities: $D\pi \rightarrow D\pi, D\pi \rightarrow D\eta$ and $D\pi \rightarrow D_s \bar{K}$ scatterings.

The separation made in Eq.~(\ref{eq:widthT2}) makes it clear that $\Gamma_k^{(U)}$ evaluates the scattering amplitude above threshold, and it is related to the unitary cut of the scattering amplitude, while $\Gamma_k^{(L)}$ evaluates it below threshold and is therefore related to the Landau cut.

Before showing results, let us mention that, as in the on-shell  reduction of the transport theory, if the internal $D$ meson is approximated by a narrow quasiparticle, only the positive branch of the spectral function $S_D(k_1^0,{\bf k}_1)$ and $\lambda=+1$ can hold the energy conservation in Eq.~(\ref{eq:widthT2U}). The same is true for Eq.~(\ref{eq:widthT2L}) but with $\lambda=-1$. Therefore in the on-shell case (o.s.), the expressions in Eqs.~(\ref{eq:widthT2U}) and (\ref{eq:widthT2L}) reduce to
\begin{align}
\left. \Gamma_k^{(U)} \right|_{\textrm{o.s.}} & = \frac{1}{2E_k} \frac{1}{\tilde{f}^{(0)}_k} \int \prod_{i=1}^3 \frac{d^3 k_i }{(2\pi)^3 2E_i}  |T (E_k+E_3,{\bf k}+{\bf k}_3)|^2 \nn \\
&\times (2\pi)^4 \delta^{(3)} ( {\bf k}+{\bf k}_3-{\bf k}_1-{\bf k}_2) \delta(E_k+E_3-E_1-E_2) \tilde{f}^{(0)} (E_1) f^{(0)} (E_3)  \tilde{f}^{(0)} (E_2)      \label{eq:widthT2Uos} , \\
\left. \Gamma_k^{(L)} \right|_{\textrm{o.s.}} & = \frac{1}{2E_k} \frac{1}{\tilde{f}^{(0)}_k}  \int \prod_{i=1}^3 \frac{d^3 k_i }{(2\pi)^3 2E_i} |T (E_k-E_3,{\bf k}+{\bf k}_3)|^2 \nn \\
&\times (2\pi)^4 \delta^{(3)} ( {\bf k}+{\bf k}_3-{\bf k}_1-{\bf k}_2) \delta(E_k-E_3-E_1+E_2) \tilde{f}^{(0)} (E_1) \tilde{f}^{(0)} (E_3)  {f}^{(0)} (E_2)   \ . \label{eq:widthT2Los}
 \end{align}

Given the special kinematics of $\Gamma_k^{(L)}$, one cannot express the energy-momentum conservation in terms of a single $\delta^{(4)}$ function. Only in the particular case of elastic (el) scattering one can make a change of variables ${\bf k}_2 \leftrightarrow -{\bf k}_3$ in $\Gamma_k^{(L)}$ to arrive to
\begin{align}
\left. \Gamma_k^{(U)} \right|^{\textrm{el}}_{\textrm{o.s.}}  & = \frac{1}{2E_k} \frac{1}{\tilde{f}^{(0)}_k}  \int \prod_{i=1}^3 \frac{d^3 k_i }{(2\pi)^3 2E_i}
(2\pi)^4 \delta^{(4)} ( k+k_3-k_1-k_2) |T (E_k+E_3,{\bf k}+{\bf k}_3)|^2 \nn \\
& \times \tilde{f}^{(0)} (E_1) \tilde{f}^{(0)} (E_2) f^{(0)} (E_3)  \label{eq:widthT2Uosalt} , \\
\left. \Gamma_k^{(L)} \right|^{\textrm{el}}_{\textrm{o.s.}}  & = \frac{1}{2E_k} \frac{1}{\tilde{f}^{(0)}_k}  \int \prod_{i=1}^3 \frac{d^3 k_i }{(2\pi)^3 2E_i}
(2\pi)^4 \delta^{(4)} ( k+k_3-k_1-k_2) |T (E_k-E_2,{\bf k}-{\bf k}_2)|^2 \nn \\ & \times \tilde{f}^{(0)} (E_1) \tilde{f}^{(0)} (E_2) f^{(0)} (E_3)  \ . \label{eq:widthT2Losalt}
 \end{align}
 
Equations (\ref{eq:widthT2Uosalt}) and (\ref{eq:widthT2Losalt}) can be potentially useful when the $D$ meson is treated as a narrow quasiparticle and inelastic collisions are neglected. Unless otherwise stated, we do not assume this.

Coming back to the general result of Eqs.~(\ref{eq:widthT2U}) and (\ref{eq:widthT2L}), where the full spectral function of the internal $D$ meson is kept, it is possible to analytically check that $\Gamma_k^{(U)}$ in Eq.~(\ref{eq:widthT2U}) is equal to the combination $\Gamma_k^{(1)}+\Gamma_k^{(3)}$ in Eqs.~(\ref{eq:Gammak1}) and (\ref{eq:Gammak3}), while $\Gamma_k^{(L)}$ in Eq.~(\ref{eq:widthT2L}) exactly coincides with $\Gamma_k^{(2)}+\Gamma_k^{(4)}$ in Eqs.~(\ref{eq:Gammak2}) and (\ref{eq:Gammak4}). To do that, one needs to apply the unitarity condition (or optical theorem in the coupled-channel case)
\be \textrm{Im } T_{D\pi \rightarrow D\pi}(E,{\bf p}) = \sum_a T^*_{D\pi \rightarrow a}(E,{\bf p}) \ \textrm{Im } G_a^R (E,{\bf p}) T_{a \rightarrow D\pi}(E,{\bf p}) \ , \label{eq:optitheo} \ee
which follows from the $T$-matrix equation at finite temperature, together with~\cite{Montana:2020vjg}
\be G_{a=D_i,\Phi_j}^R (E,{\bf p})=\int \frac{d^3q}{(2\pi)^3} \int d\omega \int d\omega' \frac{S_{D_i}(\omega,{\bf q}) S_{\Phi_j}(\omega',{\bf p}-{\bf q})}{E-\omega-\omega'+i\epsilon} [1+f^{(0)}(\omega)+f^{(0)}(\omega')] \ , \ee
where the spectral function of the light meson $S_\Phi(\omega',{\bf p}-{\bf q})$ is to be taken in the narrow limit.

Notice that the sum over intermediate states ($a$) in the optical theorem in Eq.~(\ref{eq:optitheo}) is to be taken as a sum over species $D_i$ and $\Phi_j$ restricted to the physical states which couple to $D\pi$. If only elastic collisions $D\pi \rightarrow D\pi$ were used, then the optical theorem is necessarily violated. The effect of inelastic processes has been normally ignored in the literature.

\begin{figure}[ht]
  \centering
  \includegraphics[width=75mm]{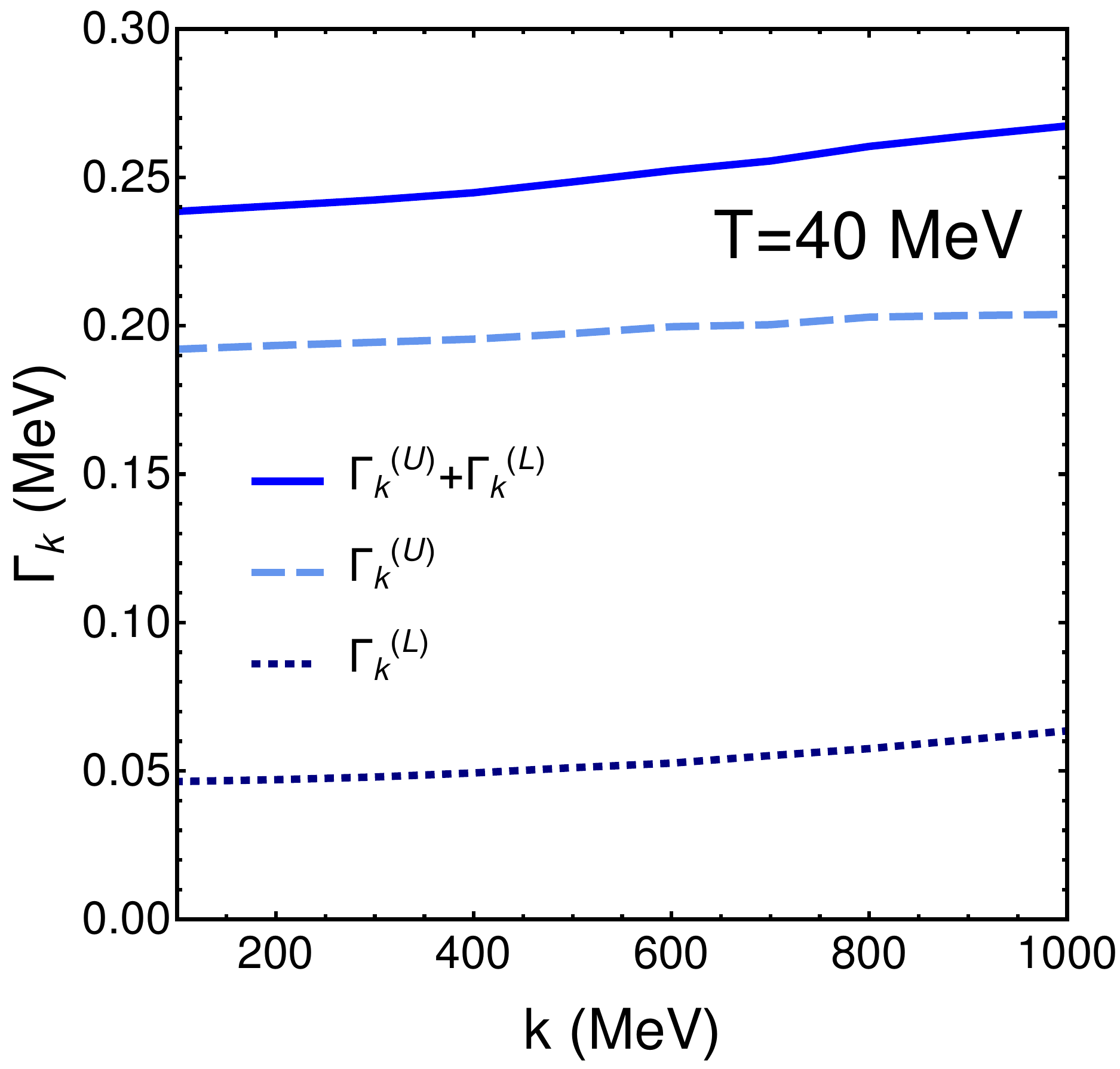}
   \includegraphics[width=75mm]{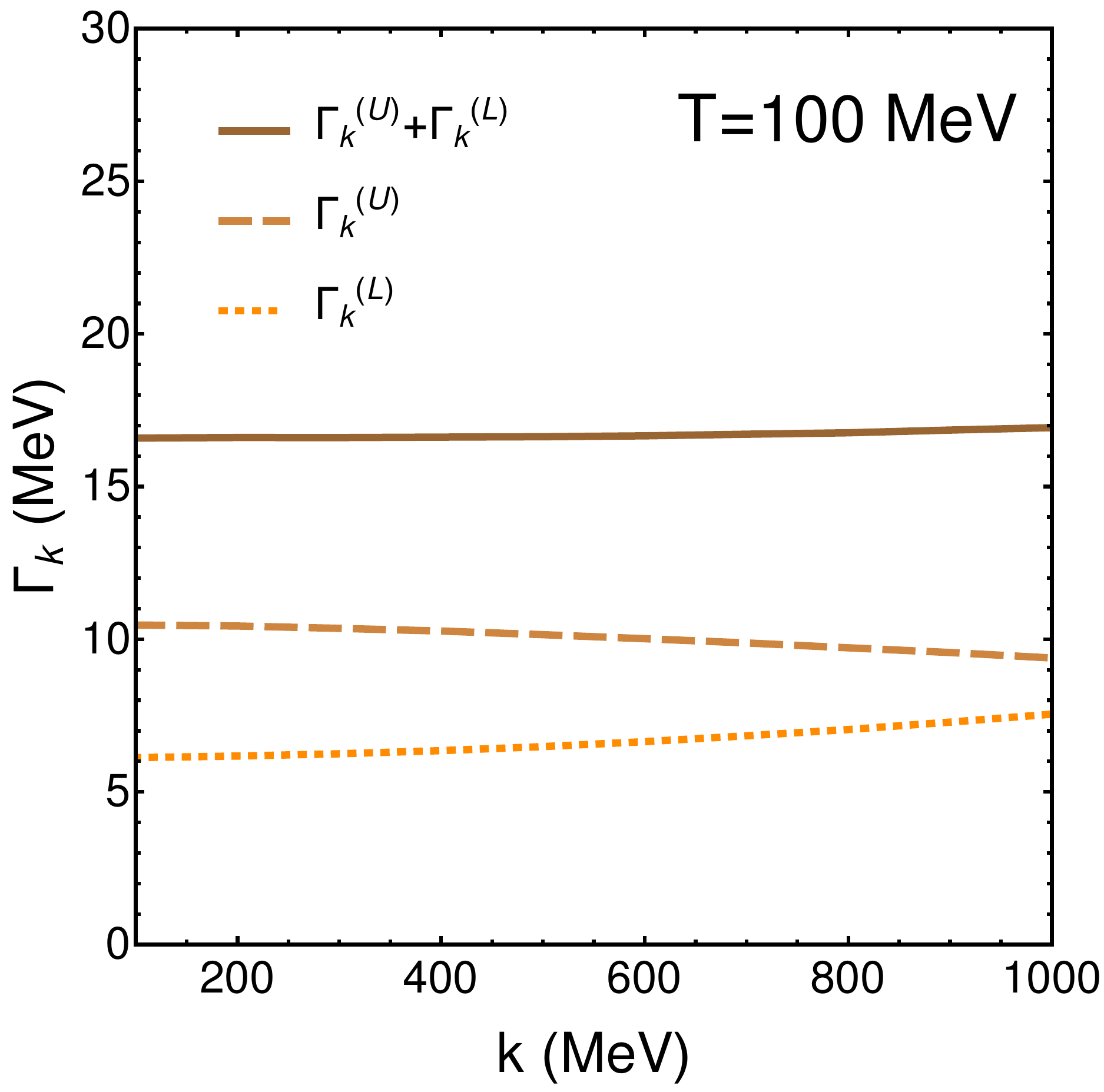}
  \includegraphics[width=75mm]{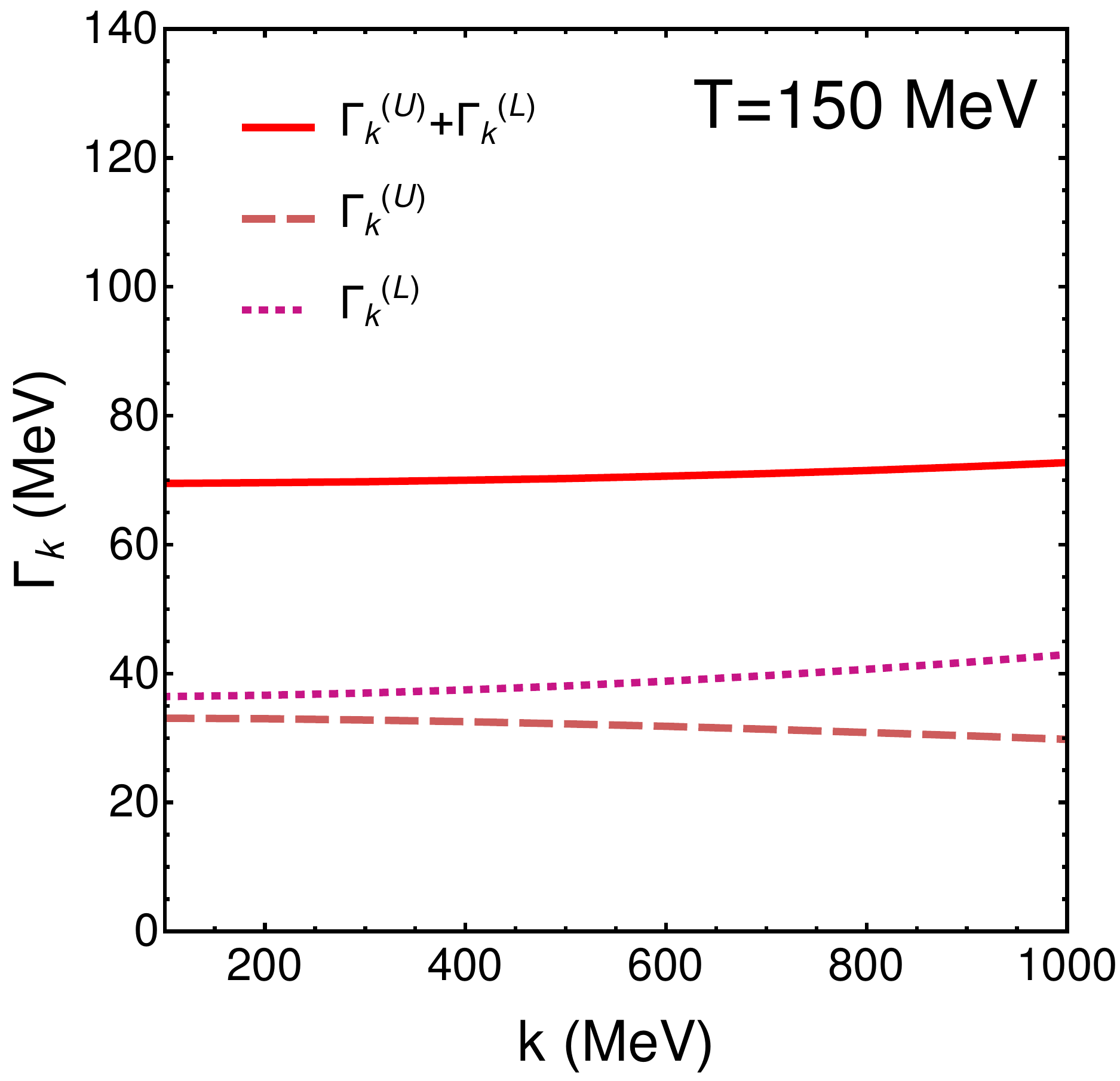}
  \caption{Thermal width of $D$ mesons generated with the interaction with pions as computed  from $\Gamma^{(U)}$ (Eq.~(\ref{eq:widthT2U})) using dashed lines, and $\Gamma^{(L)}$ (Eq.~(\ref{eq:widthT2L})) using dotted lines. The full calculation that includes both $\Gamma^{(U)}$ and $\Gamma^{(L)}$ is shown with solid lines. Inelastic channels ($D\pi \rightarrow D\eta$ and $D\pi \rightarrow D_s \bar{K}$) are also included.}
    \label{fig:GammakKinetic}
\end{figure}

We present the results of Eqs.~(\ref{eq:widthT2U}) and (\ref{eq:widthT2L}) for temperatures $T=40,100,150$ MeV in Fig.~\ref{fig:GammakKinetic}. We separate the contributions of the unitary and Landau cuts for each temperature, and obtain a similar result to that in Fig.~\ref{fig:Gammak} (where the integration over $\textrm{Im } T_{D\pi \rightarrow D\pi}$ was employed). For consistency with the coupled-channels optical theorem, we have included the three channels $D\pi \rightarrow D\pi, D\pi \rightarrow D\eta$ and $D\pi \rightarrow D_s \bar{K}$.

\subsection{Quantification of different effects}

The differences between the two approaches, as well as the analysis of several other effects are summarized in the following.

\subsubsection{Effect of truncation}

As stated, one can analytically prove that the two alternative methods to extract $\Gamma_k$, first via Eq.~(\ref{eq:Gammak4terms}) and second through Eq.~(\ref{eq:widthT2}), are equivalent. In addition, we have stated that this equivalence can also be checked via direct application of the optical theorem in Eq.~(\ref{eq:optitheo}).

However, the numerical implementation can introduce small differences when a UV cutoff is employed. We name this the effect of truncation. This can be easily understood by looking at Eq.~(\ref{eq:optitheo}). The first method to compute $\Gamma_k$ uses $\textrm{Im } T_{D \pi \rightarrow D\pi}$, and a UV cutoff in $|{\bf p}|$ simply truncates the left-hand side of Eq.~(\ref{eq:optitheo}) at that momentum. 
On the other hand, the second method employs the right-hand side of Eq.~(\ref{eq:optitheo}), where the same cutoff is imposed on $|T_{D\pi \rightarrow a}|^2$, but the term $\textrm{Im } G_a^R$ is calculated analytically to perform the integrations. Therefore, the way in which a UV cutoff is imposed in the numerical calculations does not ensure that the truncation effect is the same for the two methods.

\begin{figure}[ht]
  \centering
  \includegraphics[width=75mm]{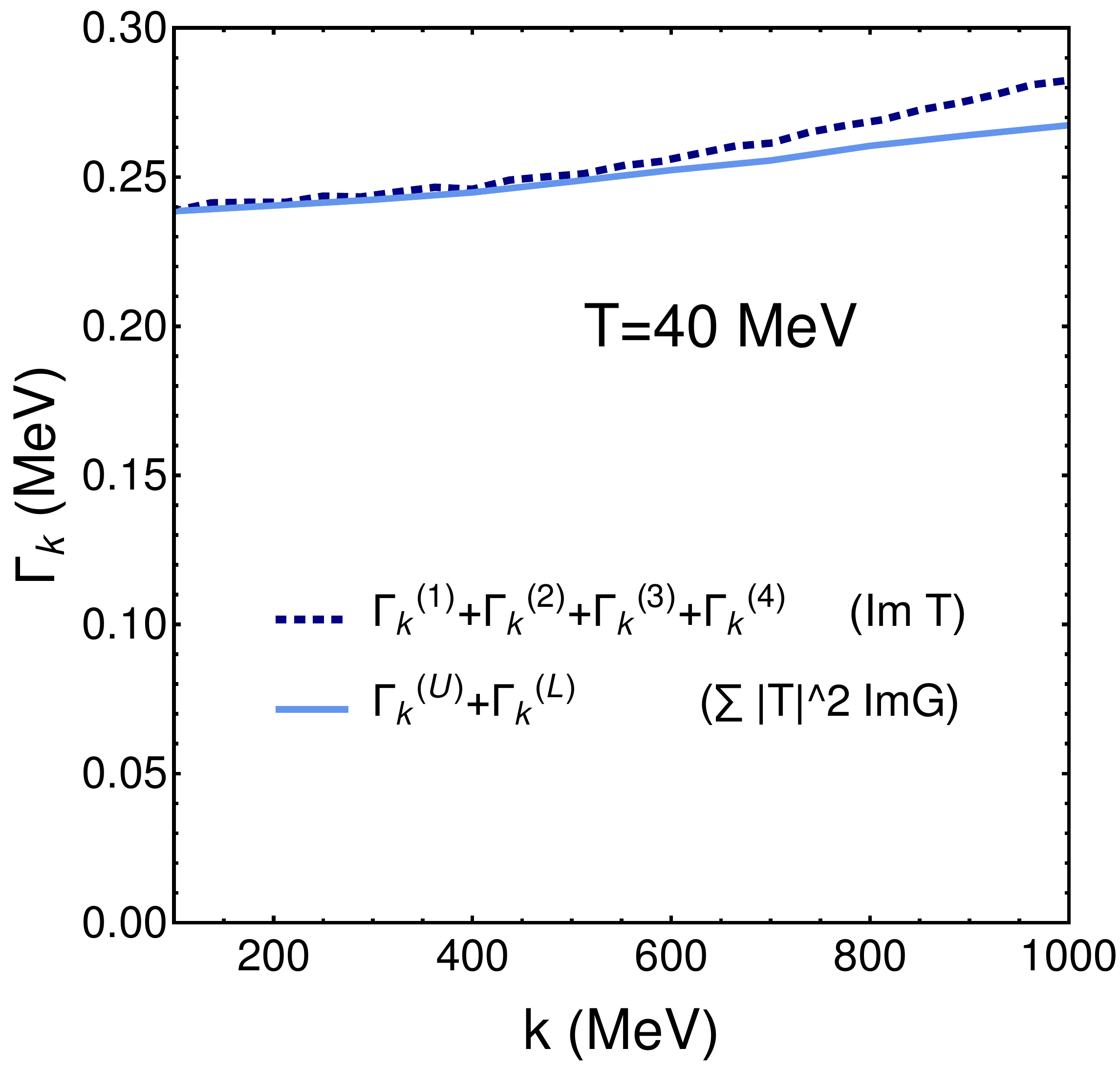}
  \includegraphics[width=75mm]{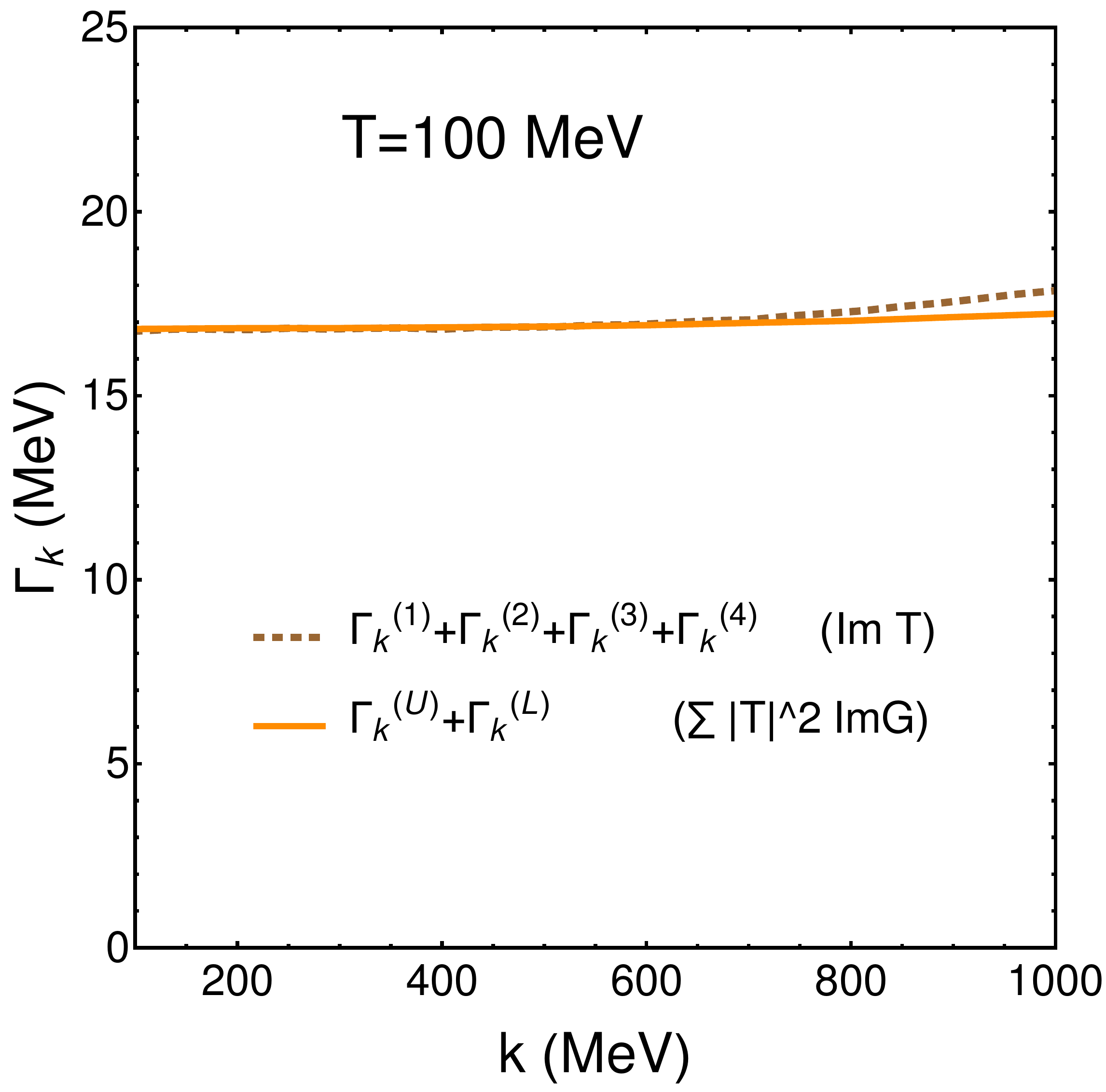}
  \includegraphics[width=75mm]{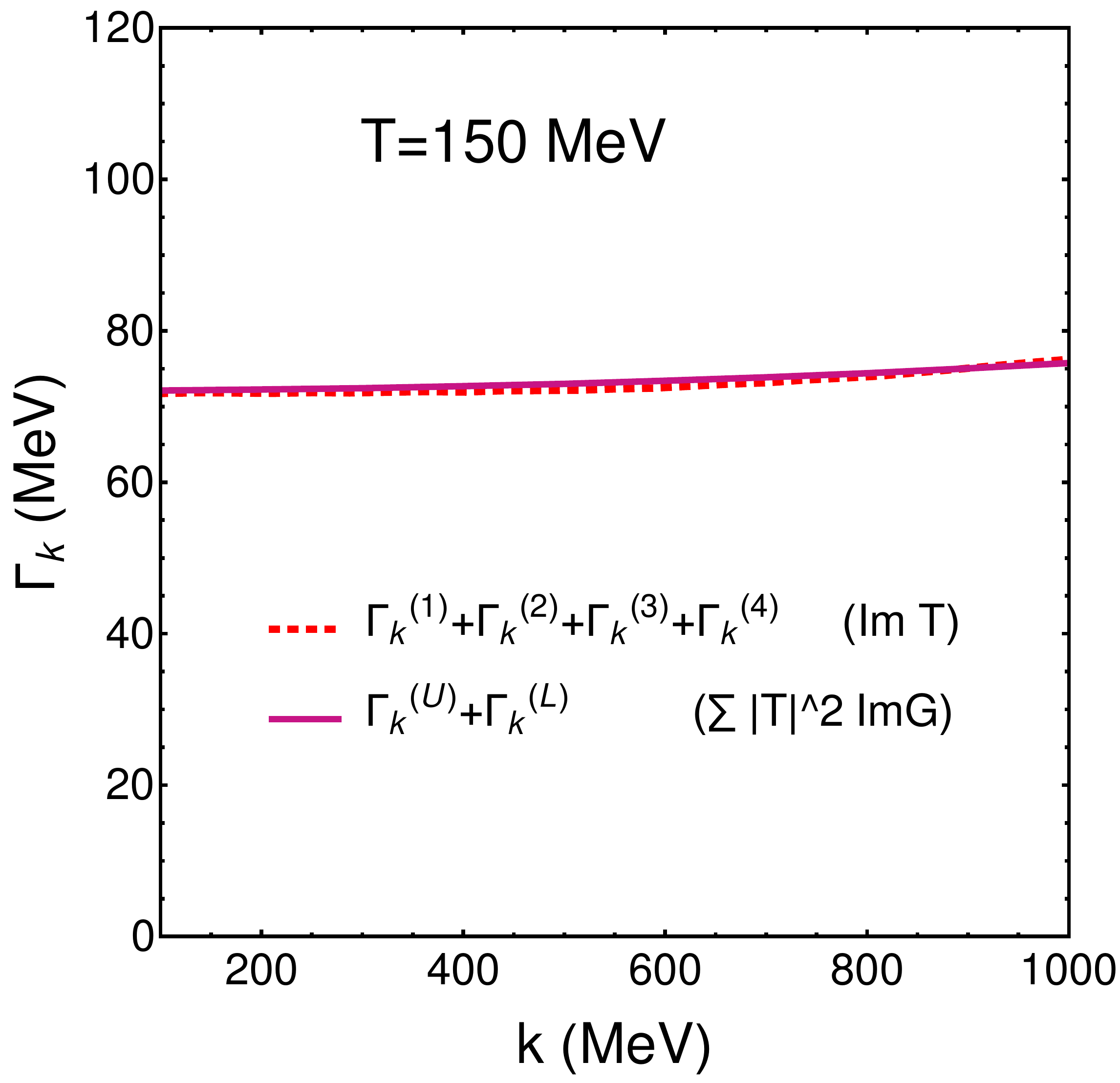}
  \caption{Thermal width of $D$ mesons in a thermal pion gas. Comparison between the two methods described in the text.}
  \label{fig:GammaTall}
\end{figure}

A first comparison between Figs.~\ref{fig:Gammak} and~\ref{fig:GammakKinetic} does not show appreciable differences. In Fig.~\ref{fig:GammaTall} we show the total $\Gamma_k$ from the two methods in a single plot for better comparison. Note that the second method includes the three channels (elastic and inelastic) involving pions. Both methods compare very well for low momentum at all temperatures. We have checked that the good comparison persists between unitary and Landau cuts separately. We only obtain deviations at high momentum (hence cutoff effects). Nevertheless, the differences in $\Gamma_k$ are at most $5 \%$, and only for high momenta (which are in any case suppressed when folded with the Bose distribution function).

\subsubsection{Off-shell effects}

  We now describe the differences between the use of the on-shell and off-shell approaches. We have extensively described how to implement off-shell effects by keeping the full spectral function of the internal $D$-meson propagator, as opposed to using the narrow limit. To determine the differences we use the second method to compute $\Gamma_k$, including the three channels involving pions.

\begin{figure}[ht]
  \centering
  \includegraphics[width=75mm]{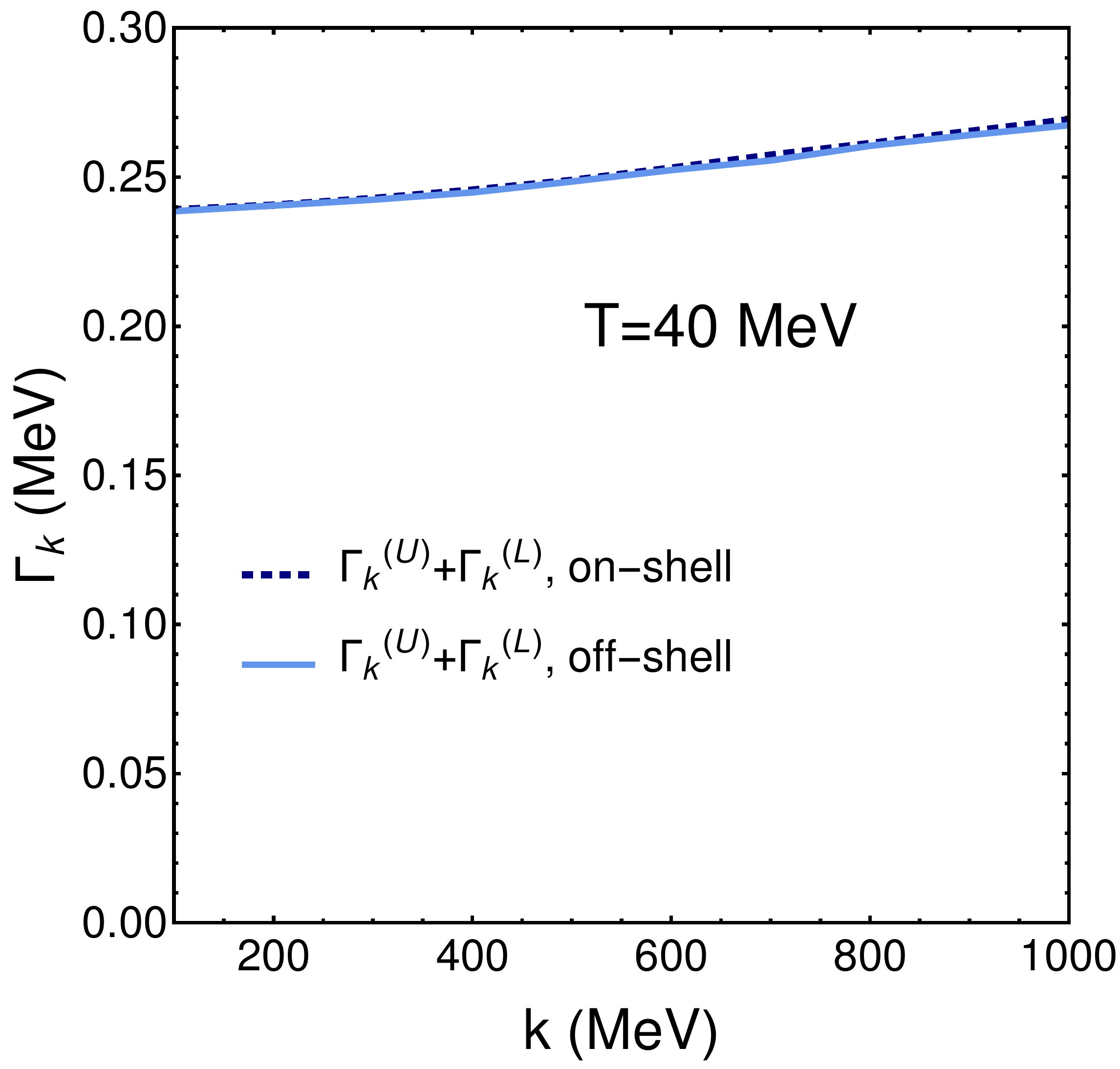}
   \includegraphics[width=75mm]{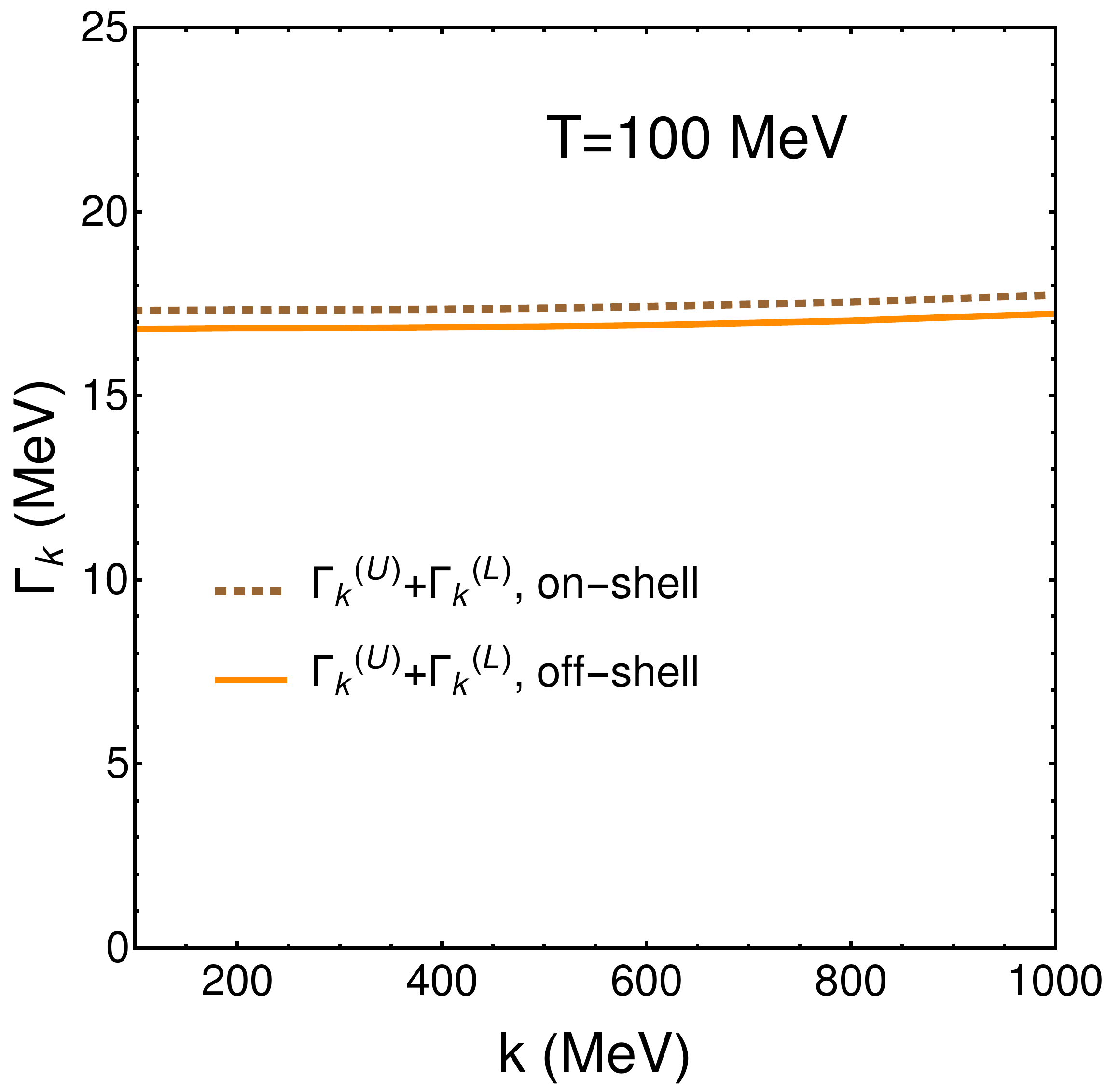}
  \includegraphics[width=75mm]{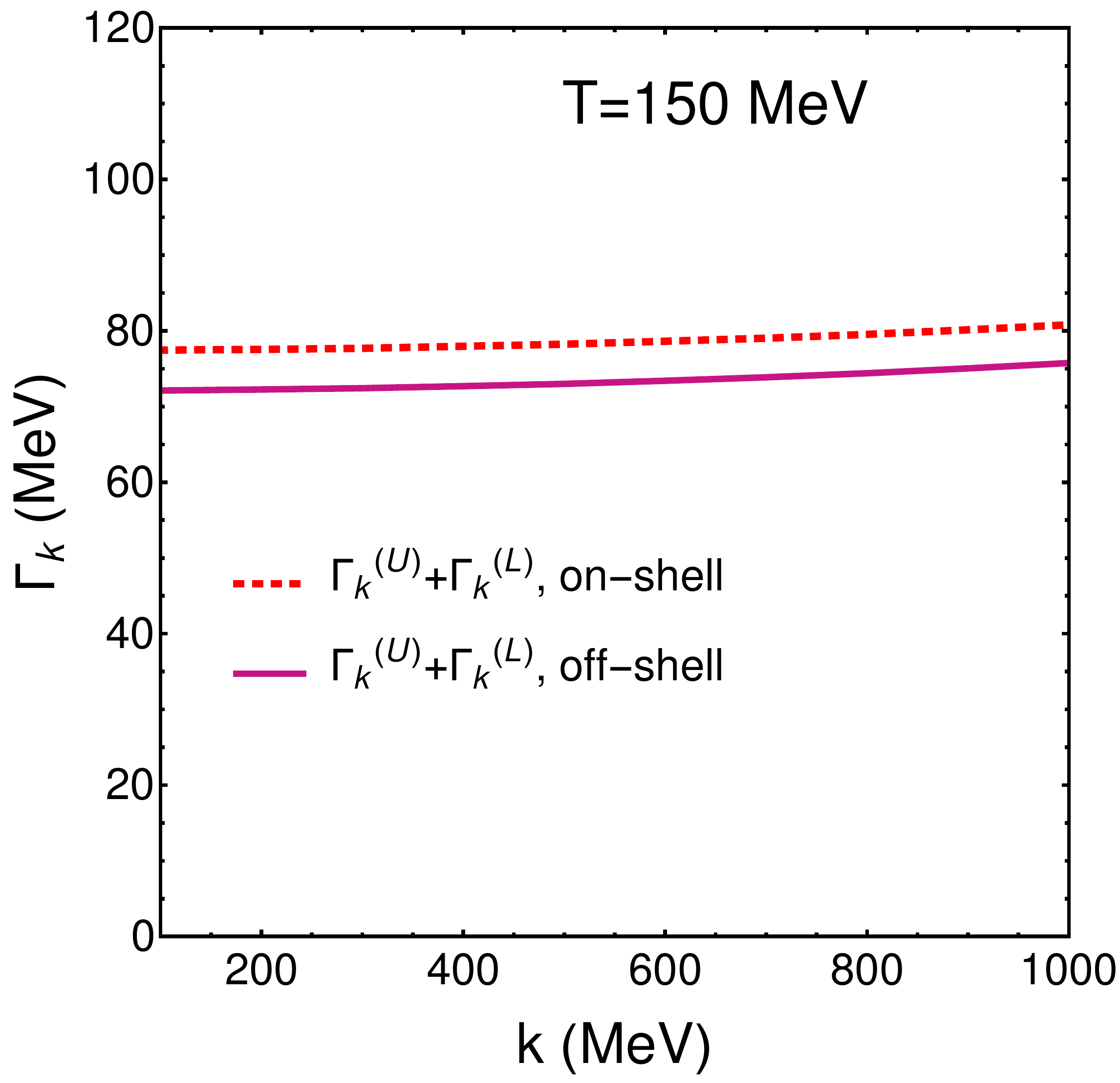}
  \caption{Thermal width of $D$ mesons in a thermal pion gas. Comparison between the off-shell calculation (solid lines) (Eqs.~(\ref{eq:widthT2U})+(\ref{eq:widthT2L})) and the on-shell  one (dotted lines) (Eqs.~(\ref{eq:widthT2Uos})+(\ref{eq:widthT2Los})).}
  \label{fig:Gammaonoff}
\end{figure}

  The results are presented in Fig.~\ref{fig:Gammaonoff}. The off-shell calculation is performed through Eqs.~(\ref{eq:widthT2U})+(\ref{eq:widthT2L}), and the spectral function $S_D$ is taken from the update of the results in Refs.~\cite{Montana:2020lfi,Montana:2020vjg}. For the on-shell calculation we employ Eqs.~(\ref{eq:widthT2Uos})+(\ref{eq:widthT2Los}) where the intermediate $D$ meson is taken on shell (narrow limit). In both cases we use the same temperature dependent scattering amplitudes. 

  As expected, the effects of the spectral function width become more apparent at higher temperatures. When $T \rightarrow 0$ the thermal width goes to zero, and the narrow quasiparticle approximation becomes exact in this limit. In any case, the quasiparticle peak is rather narrow at all temperatures considered, as was already reported in Ref.~\cite{Montana:2020vjg}, and the off-shell effects are generically small. These effects are of the order of 10\% for $\Gamma_k$ at the highest temperature $T=150$ MeV, and rather independent of the external momentum $k$.

\subsubsection{Effect of inelastic channels}

\begin{figure}[ht]
  \centering
  \includegraphics[width=75mm]{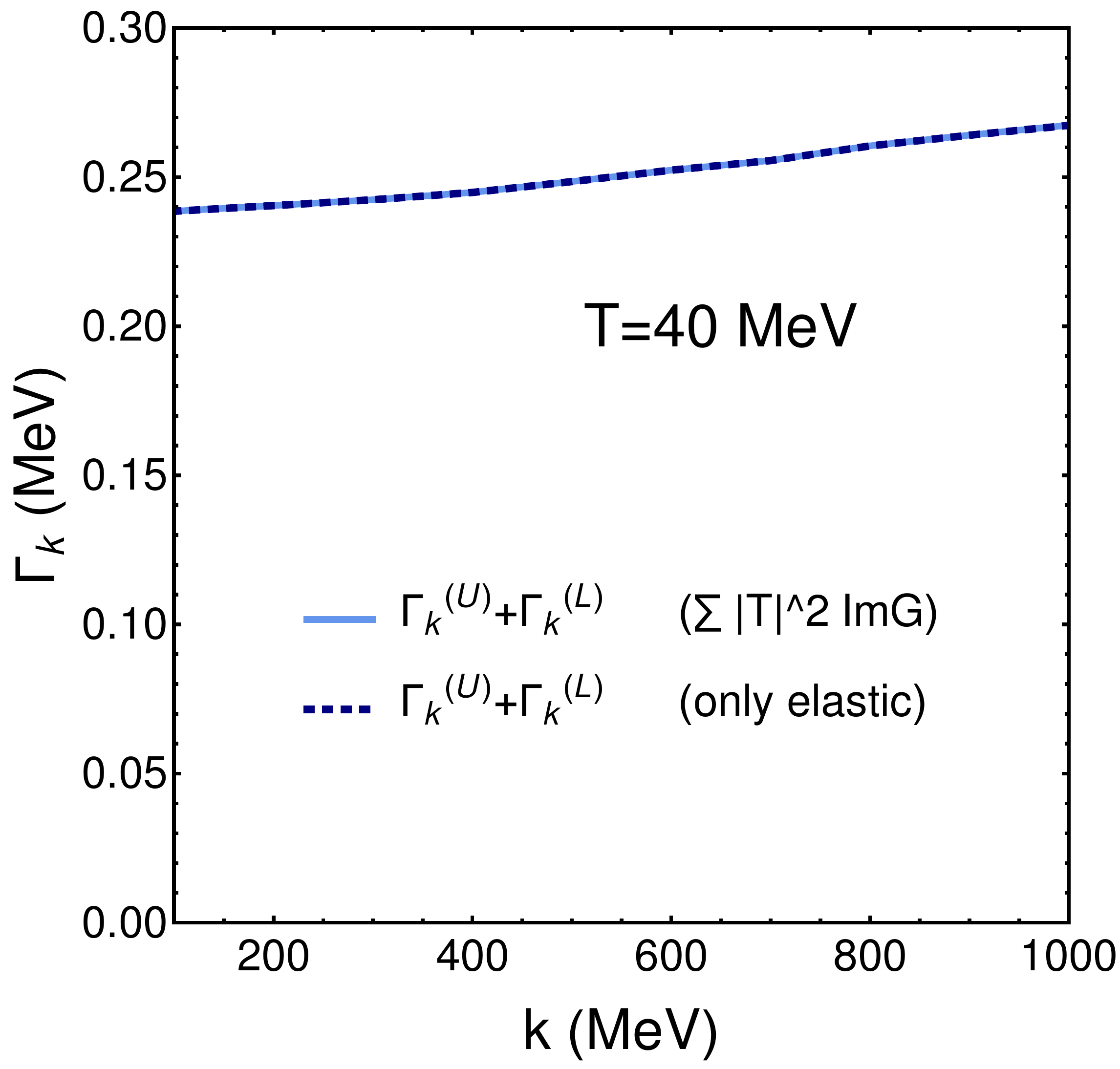}
   \includegraphics[width=75mm]{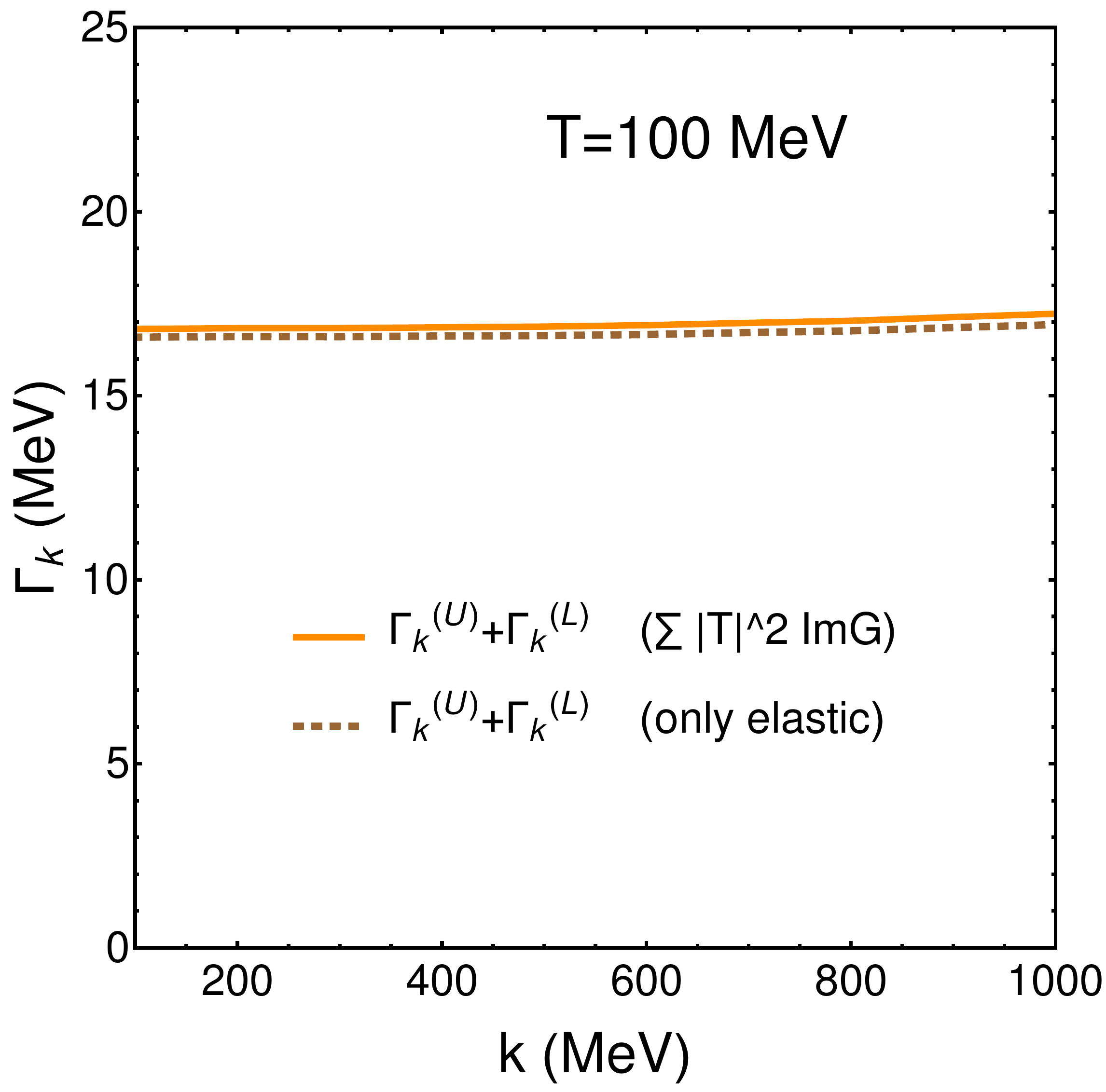}
  \includegraphics[width=75mm]{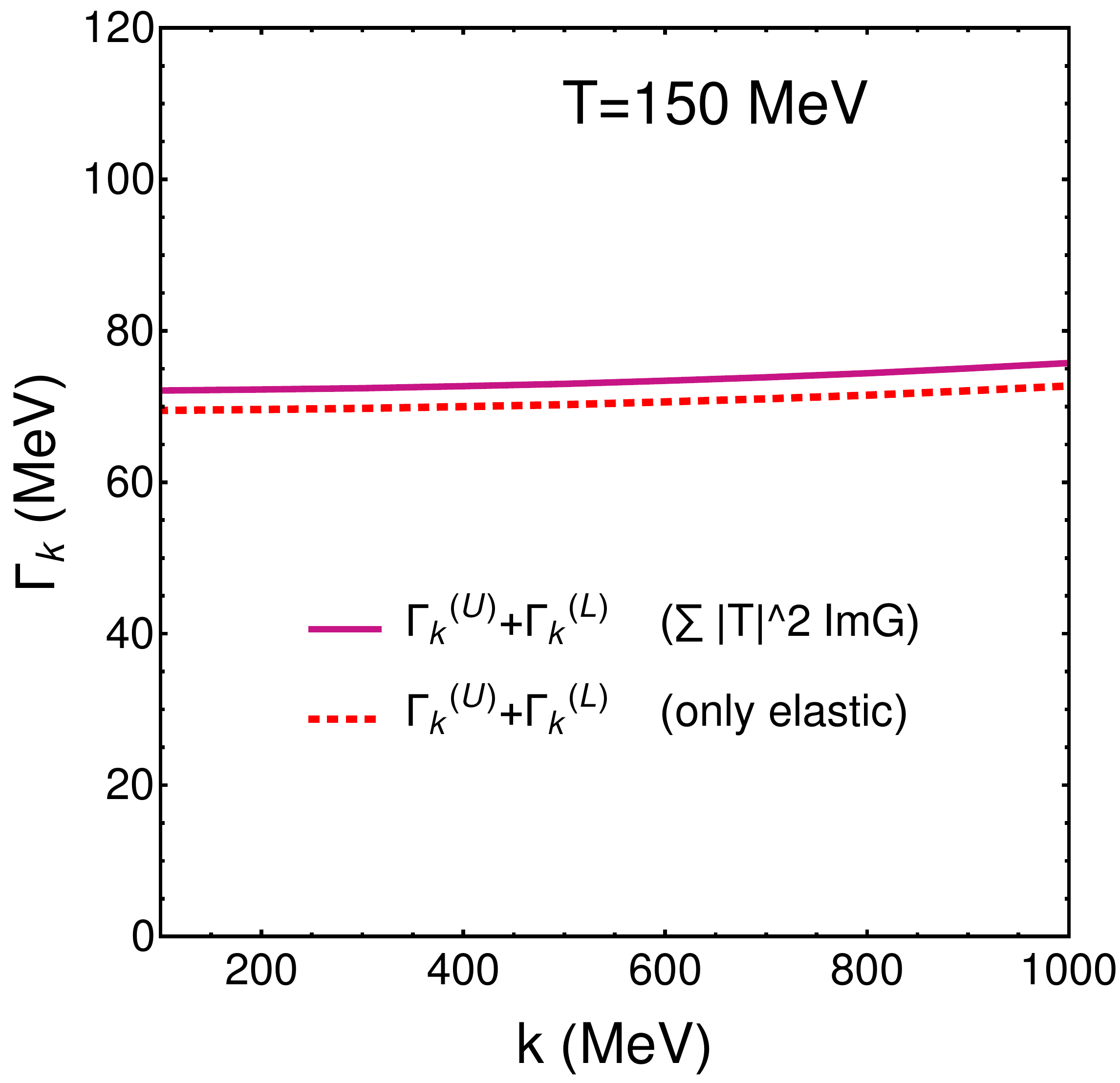}
  \caption{Thermal width of $D$ mesons in a thermal pion gas computed via the off-shell kinetic formulas in Eqs.~(\ref{eq:widthT2U}) and (\ref{eq:widthT2L}). The calculations are done using elastic scattering (dotted lines) or adding also inelastic channels (solid lines), following the optical theorem in coupled channels.}
  \label{fig:GammaTIne}
\end{figure}

We now discuss the effect of inelastic channels in the second method to compute $\Gamma_k$, given by Eqs.~(\ref{eq:widthT2}),(\ref{eq:widthT2U}),(\ref{eq:widthT2L}). While their inclusion is strictly required to account for the coupled-channel optical theorem, in the practice, their effect is small. This can be seen in Fig.~\ref{fig:GammaTIne} for the case of the $D$-meson thermal width, only due to the pions of the medium. In that plot we show in solid lines the complete result with the three inelastic channels, and in dashed lines the result with only the elastic channel $D\pi \rightarrow D\pi$. The effect is rather small for all $T$ and $k$, and for the highest temperature $T=150$ MeV and momentum they are at most 5\%. Therefore, we will neglect the effect of inelastic channels in the  calculations of transport coefficients.

\subsubsection{Effects of the light mesons in the bath}

While one can safely neglect the inelastic channels, one should not forget that there are four different elastic channels for the interactions of $D$ mesons with light pseudoscalars ($D\pi, DK, D\bar{K}, D\eta$). While the contribution of the most massive mesons is also Boltzmann suppressed, they become increasingly important as the temperature is increased. 

 In order to study the effect of the contribution of the different species, we define an averaged thermal width (only function of temperature) as
\be \Gamma(T)=\frac{1}{n_D} \int d^3k \ f^{(0)} (E_k) \Gamma_k \ , \label{eq:GammaT} \ee
where $f^{(0)} (E_k)$ is the equilibrium Bose-Einstein distribution function and $n_D$ is the $D$-meson particle density. 

\begin{figure}[ht]
\centering
\includegraphics[scale=0.4]{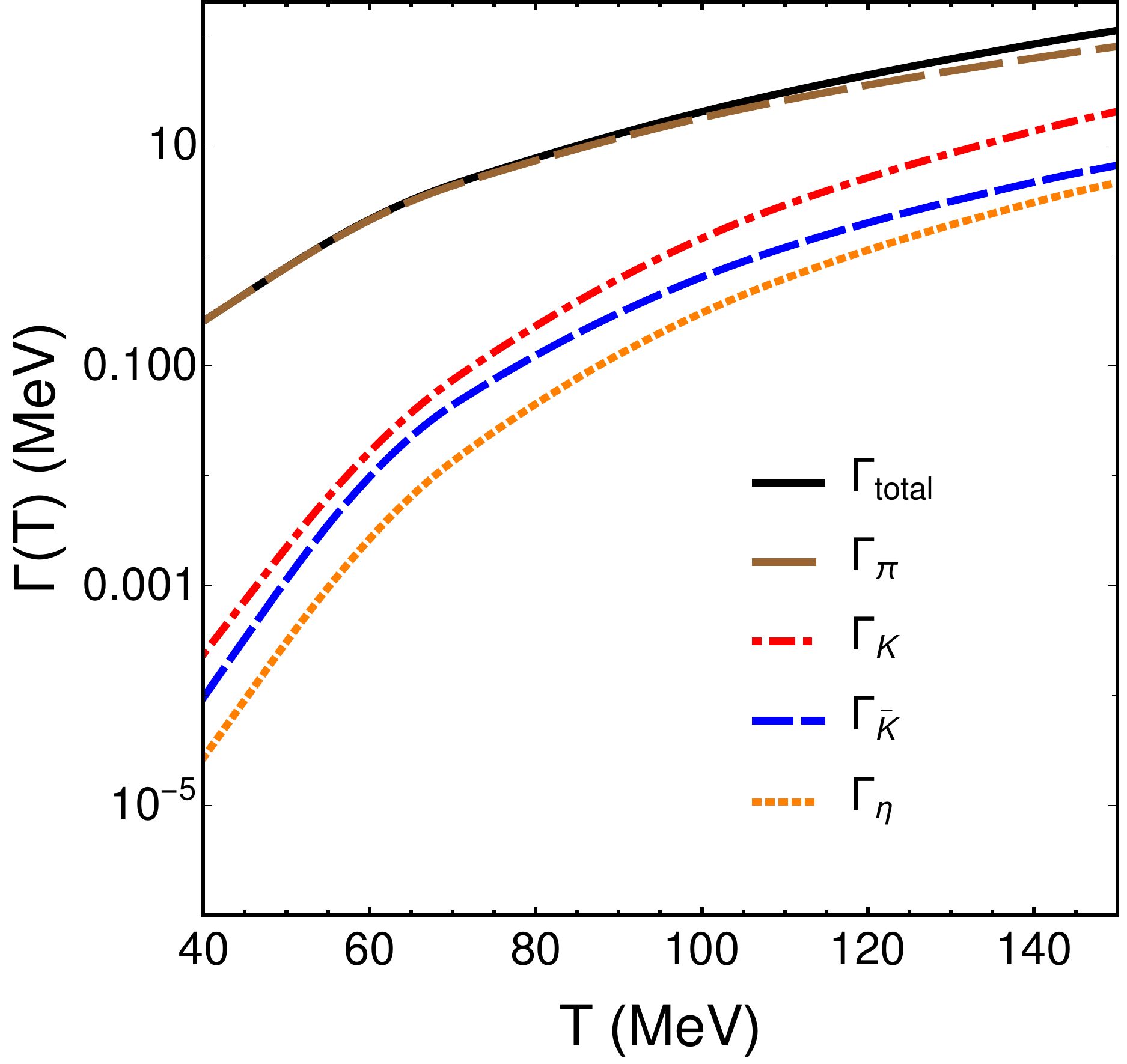}
\includegraphics[scale=0.31]{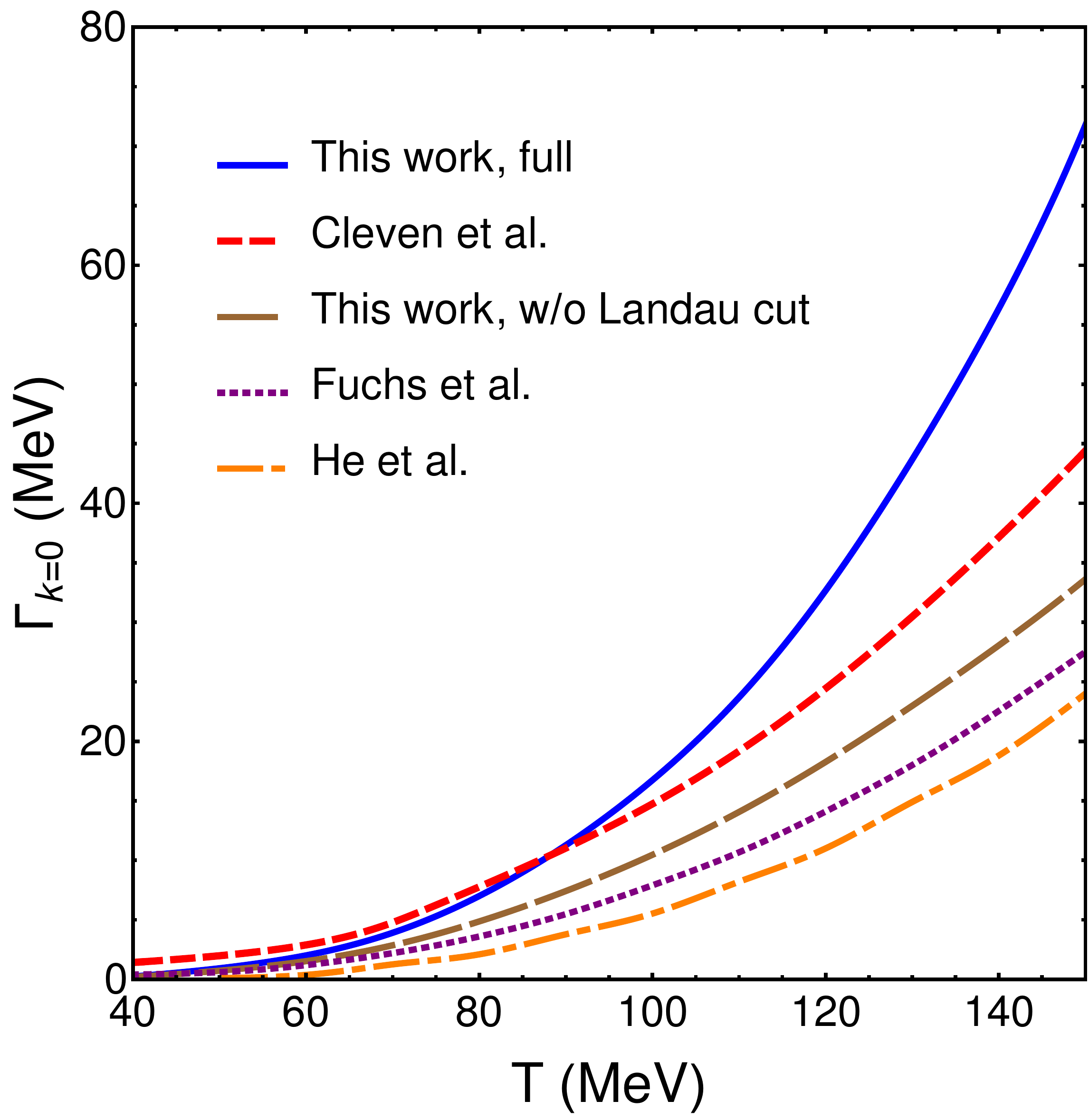}
\caption{Left panel: Contribution to the $D$-meson averaged thermal width from  a thermal bath of pions, kaons, antikaons, and $\eta$ mesons. Right panel: Comparison of the $D$-meson thermal width  at $k=0$ in a pion thermal bath for different calculations. See main text for the different sources.
\label{fig:widths}}
\end{figure}

In the left panel of Fig.~\ref{fig:widths} we show the contributions to the $D$-meson width coming from the different meson baths ($\pi,K,{\bar K},\eta$). As expected, the contribution of more massive mesons is negligible at low temperatures due to the thermal suppression factor, and only the pion term is relevant. At $T=150$ MeV the more massive mesons already contribute with several MeV to the $D$-meson decay width, but are still subdominant with respect the pion one. 

\subsubsection{Comparison with previous approaches}

Finally we compare our results (labelled as ``full''), including unitary and Landau contributions, together with thermal amplitudes, inelastic channels, and off-shell effects, with the calculations of Refs~\cite{Fuchs:2004fh, He:2011yi,Cleven:2017fun}. These are shown in the right panel of Fig.~\ref{fig:widths}. We focus on the thermal width of $D$ mesons coming from the interaction with a thermal bath of only pions, and fix the $D$-meson momentum to $k=0$. Fuchs {\it et al.} \cite{Fuchs:2004fh} use an effective interaction at lowest order between $D$ mesons and pions and extract the width from the self-energy correction due to pions. He {\it et al.} \cite{He:2011yi} use a similar interaction based on Ref.~\cite{Fuchs:2004fh}, but compute the thermal width using a formula similar to Eq.~(\ref{eq:widthT2Uosalt}). The two calculations provide similar results and they are fairly consistent with our results using the unitary cut alone (label ``w/o Landau cut''). Notice that, apart from the different interaction, we also include inelastic channels while He {\it et al.} in Ref.~\cite{He:2011yi} do not. This partially
explains why our curve is slightly larger than the other two. Then, Cleven {\it et al.} \cite{Cleven:2017fun} perform a similar calculation to ours, but the effective approach is based on $SU(4)$ chiral symmetry. Medium effects are also incorporated, including the Landau cut contributions, resulting in a $D$-meson thermal width almost twice larger that the previously discussed two approaches but still smaller than the present results for temperatures higher than 100 MeV, the difference reaching around 30\% at $T=150$~MeV. This is probably due to the fact that the small mass shift of the $D$ meson, which in our model turns out to be attractive, is ignored in the results of Ref.~\cite{Cleven:2017fun}, thereby making them to be less affected by the contributions of the subthreshold Landau cut.

\vspace{5mm}

To summarize, in this section we have analyzed several contributions to the $D$-meson width and found that the effect of off-shell dynamics, inelastic channels and truncation errors are relatively small for the calculation of the $D$-meson thermal width. However the contribution of the Landau cut is essential to describe this coefficient at finite temperatures. We have shown that this contribution appears---thanks to exact unitarity considerations---not only in the imaginary part of the retarded self-energy, but also in the collision term of the kinetic equation. Guided from the results in $\Gamma_k$, we expect that this contribution will be very important for the calculation of the $D$-meson transport coefficients as well.

\section{Off-shell $D$-meson transport coefficients~\label{sec:transport}}

In this section we study the transport coefficients of a $D$ meson, when thermal scattering amplitudes are implemented. To obtain a sensible definition of the relevant transport coefficients we need to go back to the kinetic equation described in Sec.~\ref{sec:Boltzmann} and incorporate the separation of scales between the $D$-meson heavy mass and other scales in the system, to convert the off-shell kinetic equation in Eq.~(\ref{eq:transportG}) into a Fokker-Planck equation~\cite{lifshitz1981physical,Svetitsky:1987gq}.
Once this is done, we are able to identify the so-called drag force $A$, and the diffusion coefficients $B_0,B_1,D_s$, and compute them with thermal effects incorporated.

\subsection{Reduction to an off-shell Fokker-Planck equation~\label{sec:fokker-planck}}

Let us start with the off-shell kinetic equation (see Eq.~(\ref{eq:transportG})) where, for simplicity we keep implicit the sum over scattering channels,
\begin{align}
& \left(  k^\mu - \frac{1}{2} \frac{\pa \textrm{Re } \Pi^R}{\pa k_\mu} \right) \frac{\pa}{\pa X^\mu} G_D^< (X,k) = \nn \\ 
& = \frac{1}{2} \int \prod_{i=1}^3 \frac{d^4k_i}{(2\pi)^4}  (2\pi)^4 \delta^{(4)} (k_1+k_2-k_3-k) |T (k_1^0+k_2^0+i\epsilon, {\bf k}_1 + {\bf k}_2)|^2  \nn \\
&\times  \left[ G_D^<(X,k_1) G_\Phi^<(X,k_2) G_\Phi^>(X,k_3) G_D^>(X,k) - G_D^>(X,k_1) G_\Phi^>(X,k_2) G_\Phi^<(X,k_3) G_D^<(X,k) \right]  \ . \label{eq:kinoff}
\end{align}

Inspired by previous derivations~\cite{lifshitz1981physical,Svetitsky:1987gq,Abreu:2011ic}, we define an off-shell scattering rate
\begin{align}
  W(k^0,{\bf k},k_1^0,{\bf q}) & \equiv \int \frac{d^4k_3}{(2\pi)^4} \frac{d^4k_2}{(2\pi)^4}  (2\pi)^4 \delta (k_1^0+k_2^0-k_3^0-k^0) \delta^{(3)} ({\bf k}_2-{\bf k}_3-{\bf q}) \nn \\
  \times & |T (k_1^0+k_2^0+i\epsilon, {\bf k} - {\bf q} + {\bf k}_2)|^2  G_\Phi^>(X,k_2)G_\Phi^<(X,k_3) G_D^>(X,k_1^0,{\bf k}-{\bf q}) \ , \label{eq:scatrate}
\end{align}
where we have replaced the variable ${\bf k}_1$ by the momentum loss ${\bf q} \equiv {\bf k} - {\bf k}_1$. Equation~(\ref{eq:scatrate}) describes the collision rate of a $D$ meson with energy $k^0$ and momentum ${\bf k}$ to a final $D$ meson with energy $k_1^0$ and momentum ${\bf k}-{\bf q}$. It depends on the spectral weights and the populations of the particles $1,2$ and $3$ of the binary collision, encoded in the Wightman functions. It can be interpreted as a collision loss term for a $D$ meson with momentum ${\bf k}$. In fact, the loss term of Eq.~(\ref{eq:kinoff}) can be directly written as,
\be -\frac12 \int \frac{dk_1^0}{2\pi}\frac{d^3q}{(2\pi)^3} W(k^0,{\bf k},k_1^0,{\bf q}) G_D^<(X,k^0,{\bf k}) \ . \ee

The gain term of Eq.~(\ref{eq:kinoff}) can be interpreted as a loss term of an incoming $D$ meson with momentum ${\bf k}_1$, losing the same momentum amount ${\bf q}$ and ending with momentum ${\bf k}$ (notice that $k^0$ is an independent free variable). This term reads
\begin{align}
& \frac{1}{2} \int \prod_{i=1}^3 \frac{d^4k_i}{(2\pi)^4}  (2\pi)^4 \delta (k^0+k_3^0-k^0_1-k^0_2) 
\delta^{(3)} ({\bf q}+{\bf k}_3-{\bf k}_2) |T (k^0+k_3^0+i\epsilon, {\bf k}+{\bf q} +{\bf k}_3)|^2  \nn \\
&\times   G_D^<(X,k^0,{\bf k}+{\bf q}) G_\Phi^>(X,k_2) G_\Phi^<(X,k_3) G_D^>(X,k^0_1,{\bf k}) \nn \\
&= \frac12  \int \frac{dk_1^0}{2\pi}\frac{d^3q}{(2\pi)^3} W(k^0,{\bf k}+{\bf q},k_1^0,{\bf q}) G_D^<(X,k^0,{\bf k}+{\bf q}) \ . \label{eq:gainasloss}
\end{align}

Then Eq.~(\ref{eq:kinoff}) can be written as
\begin{align}
& 2 \left(  k^\mu - \frac{1}{2} \frac{\pa {\textrm Re } \Pi^R}{\pa k_\mu} \right) \frac{\pa}{\pa X^\mu} G^<_D (X,k)  \nn \\ 
& = \int \frac{dk_1^0}{2\pi} \frac{d^3q}{(2\pi)^3} [  W(k^0,{\bf k}+{\bf q}, k_1^0,{\bf q}) G_D^< (X,k^0,{\bf k}+{\bf q}) - W(k^0,{\bf k},k_1^0,{\bf q}) G_D^<(X,k^0,{\bf k})] \ .   
\end{align}

This equation is an alternative form of Eq.~(\ref{eq:kinoff}), convenient for the formal reduction to the off-shell Fokker-Planck equation. For this purpose we exploit the separation of scales between the meson masses, as the mass of the $D$ meson is much larger than the temperature, and any of the light-meson masses. Such a Brownian picture implies that the typical momentum exchanged in the elastic collision is of the order of $T$ and much smaller than the total momentum of the heavy particle ${\bf q} \ll {\bf k}$~\cite{lifshitz1981physical,Svetitsky:1987gq,Abreu:2011ic}. 

Then we can Taylor expand the combination  $W(k^0,{\bf k}+{\bf q}, k_1^0,{\bf q}) G_D^< (X,k^0,{\bf k}+{\bf q})$ around ${\bf k}$ up to second order. In doing so, we consider a homogeneous thermal bath, as the light sector is assumed to be equilibrated in much shorter time scales, so that one can employ a space-averaged Green's function~\cite{Svetitsky:1987gq}. In addition, we also set $z_k \simeq 1$ as usual.

After a few steps one obtains a Fokker-Planck equation for $G_D^< (t,k^0,{\bf k})$
\be  \frac{\pa}{\pa t} G_D^< (t,k) = \frac{\pa}{\pa k^i} \left\{ \hat{A} (k;T) k^i G_D^< (t,k) + \frac{\pa}{\pa k^j} \left[ \hat{B}_0(k;T) \Delta^{ij} + \hat{B}_1(k;T) \frac{k^i k^j}{{\bf k}^2} \right] G_D^< (t,k) \right\} \ , \label{eq:offFP} \ee
with $\Delta^{ij}=\delta^{ij}-k^ik^j/{\bf k}^2$ and we have defined
\begin{align}
 \hat{A} (k^0, {\bf k};T) & \equiv \left \langle 1 -\frac{{\bf k} \cdot {\bf k}_1}{{\bf k}^2} \right \rangle \ , \label{eq:hatA} \\
 \hat{B}_0 (k^0, {\bf k};T) & \equiv  \frac14 \left \langle {\bf k}_1^2 - \frac{({\bf k} \cdot {\bf k}_1)^2}{{\bf k}^2} \right \rangle \ , \label{eq:hatB0} \\
  \hat{B}_1 (k^0, {\bf k};T) & \equiv \frac12 \left \langle  \frac{[ {\bf k} \cdot ({\bf k}-{\bf k}_1)]^2}{{\bf k}^2} \right \rangle \ , \label{eq:hatB1}
\end{align}
where ${\bf k}_1$ has been reintroduced, replacing ${\bf q}$. The hat is used to denote off-shell transport coefficients, as they depend separately on $k^0$ and ${\bf k}$~\footnote{In our previous works we have denoted the coefficients as $F,\Gamma_0$ and $\Gamma_1$, respectively. To avoid confusion with the thermal width $\Gamma_k$ we have modified the notation to $A,B_0,B_1$, which is also  a common choice in the literature.}. The average is defined as
\begin{align}
\left\langle {\cal F}({\bf k},{\bf k}_1) \right\rangle & = \frac{1}{2k^0} \sum_{\lambda,\lambda'=\pm} \lambda \lambda' \int_{-\infty}^\infty \ dk_1^0 \int \prod_{i=1}^3 \frac{d^3k_i}{(2\pi)^3} \frac{1}{2E_22E_3}  \ S_D(k_1^0,{\bf k}_1)   \nn \\
& \times (2\pi)^4 \delta^{(3)} ({\bf k}+{\bf k}_3-{\bf k}_1-{\bf k}_2) \delta (k^0+\lambda' E_3- \lambda E_2-k^0_1) |T(k^0+ \lambda' E_3,{\bf k}+{\bf k}_3)|^2  \nn \\
& \times f^{(0)} (\lambda'E_3) \tilde{f}^{(0)} (\lambda E_2)  \tilde{f}^{(0)} (k_1^0) \ \ {\cal F}({\bf k},{\bf k}_1)  \ , \label{eq:rateoff}
\end{align}
where the spectral function of the $D$ meson is kept.  We will use the relation of Eq.~(\ref{eq:rateoff}) to compute the off-shell transport coefficients in Eqs. (\ref{eq:hatA}) to (\ref{eq:hatB1}). This approach stands at the same level of Eqs.~(\ref{eq:widthT2U}) and (\ref{eq:widthT2L}), and accounts for thermal modifications, off-shell effects, as well as the Landau cut contributions (case $\lambda'<0$). It will be denoted as ``OffShell'' in the following. This is the most complete calculation of transport coefficients used in this work.

It is important to notice that in general it is not possible to derive a Fokker-Planck equation for $f_D(t,{\bf k})$ including off-shell effects in the transport coefficients, because after $k^0$ integration in both sides, one cannot factorize the distribution function from the transport coefficients. Only in the particular case of a narrow quasiparticle it is possible to trivially integrate $k^0$ and arrive at the kinetic equation for $f_D(t,{\bf k})$, thus reproducing the previous approaches in the literature~\cite{Berrehrah:2013mua,Liu:2018syc}. For the interested reader we detail the derivation of the on-shell Fokker-Planck equation in Appendix~\ref{app:onshell}. 

We remind that, although not explicitly written in Eq.~(\ref{eq:rateoff}), there is a sum over all allowed (elastic+inelastic) channels in these expressions. However, as we have learned from the thermal width, the contribution of the inelastic processes is very small, and therefore they will be neglected in what follows. Nevertheless, all elastic channels ($D\pi$, $DK$, $D\bar{K}$, and $D\eta$) will be added when computing the coefficients.

The described ``OffShell'' approximation, based on Eq.~(\ref{eq:rateoff}), is rather general. However we already know that the quasiparticle approximation is excellent for the $D$ mesons. Therefore one can replace the $D$-meson spectral function by the expression in Eq.~(\ref{eq:Diracdelta}), and neglect the $z_k$ factor altogether. This brings two consequences: 1) The Fokker-Planck equation in Eq.~(\ref{eq:offFP}) for $G^<_D(t,k)$ can be written for $f_D(t,k)$ instead,
\be  \frac{\pa}{\pa t} f_D(t,E_k) = \frac{\pa}{\pa k^i} \left\{ A({\bf k};T) k^i f_D(t,E_k) + \frac{\pa}{\pa k^j} \left[ B_0({\bf k};T) \Delta^{ij} + B_1({\bf k};T) \frac{k^i k^j}{k^2} \right] f_D(t,E_k) \right\} \ , \label{eq:onFP} \ee
where the coefficients only depend on $|{\bf k}|$ as the quasiparticle energy is put on shell,

\begin{align}
 A ({\bf k};T) & \equiv \left \langle 1 -\frac{{\bf k} \cdot {\bf k}_1}{{\bf k}^2} \right \rangle_{ \textrm{Thermal U+L}} \ , \label{eq:hatAon} \\
B_0 ({\bf k};T) & \equiv  \frac14 \left \langle {\bf k}_1^2 - \frac{({\bf k} \cdot {\bf k}_1)^2}{{\bf k}^2} \right \rangle_{ \textrm{Thermal U+L}}\ , \label{eq:hatB0on} \\
B_1 ({\bf k};T) & \equiv \frac12 \left \langle  \frac{[ {\bf k} \cdot ({\bf k}-{\bf k}_1)]^2}{{\bf k}^2} \right \rangle_{ \textrm{Thermal U+L}} \ , \label{eq:hatB1on}
\end{align}
and 2) the scattering rate gets simplified because, as for the Boltzmann equation, only one type of process is able to conserve energy-momentum when all four particles in the collision are on their mass shell. In this approximation it is possible to write,
\begin{align}
\left\langle {\cal F} ({\bf k},{\bf k}_1) \right\rangle_{\textrm{ Thermal U+L}} & = \frac{1}{2E_k}  \int \frac{d^3k_1}{(2\pi)^4} \frac{d^3k_2}{(2\pi)^3} \frac{d^3k_3}{(2\pi)^3}  (2\pi)^4 \delta^{(4)} (k_1+k_2-k_3-k) \nn  \\
  & \times \left[ |T(E_k+E_3,{\bf k}+{\bf k}_3)|^2+|T(E_k-E_2,{\bf k}-{\bf k}_2)|^2 \right] \nn \\
&  \times \frac{1}{2E_1 2E_2 2E_3} f^{(0)} (E_3) \tilde{f}^{(0)} (E_2) \ {\cal F}({\bf k},{\bf k}_1) \ , \label{eq:ratetherm}
\end{align}
where we have only considered elastic collisions, as in Eq.~(\ref{eq:onshellkineticelastic}).

This expression looks closer to the previous calculations of the heavy-flavor transport coefficients, but the Landau contribution still remains in addition to the unitary one. Scattering amplitudes also include medium effects. This approximation to compute the transport coefficients will be denoted as ``Thermal U+L''.

One can yet consider another simplification, in which one simply sets the Landau  contribution to zero. At finite temperature there is no reason to neglect this term, but we will consider this approximation---denoted as ``Thermal U''---for the sake of comparison and to quantify the effect of the Landau cut. In any case, this approximation should be realistic at low temperatures, where the  Landau cut dissapears. The scattering rate to be used in the ``Thermal U'' approximation reads
\begin{align}
\left\langle {\cal F} ({\bf k},{\bf k}_1) \right\rangle_{\textrm{Thermal U}}  & = \frac{1}{2E_k}  \int \frac{d^3k_1}{(2\pi)^4} \frac{d^3k_2}{(2\pi)^3} \frac{d^3k_3}{(2\pi)^3}  (2\pi)^4 \delta^{(4)} (k_1+k_2-k_3-k) \nn \\
  & \times |T(E_k+E_3,{\bf k}+{\bf k}_3)|^2 \frac{1}{2E_22E_3 2E_1} f^{(0)} (E_3) \tilde{f}^{(0)} (E_2) \ {\cal F} ({\bf k},{\bf k}_1)  \ . \label{eq:ratevac}
\end{align}

We should point out that, strictly speaking the previous two approximations partially contain off-shell effects (that is, information about the $D$-meson spectral shape) in the self-consistent calculation of the $T$-matrix, but not in the explicit spectral function in the interaction rate. On the contrary, the ``OffShell'' approximation contains the complete spectral function in both instances.

Finally, to match our results to previous approaches we will simply use Eq.~(\ref{eq:ratevac}) without any thermal effects neither in the quasiparticle energies nor the scattering amplitudes. We will use vacuum amplitudes and standard relativistic expressions for the energies $E_k=\sqrt{k^2+m_{D}^2} , E_1 = \sqrt{k_1^2 + m_{D}^2}$ where $m_{D}$ is the $D$-meson vacuum mass. This approximation is denoted as ``Vacuum'', as it is the one that most resembles our previous calculations.

We summarize in Table~\ref{tab:approxs} the different approximations to compute the $D$-meson transport coefficients. It starts with the simplest one, where vacuum amplitudes without thermal corrections are used, to the most involved one, where thermal and off-shell effects are taken into account.

\begin{table}[ht]
\begin{tabular}{|c|cccc|}
\hline
\hspace{2mm} \makecell{Approximation \\ name} \hspace{2mm} &  \hspace{2mm} Interaction Rate \hspace{2mm} & \hspace{2mm} \makecell{Thermal effects \\ on $|T|^2$ and $E_k$} \hspace{2mm} & \hspace{2mm} Landau cut \hspace{2mm} & \hspace{2mm} Off-shell effects  \hspace{2mm} \\
\hline
Vacuum & Eq.~(\ref{eq:ratevac}) & \ding{55} &  \ding{55} & \ding{55} \\
Thermal U & Eq.~(\ref{eq:ratevac}) & \ding{51} & \ding{55} & \ding{55} \\
Thermal U+L & Eq.~(\ref{eq:ratetherm}) & \ding{51} & \ding{51} & \ding{55} \\
OffShell & Eq.~(\ref{eq:rateoff}) & \ding{51} &\ding{51} & \ding{51} \\
\hline
\end{tabular}
\centering
\caption{Different approximations for the computation of the $D$-meson transport coefficients used in this work. Details are given in the main text.}
\label{tab:approxs}
\end{table}

\section{Results for $D$-meson transport coefficients~\label{sec:results}}

We start the description of our numerical results with the $D$-meson drag force $A({\bf k};T)$ (or $\hat{A}(k_0,{\bf k};T)$ in the off-shell case), and the momentum diffusion coefficient $B_0({\bf k};T)$ ($\hat{B}_0 (k_0,{\bf k};T)$ for the off-shell approximation). We will present results in the so-called {\it static limit} ${\bf k} \rightarrow 0$ ($|{\bf k}|=50$ MeV in the actual computation). In this limit $B_0=B_1$, which we have checked numerically in all cases.

\begin{figure}[ht]
  \centering
\includegraphics[width=75mm]{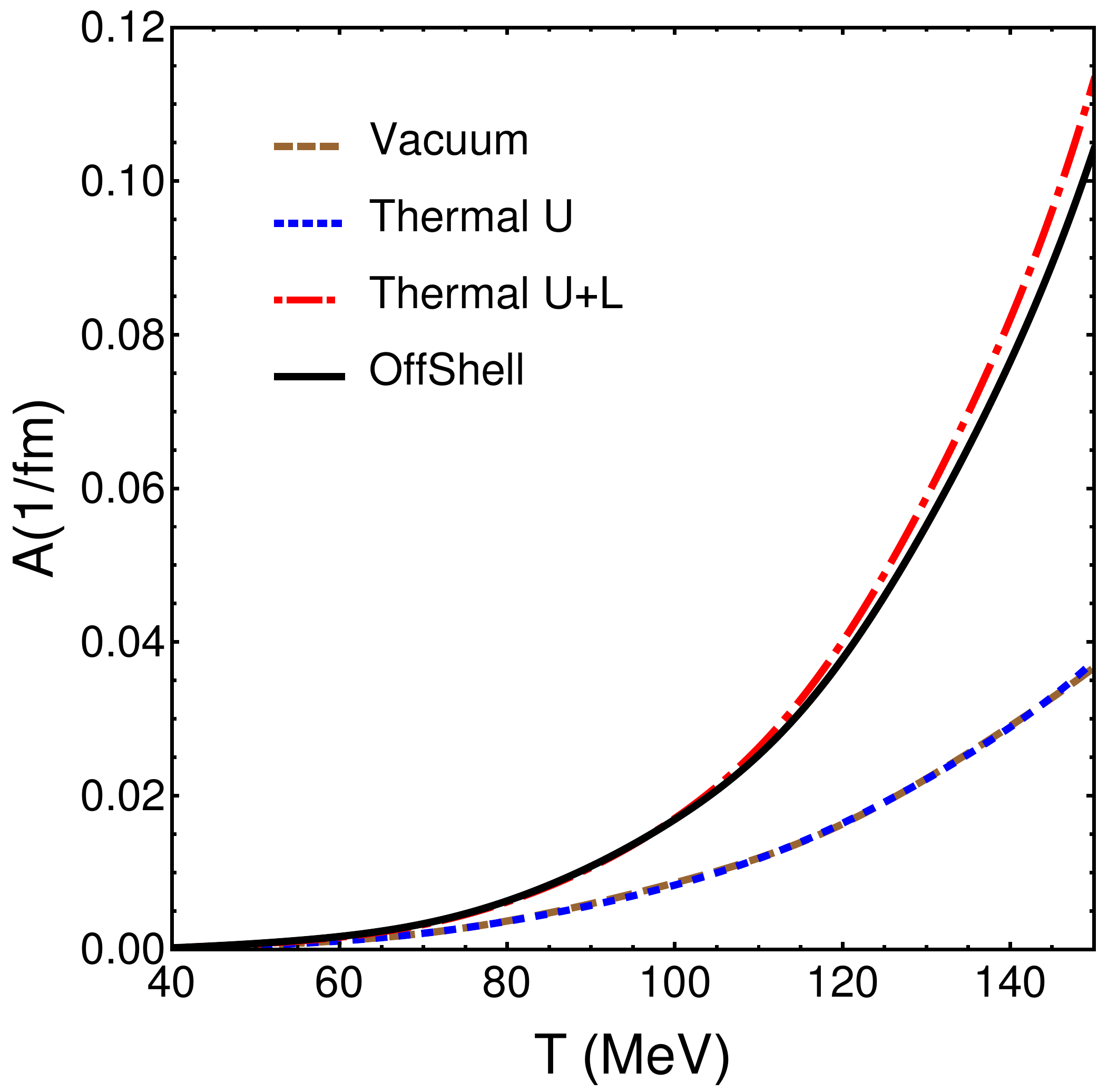}
\includegraphics[width=75mm]{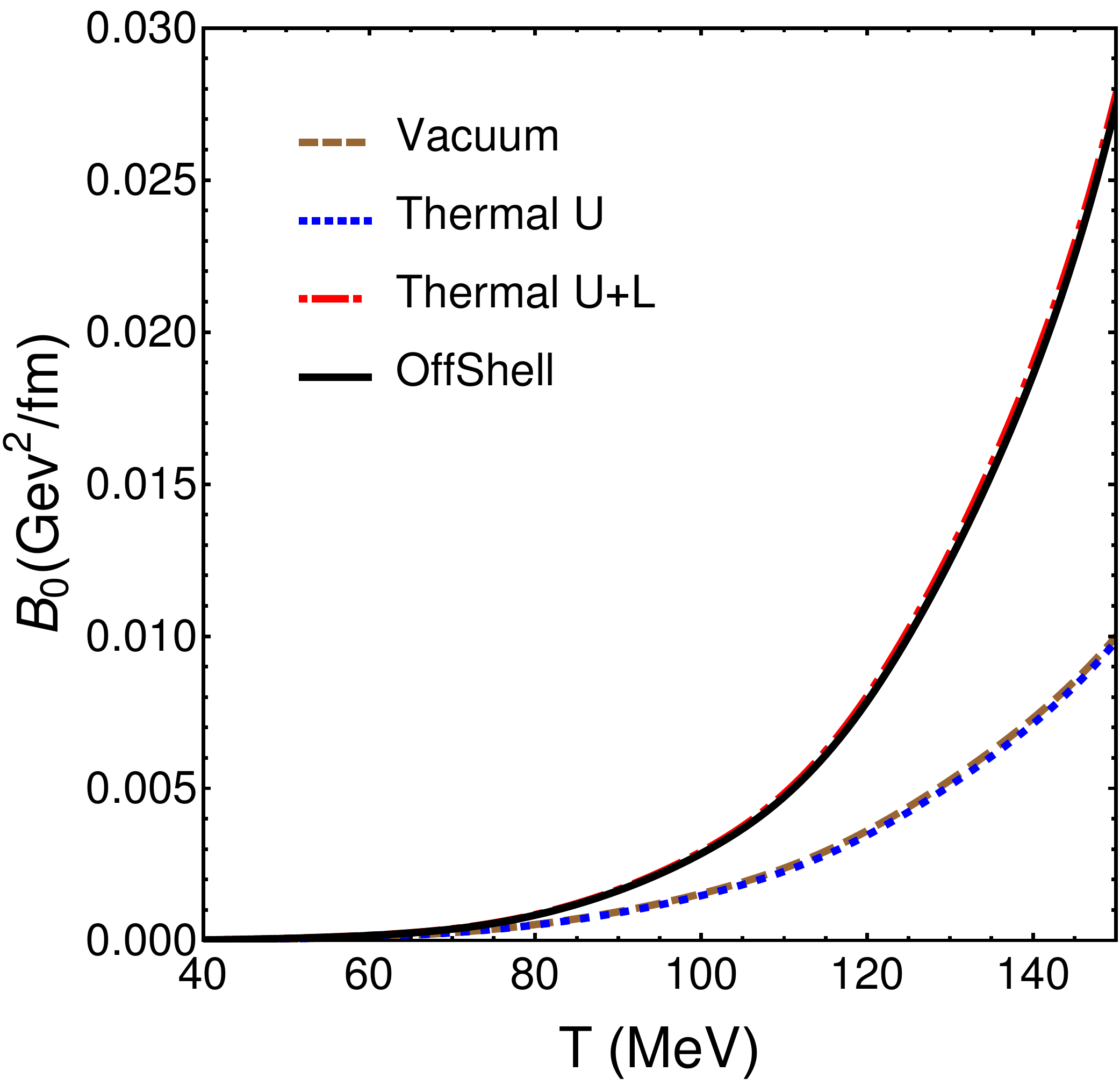}
\caption{
$D$-meson transport coefficients in the static limit ${\bf k}\rightarrow 0$ (where $B_1=B_0$) using the different approximations described in Table~\ref{tab:approxs} (see also main text for details). The curve that incorporates all the thermal and off-shell effects is the one denoted as ``OffShell''.
}

  \label{fig:transportcomp}
\end{figure}

In Fig.~\ref{fig:transportcomp} we present the drag force $A$ (left panel) and the diffusion coefficient $B_0$ (right panel) under the different approximations of Table~\ref{tab:approxs}. ``Vacuum'' corresponds to the approximation used in our previous work~\cite{Tolos:2013kva}, where vacuum scattering amplitudes and masses were employed. 

All the remaining approximations use thermal scattering amplitudes and temperature-dependent quasiparticle energies. ``Thermal U'' only incorporates the unitary cut, and the differences with respect to ``Vacuum'' are entirely due to thermally dependent interactions and masses. Rather surprisingly, we find no appreciable differences with respect to ``Vacuum'' even at high temperatures. The main difference comes when we add the Landau contribution, which is incorporated in the ``Thermal U+L'' scenario. At our top temperatures, the contribution of this cut is even more important than the one of the unitary cut. This was already anticipated in the $D$-meson thermal width in Sec.~\ref{sec:kinematic}.

Finally we present our results incorporating off-shell effects, which employ the full spectral distribution of the in-medium $D$ meson. This approximation is denoted ``OffShell'' in Table~\ref{tab:approxs}. In this case, to fix the external energy dependence we have simply set $k^0=E_k$ with $|{\bf k}|=50$ MeV (static limit).
Only a small difference can be observed in $A$ at high temperatures with respect to the ``Thermal U+L'' approximation, concluding that the genuine off-shell effects are not as important as including the Landau cut contribution (the same happened for the thermal width in Fig.~\ref{fig:Gammaonoff}). This result is not very surprising as the $D$-meson spectral function is still very narrow for the temperatures  considered here, so the quasiparticle approximation works extremely well. As in the ``Thermal U+L'' case, the ``OffShell'' approximation presents a substantial contribution of the Landau cut to the transport coefficients, absent in the vacuum case.
 
In the off-shell approximation, when the $D$ meson carries a finite thermal width, $1 \leftrightarrow 3$ processes are also allowed. In this work we have neglected those because the required production threshold is higher than the elastic one. However, it would be very interesting to analyze the Bremsstrahlung processes $D\rightarrow D+\pi+\pi$ and their role in the $D$-meson energy loss. This is left for a future work.

We finally explore the spatial diffusion coefficient $D_s (T)$~\cite{Abreu:2011ic}. This coefficient [usually normalized by the thermal wavelength, $1/(2\pi T)$] can be obtained from the static limit of the $B_0({\bf k}; T)$ coefficient,
\be 2\pi T D_s (T) = \lim_{{\bf k} \rightarrow 0} \frac{2\pi T^3}{B_0({\bf k};T)} \ . \label{eq:Dscoeff}\ee

\begin{figure}[ht]
  \centering
\includegraphics[width=75mm]{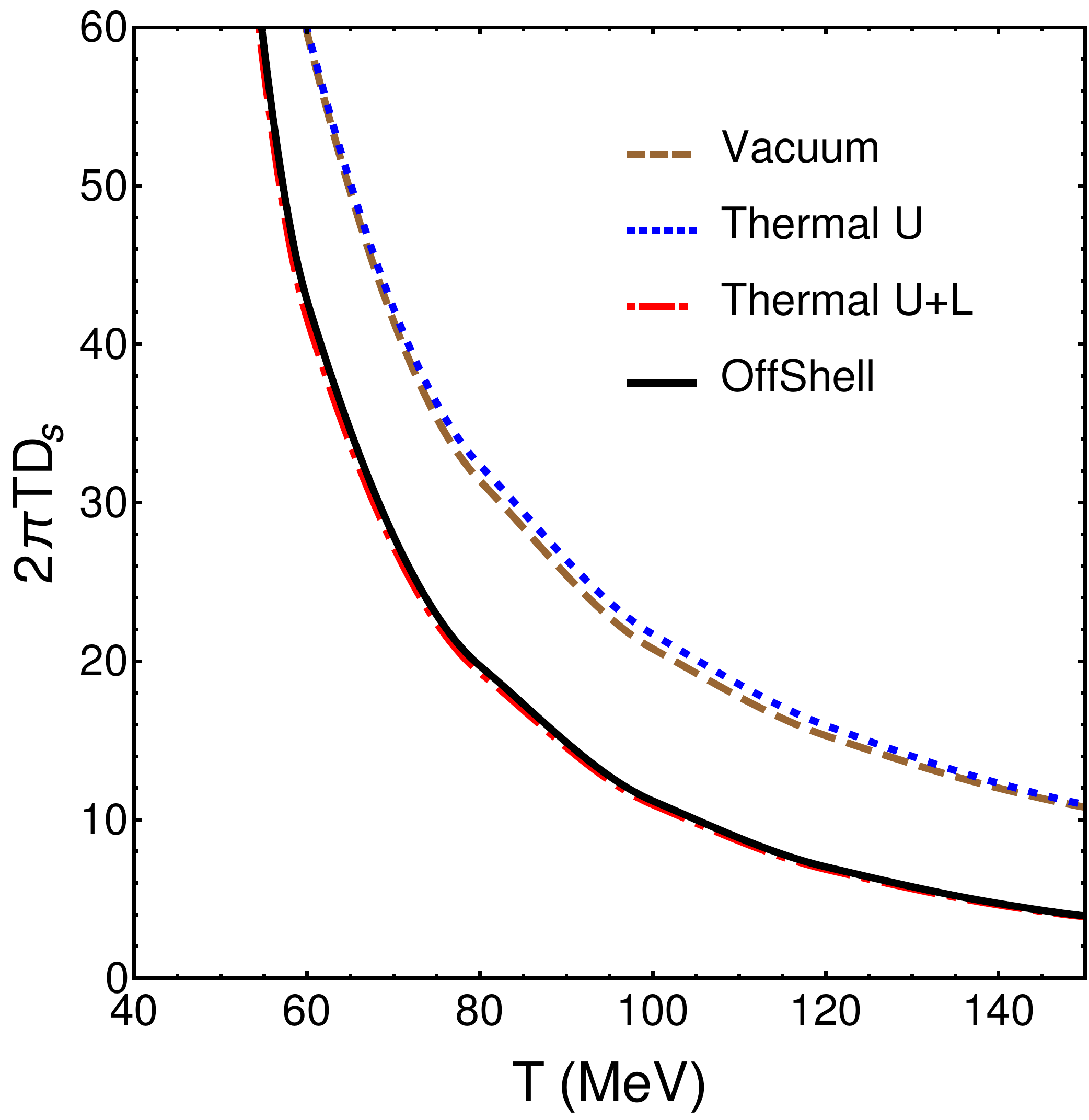}
\includegraphics[width=75mm]{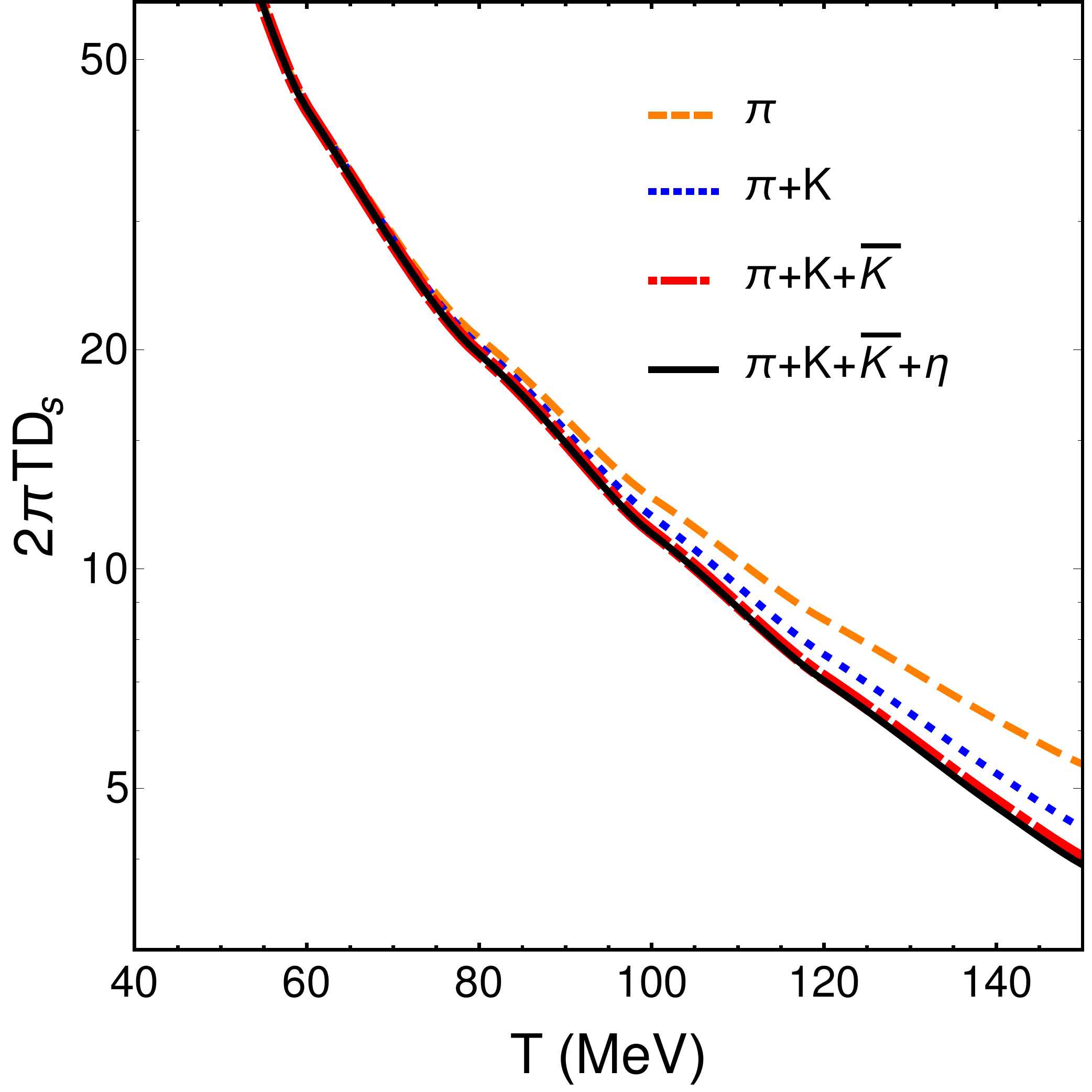}
 \caption{
 Left panel: Spatial diffusion coefficient [divided over the thermal wavelength $(2\pi T)^{-1}$] using the different approximations in Table~\ref{tab:approxs} (see also main text for details). The ``OffShell'' curve is the one incorporating all the thermal and off-shell effects. Right panel: Same coefficient in the ``OffShell'' approximation incorporating sequentially the different light mesons in the calculation.}
  \label{fig:Dscomp}
\end{figure}

This coefficient is shown in Fig.~\ref{fig:Dscomp}. In accordance with the previous transport coefficients, the main difference comes from the Landau contribution, which makes the diffusion coefficient to decrease almost by a factor of 3 close to $T_c$, which is a remarkable effect.  The results for the ``Thermal U+L'' and ``OffShell' are almost identical. 

Let us comment that an alternative way to fix the $k^0$ dependence of the off-shell transport coefficients is to define an average coefficient weighted by the $D$-meson spectral function. For example, one could define an average $\overline{B_0}$ as,
\be \overline{B_0} ({\bf k};T) = 2 \int_0^\infty dk^0 k^0 S_D(k^0,{\bf k}) \hat{B}_0 (k^0,{\bf k};T) \ . \label{eq:avB0} \ee

In the narrow quasiparticle limit this average coincides with the on-shell evaluation if one uses $z_k \simeq 1$ in addition,
\be \overline{B_0} ({\bf k};T) = \int_0^\infty dk^0 2k^0 \frac{z_k}{2 E_k} \delta (k^0-E_k) \hat{B}_0 (k^0,{\bf k};T) \simeq \hat{B}_0 (E_k,{\bf k};T)  \ . \ee 
Because in our case the quasiparticle approximation works very well the evaluation of the off-shell coefficient at $k_0 = E_k$ gives similar results to the one using Eq.~(\ref{eq:avB0}).

To conclude this section we detail the different contributions of adding the light mesons one by one. In Fig.~\ref{fig:widths} we showed how the pions provided the main contribution to the thermal decay width at temperatures below $T_c$, being the $K,\bar{K}$ and $\eta$ mesons subleading even at $T\simeq 150$ MeV. In the right panel of Fig.~\ref{fig:Dscomp} we present the spatial diffusion coefficient in the ``OffShell'' approximation, when the light mesons are sequentially added. One also observes that the contribution of the more massive states is very small due to their thermal suppresion (the negligible contribution of baryons at $\mu_B=0$ was studied in Ref.~\cite{Tolos:2013kva}). However, one should keep in mind that close to $T_c$ one could expect the excitation of many states and resonances which can collectively contribute to the transport coefficient in a substantial way (see Ref.~\cite{Noronha-Hostler:2008kkf} for the shear and bulk viscosities). Therefore, our predictions might not be trustable in the $T \simeq T_c$ region, so our results are shown up to $T = 150$ MeV.

\subsection{Comparison with other approaches}

To conclude this study we compare our results below $T_c$ to recent lattice-QCD calculations of the heavy-flavor transport coefficients for temperatures $T \gtrsim T_c$. We also include a recent calculation using Bayesian methods to analyze the HiC data under a simulation code to obtain a posterior estimation of the spatial diffusion coefficient~\cite{Ke:2018tsh}. In our case we will show our most complete calculation (``OffShell'' approximation) together with the ``Vacuum'' calculation for comparison. From the lattice-QCD side we compile the results presented in Refs.~\cite{Banerjee:2011ra,Kaczmarek:2014jga,Francis:2015daa, Brambilla:2020siz,Altenkort:2020fgs}. All these are given as functions of $T/T_c$. To compare the different results in terms of an absolute temperature, we fix $T_c=156$ MeV~\cite{Bazavov:2018mes}.

\begin{figure}[ht]
  \centering
  \includegraphics[width=75mm]{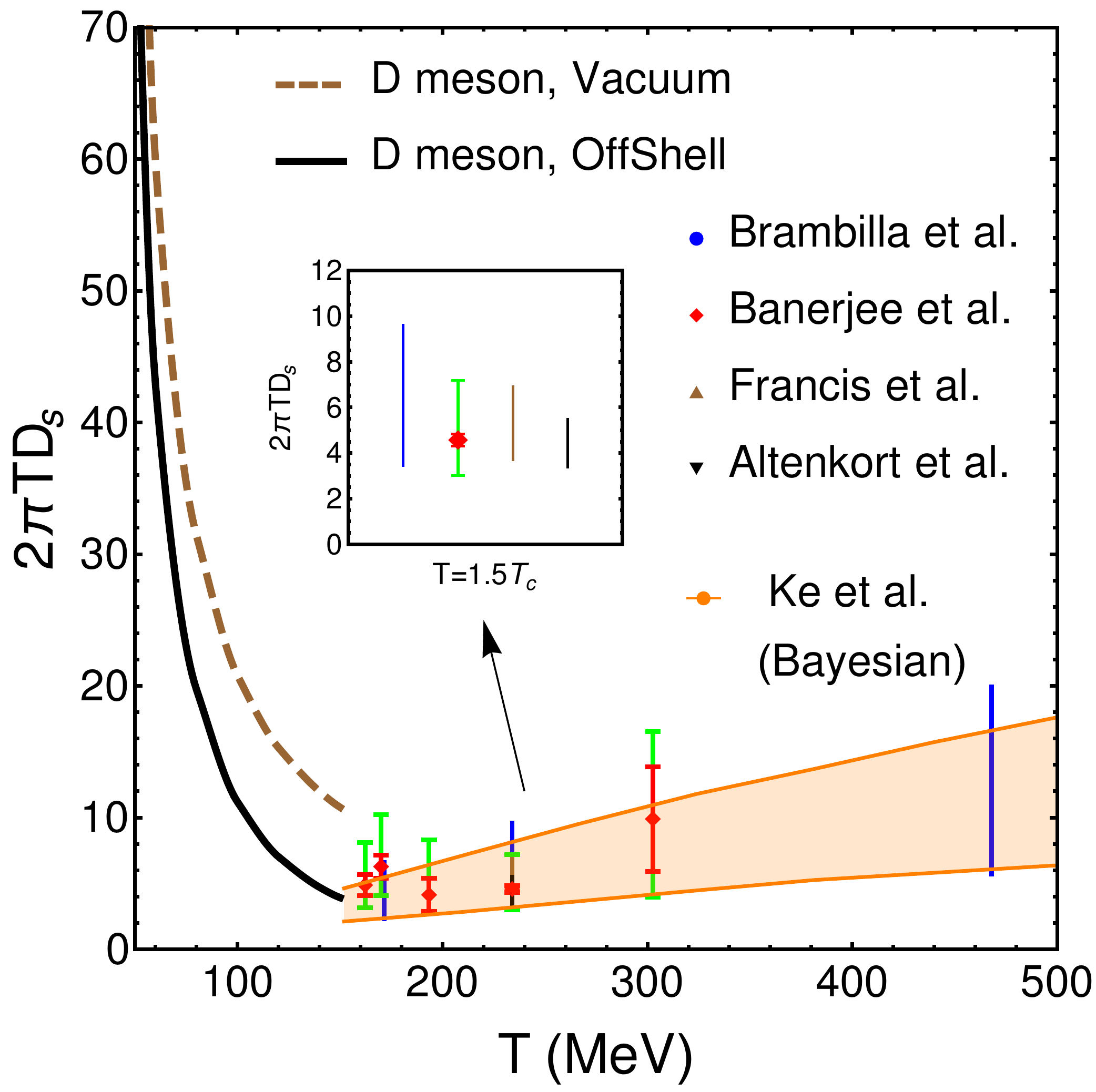}
   \includegraphics[width=75mm]{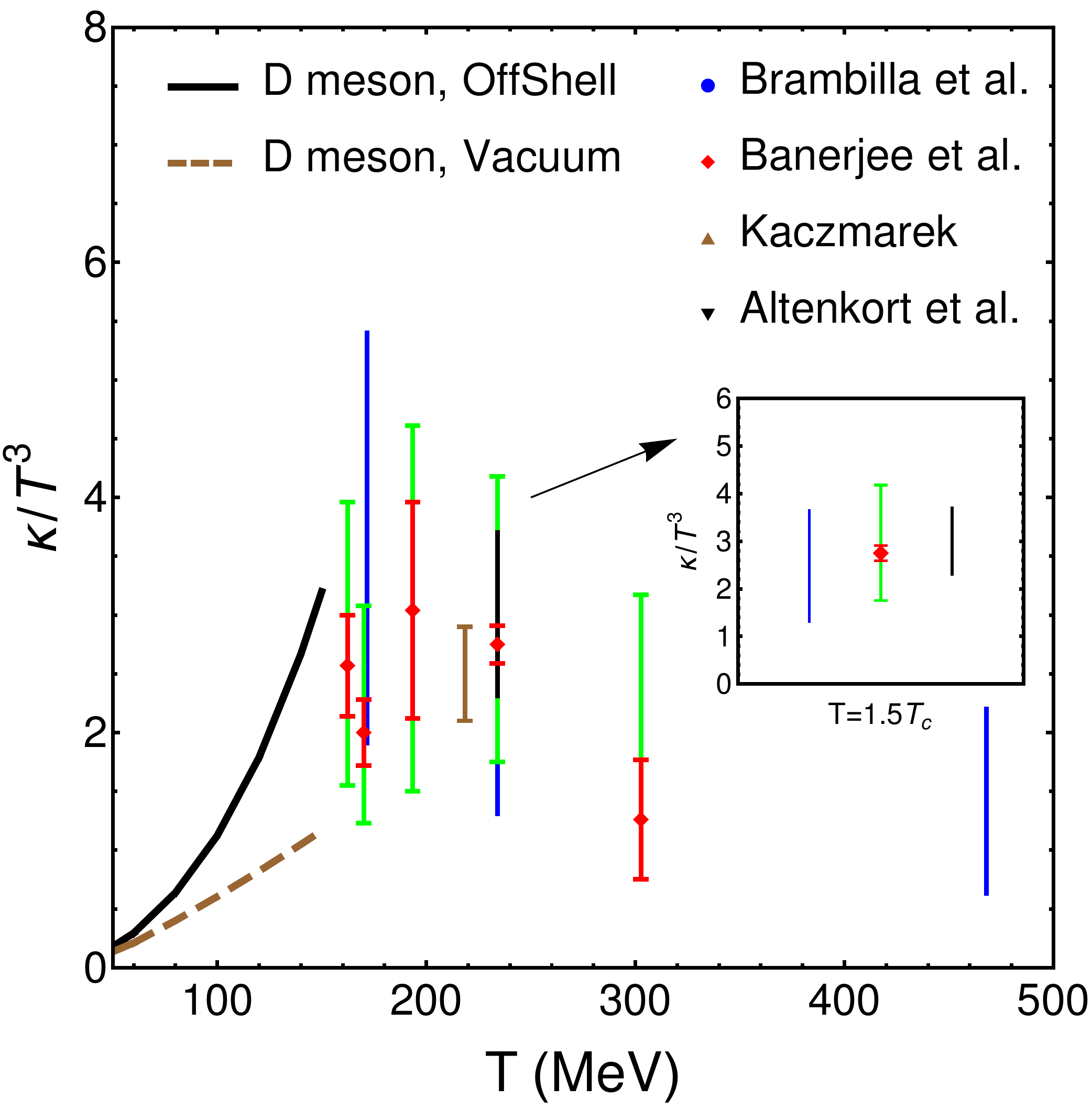}
  \caption{Left panel: Spatial diffusion coefficient (normalized by the thermal wavelength) around $T_c$. Right panel: Momentum diffusion coefficient $\kappa/T^3=2B_0/T^3$ around $T_c$.}
  \label{fig:latticecomp}
\end{figure}

In the left panel of Fig.~\ref{fig:latticecomp} we show the spatial diffusion coefficient as defined in Eq.~(\ref{eq:Dscoeff}). In the right panel we present the momentum diffusion coefficient $\kappa$ as it is usually defined in the lattice-QCD community. This coefficient is related to $B_0$ in the static limit as,
\be \kappa(T) = 2 B_0 ( {\bf k}\rightarrow 0;T)\ . \ee
In fact this coefficient is not independent of $D_s$ as $\kappa=4\pi T^3/(2\pi T D_s)$. Nevertheless we provide the results of $\kappa/T^3$ to stress the plausible matching, where a possible maximum happens at the crossover temperature.

Details of the different lattice-QCD calculations can be found in their corresponding publications. All of them are characterized by the use of the quenched approximation ($SU(3)$ pure glue plasma) and with different ranges of temperature (Refs.~\cite{Kaczmarek:2014jga,Francis:2015daa,Altenkort:2020fgs} only provide results for a characteristic temperature of $1.5T_c$). With the exception of the calculation in~\cite{Banerjee:2011ra} all results take the lattice continuum limit. The majority of the calculations use a multilevel update to reduce noise~\cite{Banerjee:2011ra,Kaczmarek:2014jga,Francis:2015daa,Brambilla:2020siz}, except for the most recent Ref.~\cite{Altenkort:2020fgs} which employs gradient flow.

We observe a very good matching around $T_c$ among our results, the lattice-QCD data and the result from a Bayesian analysis~\cite{Ke:2018tsh}, especially for the case with thermal and off-shell effects included. This is better seen for the $\kappa/T^3$ coefficient. Nevertheless, as commented before, our results are not able to capture the increase of hadronic states and resonances which are excited close to $T_c \simeq 156$ MeV. While we have shown that the individual contribution of more massive hadrons is tiny,  the total (iso)spin degeneracy of such states could compensate their thermal suppression, and produce an additional decrease of the diffusion coefficient at $T_c$~\cite{Noronha-Hostler:2008kkf}. On the other hand, the results coming from the calculations at $T>T_c$ also suffer from sizable uncertainties in the crossover region, and cannot also be fully trusted there.

We now comment on the comparison of ``Vacuum'' approximation to our previous calculation in Ref.~\cite{Tolos:2013kva} using vacuum amplitudes. In that work we reported a similar diffusion coefficient with slightly smaller values at high temperatures, e.g. at $T=150$ MeV the value $2\pi TD_s \simeq 8$ (here we obtain $2\pi TD_s \simeq 12$). In turn, the $A$ and $B_0,B_1$ coefficients were systematically larger in Ref.~\cite{Tolos:2013kva}. The differences come from several improvements with respect to that work: 1) The LECs of the effective Lagrangian are fixed here thanks to recent lattice-QCD calculations~\cite{Guo:2018tjx}, while in Ref.~\cite{Tolos:2013kva} we followed a less rigorous procedure of fixing the LECs by matching the mass and width of the $D(2400)$ resonance (whose properties reported by the PDG, in turn, have changed since then); and 2) here we adopt a full consistent coupled-channel approach, while in Ref.~\cite{Tolos:2013kva} this was done only partially (for example, the channels involving $D_s$ were not considered there). 
 
We should finally mention that for $T>T_c$ we have only shown the results coming from lattice QCD and Bayesian calculations, but there exist many theoretical calculations of these coefficients using different models or effective approaches~\cite{Moore:2004tg,vanHees:2004gq,vanHees:2007me,Das:2010tj,He:2011yi,Mazumder:2011nj,Das:2012ck,Berrehrah:2013mua,Liu:2018syc,Berrehrah:2014kba,Berrehrah:2014tva}.

\section{Conclusions~\label{sec:conclusions}}

In this work we have extended the kinetic theory description of $D$ mesons at low energy to include medium (thermal) effects and spectral properties of open-charm states~\cite{Montana:2020lfi,Montana:2020vjg}. In particular, we have derived the off-shell Boltzmann and Fokker-Planck equations from the $D$-meson effective-field theory in the $T$-matrix approximation.

As an application, we have calculated the $D$-meson thermal width (or damping rate), the drag force coefficient, the diffusion coefficients in momentum space, and the spatial diffusion coefficient. These transport coefficients were previously computed using vacuum amplitudes employing on-shell kinetic equations~\cite{Tolos:2013kva}. Here we achieve a consistent formulation between in-medium interactions with light mesons and an off-shell kinetic approach.

Due to their large vacuum mass, thermal corrections and spectral broadening of $D$ mesons are relatively small, so that the quasiparticle picture is maintained for the temperature range considered in this work. For this reason, it is still appropriate to consider a Fokker-Planck reduction of the Boltzmann equation to describe this system. We have derived an off-shell Fokker-Planck equation, where the kinetic coefficients are now interpreted as ``off-shell'' transport coefficients. We have shown that, in general, there is no off-shell Fokker-Planck extension for the $D$-meson distribution function $f_D(t,{\bf k})$ [only for the Wightman function $G^<_D(t,k^0,{\bf k})$], unless the extreme narrow limit of the spectral functions is used. In this limit a clear connection to previous approaches can be made. 

Our main observation is that the use of thermal scattering amplitudes causes the appearance of a new kinematic range in the meson-meson interaction, the so-called Landau contribution. We have found that the contribution to the transport coefficients is rather large at moderate temperatures, even for on-shell $D$ mesons. In fact, it dominates all dissipative coefficients at temperatures above 100 MeV, and at $T=150$ MeV this new contribution is as large as the standard contribution due to the unitary cut. While these new effects modify substantially the transport coefficients close to $T_c$, the final results are consistent with lattice-QCD determinations of the momentum and spatial diffusion coefficients. For a fully consistent matching with some  lattice-QCD results (which use the infinite quark mass limit) it is necessary to increase the mass of our heavy hadron. We leave for a future work the calculation using $B$ mesons~\cite{Das:2011vba,Abreu:2012et,Torres-Rincon:2014ffa,Pathak:2014nfa}, and the analysis of the validity of heavy-quark flavor symmetry at finite temperatures. As an alternative (and complementary) approach to the Fokker-Planck equation, it would be interesting to perform a study of the Langevin equation from the perspective of the Kadanoff-Baym formalism~\cite{Greiner:1998vd}.

\begin{acknowledgments}

We acknowledge Hendrik van Hees and \'Angel G\'omez-Nicola for useful discussions about the physics contained in this work. We thank Weiyao Ke for providing the numerical results from the Bayesian analysis of Ref.~\cite{Ke:2018tsh}.
 
G.M. and A.R. acknowledge support  from the Projects No. CEX2019-000918-M (Unidad de Excelencia ``Mar\'{\i}a
de Maeztu") and No. PID2020-118758GB-I00 financed by MCIN/AEI/ 10.13039/501100011033/. G.M. also acknowledges support from the FPU17/04910 Doctoral Grant from the Spanish Ministerio de Educaci\'on, Cultura y Deporte (MECD). The research of L.T. has been also supported  by  PID2019-110165GB-I00  financed by MCIN/AEI/10.13039/501100011033 and the THOR COST Action CA15213. L.T. and J.M.T.-R. acknowledge support from the DFG through Projects No. 411563442 (Hot Heavy Mesons) and No. 315477589 - TRR 211 (Strong-interaction matter under extreme conditions). 
This article is part of a project that has received funding from the
European Union's Horizon 2020 research and innovation programme under
Grant Agreement STRONG--2020 - No 824093.

\end{acknowledgments}

\appendix

\section{On-shell Fokker-Planck equation~\label{app:onshell}} 

In Sec.~\ref{sec:fokker-planck} we have obtained the (off-shell) Fokker-Planck equation for the $D$-meson (homogeneous) Wigner function $G_D^<(t,k^0,{\bf k})$. We reproduce it here again for convenience,
\begin{align} 
\frac{\pa}{\pa t} G_D^< (t,k^0,{\bf k}) &= \frac{\pa}{\pa k^i} \left\{ \hat{A}(k^0,{\bf k};T) k^i G_D^< (t,k^0,{\bf k}) \right. \nn \\ 
& \left. + \frac{\pa}{\pa k^j} \left[ \hat{B}_0(k^0,{\bf k};T) \Delta^{ij} + \hat{B}_1(k^0,{\bf k};T) \frac{k^i k^j}{{\bf k}^2} \right] G_D^< (t,k^0,{\bf k})\right\} \ , \label{eq:FPforG} 
\end{align}
with the ``off-shell'' coefficients defined in Eqs.~(\ref{eq:hatA}), (\ref{eq:hatB0}) and (\ref{eq:hatB1}),
\begin{align}
 \hat{A}(k^0,{\bf k};T) & = \frac{1}{2k^0} \int  \frac{dk_1^0}{2\pi} \frac{d^3q}{(2\pi)^3} W(k^0,{\bf k}, k_1^0,{\bf q}) \ \frac{{\bf q} \cdot {\bf k}}{{\bf k}^2} , \\
 \hat{B}_0 (k^0,{\bf k};T) & = \frac{1}{2k^0}\frac{1}{4} \int \frac{dk_1^0}{2\pi} \frac{d^3q}{(2\pi)^3} W(k^0,{\bf k}, k_1^0,{\bf q}) \ \left[ {\bf q}^2 - \frac{({\bf q} \cdot {\bf k})^2}{{\bf k}^2} \right] , \\
  \hat{B}_1 (k^0,{\bf k};T) & = \frac{1}{2k^0}\frac{1}{2} \int \frac{dk_1^0}{2\pi} \frac{d^3q}{(2\pi)^3}  W(k^0,{\bf k}, k_1^0,{\bf q}) \ \frac{({\bf q} \cdot {\bf k})^2}{{\bf k}^2} \ ,
\end{align}
where we have expressed the average in terms of the integration of a scattering rate $W$ integrated over the transferred momentum. These equations follow immediately from the Fokker-Planck reduction of the transport equation.

The scattering rate reads
\begin{align}
  W(k^0,{\bf k},k_1^0,{\bf q}) & \equiv \int \frac{d^4k_2}{(2\pi)^4} \frac{d^4k_3}{(2\pi)^4}  (2\pi)^4 \delta (k_1^0+k_2^0+k_3^0-k^0) \delta^{(3)} ({\bf k}_2+{\bf k}_3-{\bf q}) \nn \\
  & \times |T (k_1^0+k_2^0+i\epsilon, {\bf k} - {\bf q} + {\bf k}_2)|^2  G_\Phi^>(k_2^0,{\bf k}_2)G_\Phi^<(k_3^0,{\bf k}_3) G_D^>(k_1^0,{\bf k}-{\bf q}) \ .
\end{align}

We stress again that on the integration over $dk^0/(2\pi)$ of Eq.~(\ref{eq:FPforG}) it is not possible to obtain the standard Fokker-Planck equation with ``on-shell'' coefficients depending only on ${\bf k}$, due to the presence of a $D$ meson with a generic spectral function. To match the previous results and derive the ``on-shell'' version of the coefficients one needs to apply the Kadanoff-Baym ansatz of Eqs.~(\ref{eq:ansatz1}) and (\ref{eq:ansatz2}), and particularize for the narrow quasiparticle limit of the spectral function in Eq.~(\ref{eq:Diracdelta})
with $z_k \simeq 1$,
\be G^<_D (t,k^0,{\bf k}) = \frac{2\pi}{2E_k} [\delta(k^0-E_k)-\delta(k^0+E_k)] f_D(t,k^0) \ , \ee 
and similarly for particles 1, 2 and 3.

Then, after integration on the positive branch of $k^0$ one is able to obtain the Fokker-Planck equation for $f_D(t,{\bf k})$~\footnote{We slightly abuse of notation here, as it should strictly read $f(t,E_k)$.}:
\be \frac{\pa}{\pa t} f_D (t,{\bf k}) =  \frac{\pa}{\pa k^i} \left\{ k^i A({\bf k};T) f_D (t,{\bf k}) + \frac{\pa}{\pa k^j} \left[ B_0({\bf k};T) \Delta^{ij} + B_1({\bf k};T) \frac{k^i k^j}{k^2} \right] f_D (t,{\bf k}) \right\} \ , \ee
where the coefficients read
\begin{align}
A( {\bf k};T) & = \int \frac{d^3q}{(2\pi)^3} w({\bf k}, {\bf q}) \ \frac{{\bf q} \cdot {\bf k}}{{\bf k}^2} \ , \label{eq:onshellA} \\ 
 B_0 ({\bf k};T) & = \frac{1}{4} \int \frac{d^3q}{(2\pi)^3} w({\bf k},{\bf q}) \ \left[ {\bf q}^2 - \frac{({\bf q} \cdot {\bf k})^2}{ {\bf k}^2} \right] \ ,  \label{eq:onshellB0} \\
  B_1 ({\bf k};T) & = \frac{1}{2} \int \frac{d^3q}{(2\pi)^3}  w({\bf k}, {\bf q}) \ \frac{({\bf q} \cdot {\bf k})^2}{ {\bf k}^2} \ . \label{eq:onshellB1}
\end{align}
We have introduced the ``on-shell'' scattering rate
\be  w({\bf k},{\bf q}) \equiv \frac{1}{2E_k} \int \frac{dk_1^0}{2\pi} W(E_k,{\bf k},k_1^0,{\bf q}) \ ,  \ee
which in terms of the scattering amplitude reads
\begin{align} 
  w({\bf k},{\bf q}) & = \int \frac{d^3k_3}{(2\pi)^6} f_\Phi^{(0)}({\bf k}_3) \tilde{f}_\Phi^{(0)}({\bf k_3}+{\bf q}) \frac{1}{2E_k 2E_{k_3} 2E_{k+q} 2E_{k_3+q}} \nn \\
  &\times (2\pi)^4 \delta(E_k+E_{k_3}-E_{k+q}-E_{k_3+q}) \nn \\
  &\times \left[ |T(E_k+E_3,{\bf k}+{\bf k}_3)|^2+|T(E_k-E_{k_3+q},{\bf k}-{\bf k}_3-{\bf q})|^2 \right] \ . \label{eq:onshellw}
\end{align}
The expressions of the coefficients in Eqs.~(\ref{eq:onshellA}), (\ref{eq:onshellB0}) and (\ref{eq:onshellB1}) together with the on-shell scattering rate in Eq.~(\ref{eq:onshellw}) coincide with those used in our previous works~\cite{Abreu:2011ic,Tolos:2013kva,Tolos:2016slr}, apart from the Landau term arising in Eq.~(\ref{eq:onshellw}), which is the new contribution found in the present work.

\bibliographystyle{ieeetr}
\bibliography{D-meson}

\end{document}